\documentclass[journal]{IEEEtran}

\usepackage{graphics} 
\usepackage{epsfig} 
\usepackage{times} 
\usepackage{amsmath} 
\usepackage{amssymb}  
\usepackage{microtype}
\usepackage{graphicx}
\usepackage{subfig}
\usepackage{booktabs} 
\usepackage{amsfonts}
\usepackage{mathrsfs}
\usepackage{doi}
\usepackage{bm}
\usepackage{float}
\usepackage{hyperref}

\usepackage{algorithm}
\usepackage{algorithmic}
\usepackage{xcolor}
\usepackage{cite}
\usepackage{amsthm,amssymb}
\usepackage{bbm}
\usepackage{chngcntr}

\newtheorem{theorem}{Theorem}
\newtheorem{lemma}{Lemma}
\newtheorem{remark}{Remark}
\newtheorem{definition}{Definition}
\newtheorem{assumption}{Assumption}

\title{\LARGE \bf
Model-Reference Reinforcement Learning for Collision-Free Tracking Control of Autonomous Surface Vehicles}

\author{Qingrui Zhang$^{1,2}$,  Wei Pan$^{2}$, and Vasso Reppa$^{1}$
\thanks{$^{1}$Department of Maritime and Transport Technology, Delft University of Technology, Delft, the Netherlands
        {\tt\small Qingrui.Zhang@tudelft.nl; V.Reppa@tudelft.nl}}%
\thanks{$^{2}$Department of Cognitive Robotics, Delft University of Technology, Delft, the Netherlands
        {\tt\small Wei.Pan@tudelft.nl}}%
}

\begin{document}

\markboth{Journal of \LaTeX\ Class Files,~Vol.~XX, No.~XX, XX~XXXX}%
{Shell \MakeLowercase{\textit{et al.}}: Bare Demo of IEEEtran.cls for IEEE Journals}

\maketitle

\begin{abstract}
This paper presents a novel model-reference reinforcement learning algorithm for the intelligent tracking control of uncertain autonomous surface vehicles with collision avoidance. The proposed control algorithm combines a conventional control method with reinforcement learning to enhance control accuracy and intelligence. In the proposed control design, a nominal system is considered for the design of a baseline tracking controller using a conventional control approach. The nominal system also defines the desired behaviour of uncertain autonomous surface vehicles in an obstacle-free environment.  Thanks to reinforcement learning, the overall tracking controller is capable of compensating for model uncertainties and achieving collision avoidance at the same time in environments with obstacles.  In comparison to traditional deep reinforcement learning methods, our proposed learning-based control can provide stability guarantees and better sample efficiency. We demonstrate the performance of the new algorithm using an example of autonomous surface vehicles.

\end{abstract}


\section{INTRODUCTION}
Autonomous surface vehicles (ASVs) have attracted extensive research attention, due to their advantages in many applications, such as environmental monitoring \cite{Jones2019STTE}, resource exploration \cite{Majohr2006}, enhancing the efficiency and safety of waterborne transportation \cite{Levander2017IEEE,Tu2018TITS}, and many more \cite{Svec2014AR,Bloisi2017TITS}. Successful launch of ASVs in real life requires avoiding collisions with obstacles \cite{Campbel2013ARC} and accurate tracking along a desired trajectory \cite{Do2006Auto}. Both collision avoidance  and tracking control are the major research topics for ASVs in the maritime engineering \cite{Soltan2009ACC, Ramon2009TITS, Perera2012TITS, Kahveci2013Auto, Woo2019OE, Johansen2016TITS, Tu2018TITS, Li2019OE}.  However, accurate tracking control for ASVs in the presence of obstacles is challenging, as ASVs are subject to uncertain nonlinear hydrodynamics and unknown environmental disturbances \cite{Fossen2011Handbook}.  Due to the complexity of the  problem, collision avoidance and tracking control are mostly studied in a separate manner.

Collision avoidance methods for ASVs are categorized into path/motion planning approaches \cite{Greytak2009CDC, Hover2012IJRR, Gonzalez2016TITS,Chiang2018RAL} and optimization-based algorithms \cite{Hausler2012IFAC,Johansen2016TITS, Abdelaal2018OE,Meyer2019arXiv_ASV, Zhang2020TCST}.  In the path/motion planning approaches, a collision-free reference trajectory or motion is generated based on either off-line or on-line methods, e.g., A$^*$ \cite{Greytak2009CDC}, RRT$^*$ \cite{Hover2012IJRR,Chiang2018RAL}, and potential field methods \cite{Panagou2017TAC}, etc. It is assumed that the generated collision-free reference trajectory can be tracked with high accuracy by the ASV based on a well-designed control module. Thus, collision avoidance following the path/motion planning approaches may fail for uncertain systems that lack valid tracking controllers. Due to the two-module design feature, there always exists a time delay for the ASV to apply collision avoidance actions, as the inner-loop controller needs time to react to changes in the reference trajectories. Such a time delay will also downgrade the performance of the path/motion planning approaches in environments with fast-moving obstacles. 

The optimization-based algorithms can directly find a control law with collision avoidance by optimizing a certain objective function, e.g. model predictive control (MPC)  \cite{Johansen2016TITS, Abdelaal2018OE} and reinforcement learning (RL) \cite{Meyer2019arXiv_ASV}. They potentially have a better performance than the path/motion planning approaches in dynamic environments. However,  collision avoidance algorithms based on MPC suffer from high computational complexity and rely on accurate modeling of ASV systems \cite{Johansen2016TITS, Abdelaal2018OE}. They will, therefore, experience dramatic degradation in performances for uncertain ASVs. In comparison to MPC, RL can learn an intelligent collision avoidance law from data samples \cite{Fan2020IJRR,Sutton2018MIT}, which can significantly reduce the dependence on modeling efforts and thus make RL very suitable for uncertain ASVs. 

Tracking control algorithms for uncertain systems including ASVs mainly lie in four categories: 1) robust control that is the ``worst-case''  design for bounded uncertainties and disturbances \cite{Yu2012IET}; 2) adaptive control that estimates uncertainty parameters \cite{Do2006Auto,Farrell2006Book,Chowdhary2013IJACSP}; 3) disturbance observer (DO)-based control that compensates uncertainties and disturbances in terms of the observation technique \cite{Guan1991TAC,Dahmani2014TITS, Li2015TITS}; and 4) reinforcement learning (RL) that learns a control law from data samples \cite{Woo2019OE, Shi2019TNNLS}. In robust control, uncertainties and disturbances are assumed to be bounded with known boundaries \cite{Shen1995TAC}. As a consequence, robust control will lead to conservative high-gain control laws that might degrade the control performances (i.e., overshoot, settling time, and stability margins) \cite{Liu2008CDC}. Adaptive control can handle varying uncertainties with unknown boundaries, but system uncertainties are assumed to be linearly parameterized with known structure and unknown parameters \cite{Ioannou1996, Haddad2002IJACSP,Zhang2018AST}.  DO-based control can adapt to both uncertainties and disturbances with unknown structures \cite{Zhang2018TIE, Zhu2018IJRNC}.  However, the frequency information of uncertainty and disturbance signals are necessary in the DO-based control for the choice of proper control gains, otherwise, it is highly possible to end up with a high-gain control law \cite{Mondal2013ISA, Zhu2018IJRNC}. In general, comprehensive modeling and analysis of systems are essential for all model-based methods. 

In comparison to existing model-based methods, RL is capable of learning a complex tracking control law with collision avoidance from data samples using much less model information \cite{Sutton2018MIT}. It is, therefore, more promising in controlling systems subject to massive uncertainties and disturbances as ASVs  \cite{Woo2019OE, Shi2019TNNLS} and meanwhile achieving collision avoidance \cite{Meyer2019arXiv_ASV}, given the sufficiency and good quality of collected data. Nevertheless, it is challenging for model-free RL to ensure closed-loop stability, though some research attempts have been made \cite{Han2019NIPS, han2020actor}.  Model-based RL with stability guarantee has been investigated by introducing a Lyapunov constraint into the objective function \cite{Berkenkamp2017NIPS}.  However, the model-based RL with stability guarantees requires an admissible control law --- a control law that makes the original system asymptotically stable --- for the initialization. Both the Lyapunov candidate function and complete system dynamics are assumed to be Lipschitz continuous with known Lipschitz constants for the construction of the Lyapunov constraint. It is challenging to find the Lipschitz constant of an uncertain system. Therefore, the introduced Lyapunov constraint function is restrictive, as it is established based on the worst-case consideration \cite{Berkenkamp2017NIPS}. 

With the consideration of the merits and limitations of existing RL methods, we propose a novel learning-based control algorithm for uncertain ASVs with collision avoidance by combining a conventional control method with deep RL in this paper. The proposed learning-based control design, therefore, consists of two components: a baseline control law that stabilizes a nominal ASV system and a deep RL control law that compensates for system uncertainties and also achieves intelligent collision avoidance. Such a design structure has several advantages over both conventional model-based methods and pure deep RL methods. First of all, in relation to the ``model-free'' feature of deep RL, we can learn from data samples a control law that directly compensates for system uncertainties without exploiting their structures, boundaries, or frequencies \cite{Sutton1992CSM}. Intelligent collision avoidance can also be learned by the deep RL. Second, closed-loop stability is guaranteed by the overall learned control law for the tracking control in obstacle-free environments, if the baseline control law is able to stabilize the ASV system at least locally. Without introducing a restrictive Lyapunov constraint into the objective function of the policy improvement in RL as in \cite{Berkenkamp2017NIPS}, we can avoid exploiting the Lipschitz constant of the overall system and potentially produce less conservative results. Lastly, the proposed design is potentially more sample efficient than an RL algorithm learning from scratch -- that is, fewer data samples are needed for training. In RL, a system learns from mistakes, demanding a lot of trials and errors. In our design, the baseline control that can stabilize the overall system, can help to exclude unnecessary mistakes. Hence, it provides a good starting point for the RL training. A similar idea is used in \cite{Hwangbo2017RAL} for the control of quadrotors. The baseline control in \cite{Hwangbo2017RAL} is constructed based on the full accurate model of a quadrotor system, but stability analysis is missing. The design in \cite{Hwangbo2017RAL} is deployed as the inner-loop control to stabilize the attitude of quadrotors, but it is not designed for tracking control with collision avoidance. The overall contributions of this paper are summarized as below.
\begin{enumerate}
\item A new formulation method is presented for the learning-based control of ASV systems. With the new formulation, we can leverage the advantages of both model-based control methods and data-driven methods such as RL.
\item A model-reference RL algorithm is developed for the collision-free tracking control of uncertain ASVs. The proposed model-reference RL algorithm doesn't need the structures, boundaries, or frequencies of uncertainties.  It is potentially more efficient than a RL algorithm that learns from scratch. Closed-loop stability is guaranteed for the overall learning-based control law.
\item The proposed model-reference RL algorithm is analyzed systematically.  Convergence analysis is provided.  Closed-loop stability is analyzed for the tracking  control at the obstacle-free environments.
\end{enumerate}
Some of the work in this paper has been accepted to be presented in the 59th IEEE Conference on Decision and Control (CDC) that will be hosted at December, 2020. The online version of our CDC paper can be found in \cite{Zhang2020MRRL_arXiv}. In our CDC paper, the collision avoidance problem is not addressed. Mathematical proofs  are not provided for the convergence analysis. Rigorous the closed-loop stability proof is also missing in the CDC paper. Besides, in this paper, we present more details on the problem formulation and algorithm design, including the choices of the control policies in RL,  discussions of reward functions,  and descriptions of the deep neural networks, etc.  More simulation results are given in this paper.

The rest of the paper is organized as follows. In Section \ref{sec:SysDyn}, we present the ASV dynamics. The model-reference reinforcement learning control is formulated at length in Section \ref{sec:MR_DeepRL}, including the problem formulation,  basic concepts of reinforcement learning, and choices of reward functions. In Section \ref{sec:DeepRLControlDesign}, the model-reference reinforcement Learning is developed based deep neural networks. Section \ref{sec:AlgAnalysis} presents the details on the analysis  of the proposed model-reference reinforcement learning algorithm, including the convergence analysis and stability analysis. Section \ref{sec:NumSim} provides the simulation results of  the application of the algorithm to an example of ASVs. Conclusion remarks are given in Section \ref{sec:Concl}.

\section{Autonomous surface vehicle dynamics} \label{sec:SysDyn}

The full dynamics of autonomous surface vehicles (ASVs) have six degrees of freedom (DOF), including three linear motions and three rotational motions \cite{Fossen2011Handbook}. In most scenarios, we are interested in controlling the horizontal motions of (ASVs),  ignoring the vertical, rolling, and pitching motions \cite{Skjetnea2005Auto}.
\begin{figure}[tbp]
    \centering
    \includegraphics[width=0.315\textwidth]{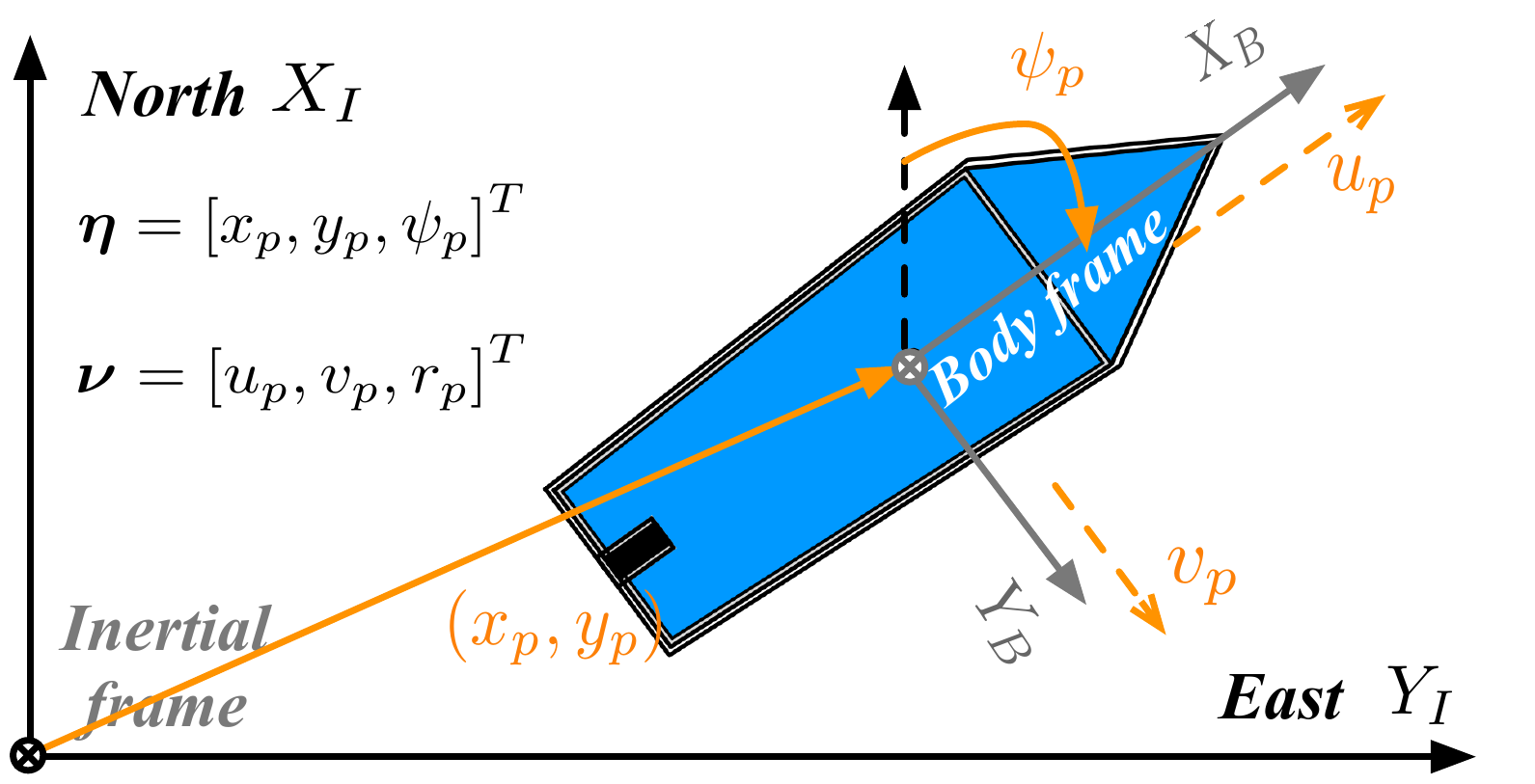}
    \caption{Coordinate systems of an autonomous surface vehicle}
    \label{fig:ship}
\end{figure}

Let  $x_p\in\mathbb{R}$ and $y_p\in\mathbb{R}$ be the horizontal position coordinates of an ASV in the inertial frame and $\psi_p\in\mathbb{R}$ the heading angle as shown in Figure \ref{fig:ship}. In the body frame,  $u_p\in\mathbb{R}$ and $v_p\in\mathbb{R}$ to represent the linear velocities in surge ($x$-axis) and sway ($y$-axis), respectively. The heading angular rate is denoted by $r_p\in\mathbb{R}$.  The general 3-DOF nonlinear dynamics of an ASV are described by
\begin{equation}
    \left\{
    \begin{array}{rcl}
       \dot{\boldsymbol{\eta}} &=& \boldsymbol{R}\left(\boldsymbol{\eta}\right)\boldsymbol{\nu}  \\
       \boldsymbol{M}\dot{\boldsymbol{\nu}}+\left(\boldsymbol{C}\left(\boldsymbol{\nu} \right)+\boldsymbol{D}\left(\boldsymbol{\nu} \right)\right)\boldsymbol{\nu} +\boldsymbol{G}\left(\boldsymbol{\nu} \right) &=& \boldsymbol{\tau}
    \end{array}
    \right. \label{eq:ASV_Dyn}
\end{equation}
where $\boldsymbol{\eta}=\left[x_p, y_p, \psi_p\right]^T\in\mathbb{R}^{3}$ is a generalized coordinate vector,  $\boldsymbol{\nu}=\left[u_p,v_p,r_p\right]^T\in\mathbb{R}^{3}$ is the speed vector, $\boldsymbol{M}$ is the inertia matrix, $\boldsymbol{C}\left(\boldsymbol{\nu} \right)$ denotes the
matrix of Coriolis and centripetal terms, $\boldsymbol{D}\left(\boldsymbol{\nu}\right)$ is the damping matrix, $ \boldsymbol{\tau}=\left[\tau_u,\tau_v,\tau_r\right]\in\mathbb{R}^{3}$ represents the control forces and moments, $\boldsymbol{G}\left(\boldsymbol{\nu} \right)=\left[\boldsymbol{g}_{1}\left(\boldsymbol{\nu} \right), \boldsymbol{g}_{2}\left(\boldsymbol{\nu} \right), \boldsymbol{g}_{3}\left(\boldsymbol{\nu} \right)\right]^T\in\mathbb{R}^{3}$ denotes unmodeled dynamics due to gravitational and buoyancy forces and moments \cite{Fossen2011Handbook},  and $\boldsymbol{R}$ is a rotation matrix given by  
\begin{equation*}
    \boldsymbol{R}\left(\boldsymbol{\eta}\right)=\left[\begin{array}{ccc}
    \cos{\psi} & -\sin{\psi} & 0 \\
    \sin{\psi} & \cos{\psi} &  0 \\
    0 & 0 & 1
    \end{array}\right] 
\end{equation*}
The inertia matrix $\boldsymbol{M}=\boldsymbol{M}^T>0$ is 
\begin{equation}
    \boldsymbol{M}=[M_{ij}]=\left[\begin{array}{ccc}
         M_{11}&   0    & 0  \\
         0     & M_{22} & M_{23} \\
         0     &  M_{32}& M_{33}
    \end{array}\right]
\end{equation}
where $M_{11}=m-X_{\dot{u}}$, $M_{22}=m-Y_{\dot{v}}$,  $M_{33}=I_z-N_{\dot{r}}$, and $M_{32}=M_{23}=mx_g-Y_{\dot{r}}$. The matrix $\boldsymbol{C}\left(\boldsymbol{\nu} \right)=-\boldsymbol{C}^T\left(\boldsymbol{\nu} \right)$ is 
\begin{equation}
    \boldsymbol{C}=[C_{ij}]=\left[\begin{array}{ccc}
         0 &   0    & C_{13}\left(\boldsymbol{\nu} \right) \\
         0 &   0 & C_{23}\left(\boldsymbol{\nu} \right) \\
         -C_{13}\left(\boldsymbol{\nu} \right)     & -C_{23}\left(\boldsymbol{\nu} \right) & 0
    \end{array}\right]
\end{equation}
where $C_{13}\left(\boldsymbol{\nu} \right)=-M_{22}v-M_{23}r$, $C_{23}\left(\boldsymbol{\nu} \right)=M_{11}u$. The damping matrix $\boldsymbol{D}\left(\boldsymbol{\nu}\right)$ is 
\begin{equation}
    \boldsymbol{D}\left(\boldsymbol{\nu}\right)=[D_{ij}]=
    \left[\begin{array}{ccc}
         D_{11}\left(\boldsymbol{\nu} \right) &   0    & 0\\
         0     & D_{22}\left(\boldsymbol{\nu} \right)& D_{23}\left(\boldsymbol{\nu} \right) \\
         0 & D_{32}\left(\boldsymbol{\nu} \right) & D_{33}\left(\boldsymbol{\nu} \right) 
    \end{array}\right]
\end{equation}
where $D_{11}\left(\boldsymbol{\nu} \right)=-X_u-X_{\vert u\vert u}\vert u\vert -X_{uuu}u^2$, $D_{22}\left(\boldsymbol{\nu} \right)=-Y_v-Y_{\vert v\vert v}\vert v\vert -Y_{\vert r\vert v}\vert r\vert $, $D_{23}\left(\boldsymbol{\nu} \right)=-Y_r-Y_{\vert v\vert r}\vert v\vert -Y_{\vert r\vert r}\vert r\vert $, 
$D_{32}\left(\boldsymbol{\nu} \right)=-N_v-N_{\vert v\vert v}\vert v\vert -N_{\vert r\vert v}\vert r\vert $, $D_{33}\left(\boldsymbol{\nu} \right)=-N_r-N_{\vert v\vert r}\vert v\vert -N_{\vert r\vert r}\vert r\vert $, and $X_{\left(\cdot\right)}$, $Y_{\left(\cdot\right)}$, and $N_{\left(\cdot\right)}$ are hydrodynamic coefficients whose definitions can be found in \cite{Fossen2011Handbook}. Accurate numerical models of the nonlinear dynamics (\ref{eq:ASV_Dyn}) are rarely available. Major uncertainty sources come from $\boldsymbol{M}$, $\boldsymbol{C}\left(\boldsymbol{\nu} \right)$, and $\boldsymbol{D}\left(\boldsymbol{\nu}\right)$ due to hydrodynamics, and  $\boldsymbol{G}\left(\boldsymbol{\nu} \right)$ due to gravitational and buoyancy forces and moments. 


\section{Problem formulation} \label{sec:MR_DeepRL}
In this section, we will formulate the model-reference control structure, introduce the reinforcement learning theory, and define reward functions for reinforcement learning.
\subsection{Model-reference control formulation} \label{subsec:MR_Form}
Let $\boldsymbol{x}=\left[\boldsymbol{\eta}^T, \boldsymbol{\nu}^T\right]^T$ and $\boldsymbol{u}=\boldsymbol{\tau}$, so (\ref{eq:ASV_Dyn})  can be rewritten as 
\begin{equation}
    \dot{\boldsymbol{x}} = \left[\begin{array}{cc}
    0 &  \boldsymbol{R}\left(\boldsymbol{\eta}\right) \\
    0 & \boldsymbol{A}\left(\boldsymbol{\nu}\right)
    \end{array}\right]\boldsymbol{x} + \left[\begin{array}{c}
    0 \\
    \boldsymbol{B}
    \end{array}\right]\boldsymbol{u} + \left[\begin{array}{c}
    0 \\
    \boldsymbol{M}^{-1}\boldsymbol{G}\left(\boldsymbol{\nu} \right)
    \end{array}\right]\label{eq:ASV_Dyn2}
\end{equation}
where $\boldsymbol{A}\left(\boldsymbol{\nu}\right)=\boldsymbol{M}^{-1}\left(\boldsymbol{C}\left(\boldsymbol{\nu} \right)+\boldsymbol{D}\left(\boldsymbol{\nu} \right)\right)$, and $\boldsymbol{B}=\boldsymbol{M}^{-1}$. Assuming that an accurate model (\ref{eq:ASV_Dyn2}) is not available, it is possible to get a nominal model expressed as
\begin{equation}
    \dot{\boldsymbol{x}}_m = \left[\begin{array}{cc}
    0 &  \boldsymbol{R}\left(\boldsymbol{\eta}\right) \\
    0 & \boldsymbol{A}_m
    \end{array}\right]\boldsymbol{x}_m + \left[\begin{array}{c}
    0 \\
    \boldsymbol{B}_m
    \end{array}\right]\boldsymbol{u}_m \label{eq:ASV_Dyn_nom}
\end{equation}
where $\boldsymbol{A}_m$ and $\boldsymbol{B}_m$ are the known system matrices, and the unmodelled dynamics $\boldsymbol{G}\left(\boldsymbol{\nu} \right)$ ignored. Note that $\boldsymbol{A}_m$ and $\boldsymbol{B}_m$ are different from $\boldsymbol{A}\left(\boldsymbol{\nu}\right)$ and $\boldsymbol{B}$, respectively. In $\boldsymbol{A}_m$ and $\boldsymbol{B}_m$, we will ignore all unknown  nonlinear terms, and obtain a linear nominal model. Assume that there exists a control law $\boldsymbol{u}_m$ allowing the states of the nominal system (\ref{eq:ASV_Dyn_nom}) to converge to a reference signal $\boldsymbol{x}_r$, i.e., $\Vert\boldsymbol{x}_m-\boldsymbol{x}_r\Vert_2\to{0}$ as $t\to\infty$.

\begin{figure}
    \centering
    \includegraphics[width=0.45\textwidth]{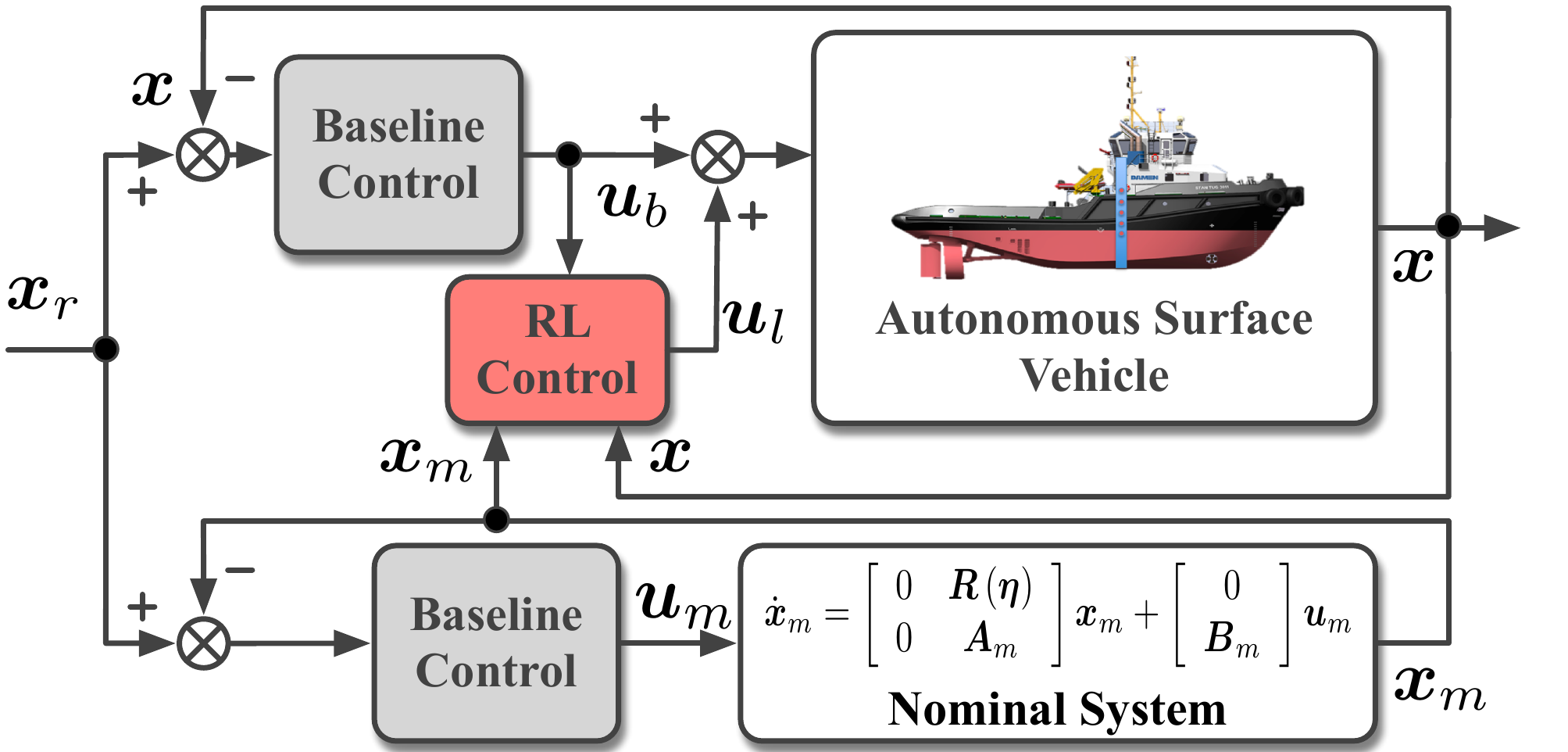}
    \caption{Model-reference reinforcement learning control}
    \label{fig:Cntrlblock}
\end{figure}

The objective of the work in this paper is to design a controller allowing the state  $\boldsymbol{x}$ to track state trajectories of the nominal model (\ref{eq:ASV_Dyn_nom}) and avoid collisions with obstacles having known states  $\boldsymbol{x}_{{o}_{i}}$, where $i\in\left\{1,\ldots,N_o\right\}$ indicates the $i$-th obstacle.  As shown in Figure \ref{fig:Cntrlblock}, the overall control structure for the ASV system (\ref{eq:ASV_Dyn2}) is 
\begin{equation}
    \boldsymbol{u} =\boldsymbol{u}_b + \boldsymbol{u}_{l}\label{eq:entireCntrl}
\end{equation}
where $\boldsymbol{u}_b$ is a baseline control designed based on (\ref{eq:ASV_Dyn_nom}), and $\boldsymbol{u}_{l}$ is a control law from the deep RL module whose design is provided in Section \ref{subsec:Training}.

\begin{remark}
The baseline control $\boldsymbol{u}_b$ is employed to ensure the basic tracking performance without obstacles, (i.e., local stability of the tracking control), while $\boldsymbol{u}_{l}$ is introduced to compensate for all system uncertainties and achieve collision avoidance. The baseline control $\boldsymbol{u}_b$ in (\ref{eq:entireCntrl}) can be designed using any existing method based on the nominal model (\ref{eq:ASV_Dyn_nom}). One potential choice for the design of  $\boldsymbol{u}_b$ is the nonlinear backstepping control \cite{Zhang2018TIE}. Hence, we ignore the design process of $\boldsymbol{u}_b$, and focus on the development of $\boldsymbol{u}_{l}$ with RL. 
\end{remark}
\subsection{Markov decision process} \label{subsec:MDP}
For the formulation of RL, the ASV dynamics (\ref{eq:ASV_Dyn2}) and (\ref{eq:ASV_Dyn_nom}) are characterized using another mathematical model called Markov decision process that is  denoted by a tuple $\mathcal{MDP}:=\big\langle \mathcal{S},\;\mathcal{U},\;\mathcal{P},\;R,\;\gamma\big\rangle$, where $\mathcal{S}$ is the state space, $\mathcal{U}$ specifies the action/input space, $\mathcal{P}:\mathcal{S}\times\mathcal{U}\times\mathcal{S}\rightarrow \mathbb{R}$ defines a transition probability, $R:\mathcal{S}\times\mathcal{U}\rightarrow \mathbb{R}$ is a reward function, and $\gamma\in\left[0,\;1\right)$ is a discount factor.  
Note that the state vector $\boldsymbol{s}\in\mathcal{S}$ contains all available signals affecting the learned control $\boldsymbol{u}_l$. In this paper, such signals include $\boldsymbol{x}$, $\boldsymbol{x}_{m}$, $\boldsymbol{x}_{r}$, $\boldsymbol{u}_{b}$,  and $\boldsymbol{x}_{{o}_{i}}$, where $\boldsymbol{x}_{m}$ represents the desired behaviour of the system (\ref{eq:ASV_Dyn2}) and $\boldsymbol{u}_{b}$ is a function of $\boldsymbol{x}$ and $\boldsymbol{x}_{r}$, and $\boldsymbol{x}_{{o}_{i}}$ are obstacle states in the neighbourhood (e.g., the position and velocity of the $i$-th obstacle). Hence, $\boldsymbol{s}=\left\{\boldsymbol{x}_{m},\boldsymbol{x}, \boldsymbol{u}_{b},  \cup_{i}^{N_{o}} \boldsymbol{x}_{{o}_{i}}\right\}$,  where $\cup_{i}^{N_{o}} \boldsymbol{x}_{{o}_{i}}$ denotes the states of $N_{o}$ obstacles detected by the ASV. More details on obstacles will be discussed in Section \ref{subsec:Reward}.

Since RL learns the control policies using data samples, it is assumed that we can sample input and state data from system (\ref{eq:ASV_Dyn2}) at discrete time steps. The sample time step is fixed and denoted by $\delta t$.  Without loss of generality, let $\boldsymbol{x}_{t}$, $\boldsymbol{u}_{b,t}$, and $\boldsymbol{u}_{l,t}$ be the ASV state, the baseline control action, and the control action from RL at the time step $t$, respectively. The union of obstacles detected by the ASV is characterized by $\cup_{i}^{N_{o}} \boldsymbol{x}_{{o}_{i}, t}$. The state signal $\boldsymbol{s}$ at the time step $t$ is, therefore, denoted by  $\boldsymbol{s}_t=\left\{\boldsymbol{x}_{m,t},\boldsymbol{x}_{t}, \boldsymbol{u}_{b,t},   \cup_{i}^{N_{o}} \boldsymbol{x}_{{o}_{i}, t}\right\}$. 

\subsection{Reinforcement learning} \label{subsec:RL}
For standard RL, the objective is to maximize an expected accumulated return  described by a value function $V_{\boldsymbol{\pi}}\left(\boldsymbol{s}_t\right)$ with
\begin{equation}
V_{\boldsymbol{\pi}}\left(\boldsymbol{s}_t\right)
=\sum_{t}^{\infty}\sum_{\boldsymbol{u}_{l,t}}\boldsymbol{\pi}\left(\boldsymbol{u}_{l,t}|\boldsymbol{s}_t\right)\sum_{\boldsymbol{s}_{t+1}}\mathcal{P}_{t+1|t}\big(R_t+\gamma V_{\boldsymbol{\pi}}(\boldsymbol{s}_{t+1}) \big) 
\label{eq:V_Func}
\end{equation}
where $\mathcal{P}_{t+1|t}=\mathcal{P}\left(\boldsymbol{s}_{t+1}\left|\boldsymbol{s}_t,\boldsymbol{u}_{l,t}\right.\right)$ is the transition probability of the ASV system, $R_{t}=R(\boldsymbol{s}_t,\boldsymbol{u}_{l,t})$ is the reward function, $\gamma\in\left[0,1\right)$ is a constant discount factor, and $\boldsymbol{\pi}\left(\boldsymbol{u}_{l,t}|\boldsymbol{s}_t\right)$ is called control policy in RL. A policy in RL, denoted by $\boldsymbol{\pi}\left(\boldsymbol{u}_{l,t}|\boldsymbol{s}_t\right)$, is the probability of choosing an action $\boldsymbol{u}_{l,t}\in\mathcal{U}$ at a state $\boldsymbol{s}_t\in\mathcal{S}$ \cite{Sutton2018MIT}.  In this paper,  a Gaussian policy is used, which is 
\begin{equation}
    \boldsymbol{\pi}\left(\boldsymbol{u}_l|\boldsymbol{s}\right)=\mathcal{N}\left(\boldsymbol{u}_l\left(\boldsymbol{s}\right), \boldsymbol{\sigma}\right) \label{eq:Policy}
\end{equation}
where $\mathcal{N}\left(\cdot, \cdot\right)$ denotes a Gaussian distribution with ${\boldsymbol{u}}_l\left(\boldsymbol{s}\right)$ as the mean value and $\boldsymbol{\sigma}$ as the covariance matrix. The covariance matrix $\boldsymbol{\sigma}$ controls the exploration performance at the learning stage.  For the algorithm design, we also introduce an action-value function (a.k.a., Q-function) defined by
\begin{equation}
Q_{\boldsymbol{\pi}}\left(\boldsymbol{s}_t,\boldsymbol{u}_{l,t}\right)=R_t+\gamma \mathbb{E}_{\boldsymbol{s}_{t+1}}\left[V_{\boldsymbol{\pi}}(\boldsymbol{s}_{t+1}) \right]\label{eq: Action-Value Func}
\end{equation}
where  $\mathbb{E}_{\boldsymbol{s}_{t+1}}\left[\cdot\right]=\sum_{\boldsymbol{s}_{t+1}}\mathcal{P}_{t+1|t}\left[\cdot\right]$ is an expectation operator over the distribution of $\boldsymbol{s}_{t+1}$. Maximizing  $Q_{\boldsymbol{\pi}}\left(\boldsymbol{s}_t,\;\boldsymbol{u}_{l,t}\right)$ is equivalent to maximizing $V_{\boldsymbol{\pi}}(\boldsymbol{s}_t)$. In the sequel, we will focus the maximization of $Q_{\boldsymbol{\pi}}\left(\boldsymbol{s}_t,\boldsymbol{u}_{l,t}\right)$ instead of $V_{\boldsymbol{\pi}}(s_t)$.

In this paper, the deep RL is resolved  based on the soft actor-critic (SAC) algorithm which provides both sample efficient learning and convergence \cite{Haarnoja2018SAC1}. In SAC, an entropy term is added to regulate the exploration performance at the training stage, thus resulting in a modified Q-function in (\ref{eq:Q_SAC}). 
\begin{align}
     Q_{\boldsymbol{\pi}}\left(\boldsymbol{s}_t,\boldsymbol{u}_{l,t}\right)=& R_t+\gamma \mathbb{E}_{\boldsymbol{s}_{t+1}}\left[V_{\boldsymbol{\pi}}(\boldsymbol{s}_{t+1}) \right.\nonumber \\
     &\left.+\alpha\mathcal{H}\left(\boldsymbol{\pi}\left(\boldsymbol{u}_{l,t+1}|\boldsymbol{s}_{t+1}\right)\right)\right]\label{eq:Q_SAC}
\end{align}
where $\mathcal{H}\left(\boldsymbol{\pi}\left(\boldsymbol{u}_{l,t}|\boldsymbol{s}_{t}\right)\right)=-\sum_{\boldsymbol{u}_{l,t}}\boldsymbol{\pi}\left(\boldsymbol{u}_{l,t}|\boldsymbol{s}_t\right)\ln\left(\boldsymbol{\pi}\left(\boldsymbol{u}_{l,t}|\boldsymbol{s}_{t}\right)\right)=-\mathbb{E}_{\boldsymbol{\pi}}\left[\ln\left(\boldsymbol{\pi}\left(\boldsymbol{u}_{l,t}|\boldsymbol{s}_{t}\right)\right)\right]$ is the entropy of the policy, and $\alpha$ is a temperature parameter \cite{Haarnoja2018SAC1}. Hence, SAC aims to solve the following optimization problem.
\begin{align}
     \boldsymbol{\pi}^*=& \arg\max_{\boldsymbol{\pi}\in \Pi} \left(R_t+\gamma \mathbb{E}_{\boldsymbol{s}_{t+1}}\left[V_{\boldsymbol{\pi}}(\boldsymbol{s}_{t+1}) \right.\right.\nonumber \\
     &\big.\left.+\alpha\mathcal{H}\left(\boldsymbol{\pi}\left(\boldsymbol{u}_{l,t+1}|\boldsymbol{s}_{t+1}\right)\right)\right]\big)\label{eq:SAC_Obj}
\end{align}
where $\Pi$ denotes a policy set.
\begin{remark}
Once the optimization problem (\ref{eq:SAC_Obj}) is resolved, we will have  $\boldsymbol{\pi}^*\left(\boldsymbol{u}_l|\boldsymbol{s}\right)=\mathcal{N}\left({\boldsymbol{u}}_l^*\left(\boldsymbol{s}\right), \boldsymbol{\sigma}^*\right)$ according to  (\ref{eq:Policy}).  The variance $\boldsymbol{\sigma}^*$ will be close to $0$.Thus, the stochastic policy will converge to a deterministic one in the end. The mean value function ${\boldsymbol{u}}_l^*\left(\boldsymbol{s}\right)$ will be the learned optimal control law that is eventually used to compensate system uncertainties and avoid collisions. Notably, the optimal control law ${\boldsymbol{u}}_l^*\left(\boldsymbol{s}\right)$ will be learned instead of designed. In the real implementation, ${\boldsymbol{u}}_l^*\left(\boldsymbol{s}\right)$ are approximated using deep neural networks that will be discussed in Section \ref{subsec:DNN}. The learning process is to find the optimal parameters of the deep neural networks used to approximate the optimal control law ${\boldsymbol{u}}_l^*\left(\boldsymbol{s}\right)$.
\end{remark}

Training/learning process of SAC will repeatedly execute policy evaluation and policy improvement. In the policy evaluation, the Q-value in (\ref{eq:Q_SAC}) is computed by applying a Bellman operation $Q_{\boldsymbol{\pi}}\left(\boldsymbol{s}_t,\boldsymbol{u}_{l, t}\right)=\mathcal{T}^{\boldsymbol{\pi}}Q_{\boldsymbol{\pi}}\left(\boldsymbol{s}_t,\boldsymbol{u}_{l, t}\right)$ where
\begin{align}
    \mathcal{T}^{\boldsymbol{\pi}}Q_{\boldsymbol{\pi}}\left(\boldsymbol{s}_t,\boldsymbol{u}_{l, t}\right)&=R_t+\gamma \mathbb{E}_{\boldsymbol{s}_{t+1}}\left\{\mathbb{E}_{\boldsymbol{\pi}}\left[Q_{\boldsymbol{\pi}}\left(\boldsymbol{s}_{t+1},\boldsymbol{u}_{l,t+1}\right) \right. \right. \nonumber \\
    &\left.\left.-\alpha\ln\left(\boldsymbol{\pi}\left(\boldsymbol{u}_{l,t+1}|\boldsymbol{s}_{t+1}\right)\right)\right]\right\} \label{eq:BellmanOp}
\end{align}
In the policy improvement, the policy is updated by 
\begin{equation}
    \boldsymbol{\pi}_{new} = \arg \min_{\boldsymbol{\pi}'\in \Pi}\mathscr{D}_{KL}\left(\boldsymbol{\pi}'\left(\cdot\vert\boldsymbol{s}_{t}\right) \Big\Vert {Z^{{\boldsymbol{\pi}}_{ old}}}{e^{\frac{1}{\alpha}Q^{{\boldsymbol{\pi}}_{ old}}\left(\boldsymbol{s}_{t}, \cdot\right)}}\right) \label{eq:KL_pi}
\end{equation}
where $\boldsymbol{\pi}_{ old}$ denotes the policy from the last update, $Q^{{\boldsymbol{\pi}}_{ old}}$ is the Q-value of $\boldsymbol{\pi}_{ old}$, $\mathscr{D}_{KL}$ denotes the Kullback-Leibler (KL) divergence, and $Z^{{\pi}_{old}}$ is a normalization factor. Via mathematical manipulations, the objective  is transformed into 
\begin{equation}
    \boldsymbol{\pi}^* = \arg \min_{\boldsymbol{\pi}\in \Pi} \mathbb{E}_{{\boldsymbol{\pi}}}\Big[\alpha\ln\left(\boldsymbol{\pi}\left(\boldsymbol{u}_{l,t}|\boldsymbol{s}_{t}\right)\right)-Q\left(\boldsymbol{s}_{t}, \boldsymbol{u}_{l, t}\right)\Big] \label{eq:PI_Q}
\end{equation}
More details on how (\ref{eq:PI_Q}) is obtained can be found in \cite{ Haarnoja2018SAC1}. 
\begin{figure}[tbp]
    \centering
    \includegraphics[width=0.45\textwidth]{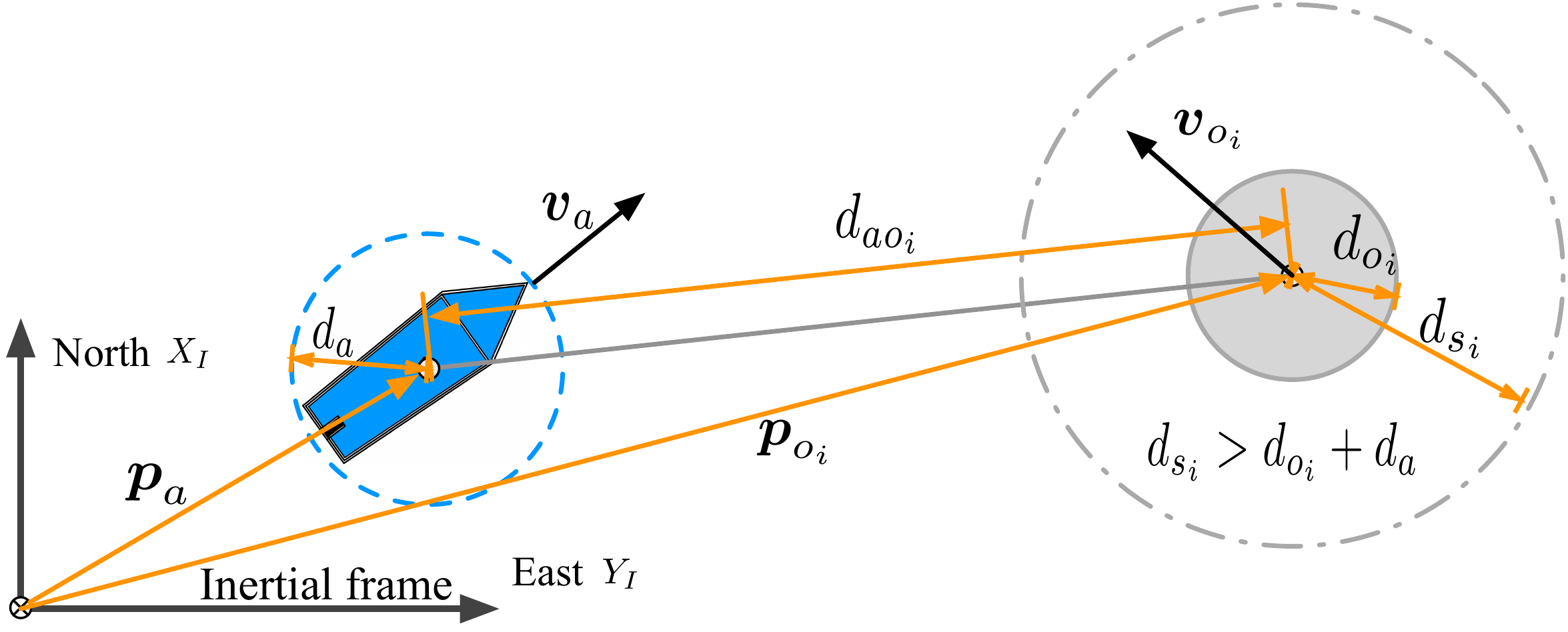}
    \caption{ Variables for collision avoidance  ($d_{a}$: size  of the ASV; $d_{{o}_{i}}$:  size of the $i$-th obstacle; $d_{{s}_{i}}$: radius of the safe region; $d_{a{o}_{i}}$: relative distance between the ASV and the $i$-th obstacle)}
    \label{fig:Obstacles}
\end{figure}
\begin{figure*}[tbp]
    \centering
    \includegraphics[width=0.825\textwidth]{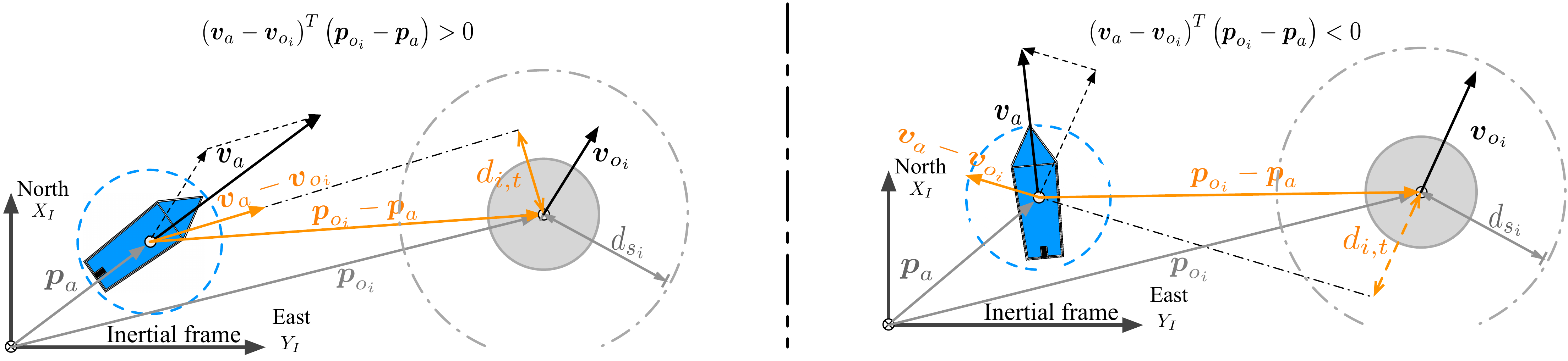}
    \caption{Illustration of $d_{i,t}$ (Note that $d_{i,t}$ is only useful when $\left(\boldsymbol{v}_a-\boldsymbol{v}_{{o}_{i}}\right)^{T}\left(\boldsymbol{p}_{{o}_{i}}-\boldsymbol{p}_a\right)>0$, otherwise collision is avoided)}
    \label{fig:Obstacles_dt}
\end{figure*}

\subsection{Reward functions} 
\label{subsec:Reward}
In our design, two objectives are defined for the ASV: trajectory tracking and collision avoidance. For the trajectory tracking, we aim to allow system (\ref{eq:ASV_Dyn2}) to track the nominal system (\ref{eq:ASV_Dyn_nom}), so  the tracking reward $R_{t,1}$ is defined as
\begin{equation}
   R_{t,1} = -\left(\boldsymbol{x}_{t}-\boldsymbol{x}_{m,t}\right)^T\boldsymbol{H}_1 \left(\boldsymbol{x}_{t}-\boldsymbol{x}_{m,t}\right)-\boldsymbol{u}_{l, t}^T\boldsymbol{H}_2\boldsymbol{u}_{l, t} \label{eq:ASV_Reward_1}
\end{equation}
where $\boldsymbol{H}_1 > 0$ and $\boldsymbol{H}_2>0$ are positive definite matrices. 

\begin{figure}[tbp]
    \centering
    \includegraphics[width=0.375\textwidth]{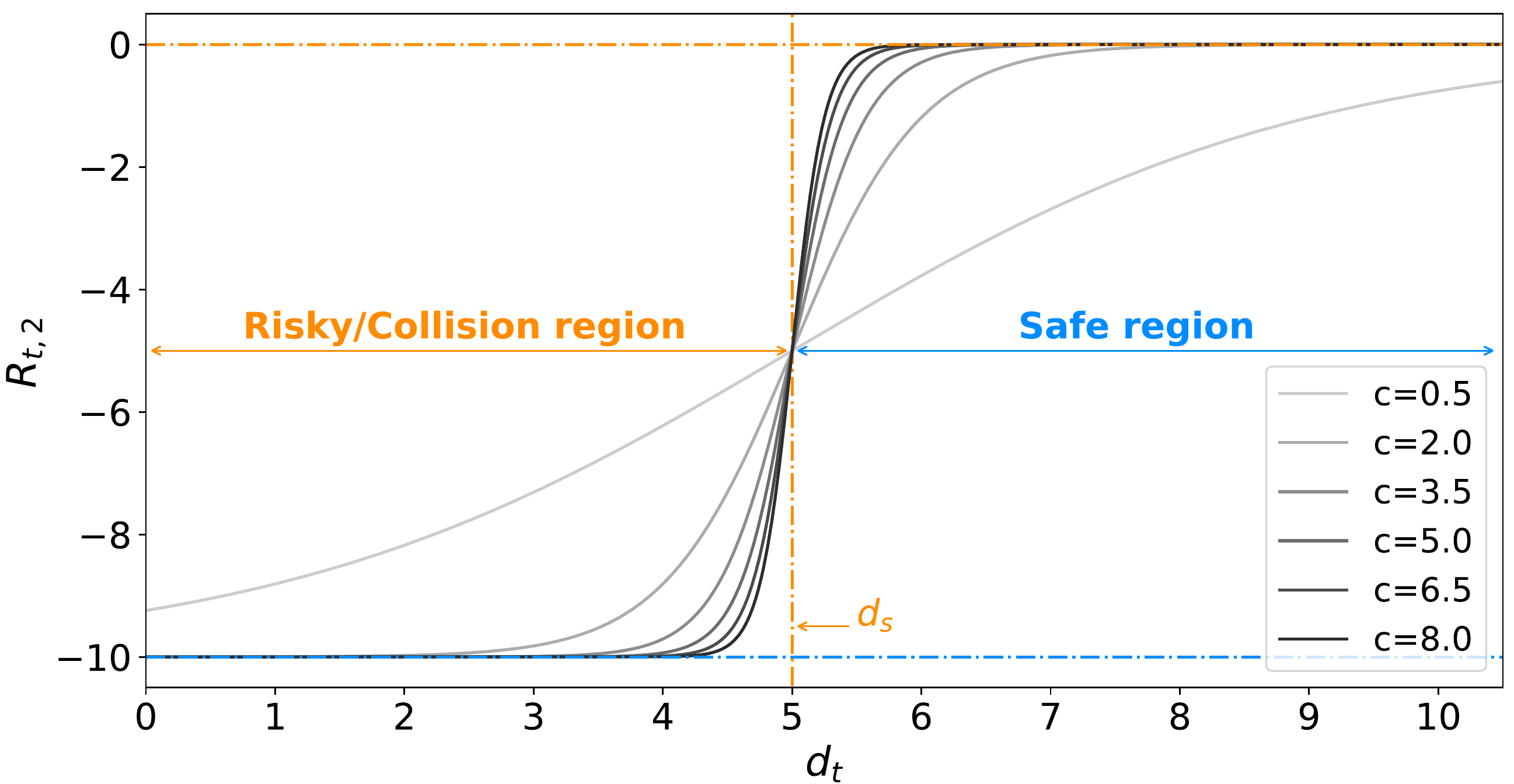}
    \caption{The impact of  $c_i$ on $R_{t,2}$ with one obstacle and $q_c=10$}
    \label{fig:Reward2}
\end{figure}

The second objective is to avoid obstacles along the trajectory of the ASV.  Figure \ref{fig:Obstacles} shows variables used for the definition of the reward function for collision avoidance. All obstacles are assumed to be inscribed  in a circle. The maximum size of the $i$-th obstacle is defined by $d_{{o}_{i}}$ as shown in Figure \ref{fig:Obstacles}. The size  of the ASV is denoted by $d_{a}$.  We introduce a safe region with a radius of $d_{{s}_{i}}$ for the $i$-th agent, where  $d_{{s}_{i}}> d_{{o}_{i}}+d_{a}$. The relative distance between the $i$-th obstacle and the ASV is defined as $d_{a{o}_{i}}$. If $d_{a{o}_{i}}\leq d_{d}$, the $i$-th obstacle is visible to the ASV, where $ d_{d}$ is the radius of the detection region of the ASV. Note that the obstacles could be either static or moving, so the state vector the $i$-th obstacle is written as $\boldsymbol{x}_{{o}_{i}}=\left[\boldsymbol{p}_{{o}_{i}}^T, \boldsymbol{v}_{{o}_{i}}^T\right]^T$, where $\boldsymbol{p}_{{o}_{i}}$ is the position of the $i$-th obstacle, and $\boldsymbol{v}_{{o}_{i}}$ is the velocity of the $i$-th obstacle.  Let $\boldsymbol{p}_a=\left[x_p,y_p\right]^T$ and $\boldsymbol{v}_a=\left[u_p,v_p\right]^T$ be the position and velocity of the ASV, respectively. For the $i$-th visible obstacle at the time step $t$, define the following variable shown in Figure \ref{fig:Obstacles_dt}.
\begin{equation}
d_{i, t}  = \frac{\Vert\left(\boldsymbol{v}_a-\boldsymbol{v}_{{o}_{i}}\right)^{\times}\left(\boldsymbol{p}_{{o}_{i}}-\boldsymbol{p}_a\right)\Vert_2}{\Vert\boldsymbol{v}_a-\boldsymbol{v}_{{o}_{i}}\Vert_2} \label{eq:ClosedPoint}
\end{equation}
where ``$^\times$'' denotes the cross product operation, $d_{i, t} $ represents the closest possible distance between the ASV and the obstacle, if the ASV keeps its current moving direction relative to the obstacle.  Note that $d_{i, t}$ is only meaningful, if $\left(\boldsymbol{v}_a-\boldsymbol{v}_{{o}_{i}}\right)^{T}\left(\boldsymbol{p}_{{o}_{i}}-\boldsymbol{p}_a\right)>0$. If $\left(\boldsymbol{v}_a-\boldsymbol{v}_{{o}_{i}}\right)^{T}\left(\boldsymbol{p}_{{o}_{i}}-\boldsymbol{p}_a\right)>0$, it implies that the ASV moves towards the obstacle, otherwise, the ASV moves away from the obstacle. Therefore, the reward function for collision avoidance is defined to be 
\begin{equation} 
   R_{t, 2} = \left\{\begin{array}{cc}
   -\sum_{i=1}^{N_{o}} \frac{q_{c,i}\mathbbm{1}_{{o}_{i}}\left(\boldsymbol{x}_{{o}_{i}},\boldsymbol{p}_a,\boldsymbol{v}_a\right)}{1+\exp\left(c_i \left(d_{i, t}-d_{{s}_{i}}\right)\right)}\text{, } & d_{a{o}_{i}}\leq d_{d}\\
   0\text{, } &  \text{otherwise}
  \end{array} \right.\label{eq:ASV_Reward_2}
\end{equation}
where $q_{c,i}>0$ is the maximum possible cost for collisions, $c_i >0$ is a design parameter, and $\mathbbm{1}_{{o}_{i}}\left(\boldsymbol{x}_{{o}_{i}},\boldsymbol{p}_a,\boldsymbol{v}_a\right)$ is 
\begin{equation*}
\mathbbm{1}_{{o}_{i}}\left(\boldsymbol{x}_{{o}_{i}},\boldsymbol{p}_a,\boldsymbol{v}_a\right)=
\left\{\begin{array}{ccl}
  1\text{, } & \left(\boldsymbol{v}_a-\boldsymbol{v}_{{o}_{i}}\right)^{T}\left(\boldsymbol{p}_{{o}_{i}}-\boldsymbol{p}_a\right)>0\\
   0\text{, } &  \text{otherwise}
  \end{array} \right. 
\end{equation*}
The parameter $c_i>0$ adjusts the sensitivity of collision avoidance of RL in related to the $i$-th obstacle. The influence of  $c_i>0$  on $   R_{t, 2} $ is illustrated in Figure \ref{fig:Reward2}.

The overall reward function is, therefore, defined to be
\begin{equation}
R_{t}=R_{t,1}+R_{t,2} \label{eq:ASV_Reward}
\end{equation}

\section{Model-Reference Deep Reinforcement Learning Design and Implementation} \label{sec:DeepRLControlDesign}
In this section, we will present the design and practical implementation of the model-reference deep RL control.  


\subsection{Deep neural networks} \label{subsec:DNN}
\begin{figure}[bp]
    \centering
    \includegraphics[width=0.45\textwidth]{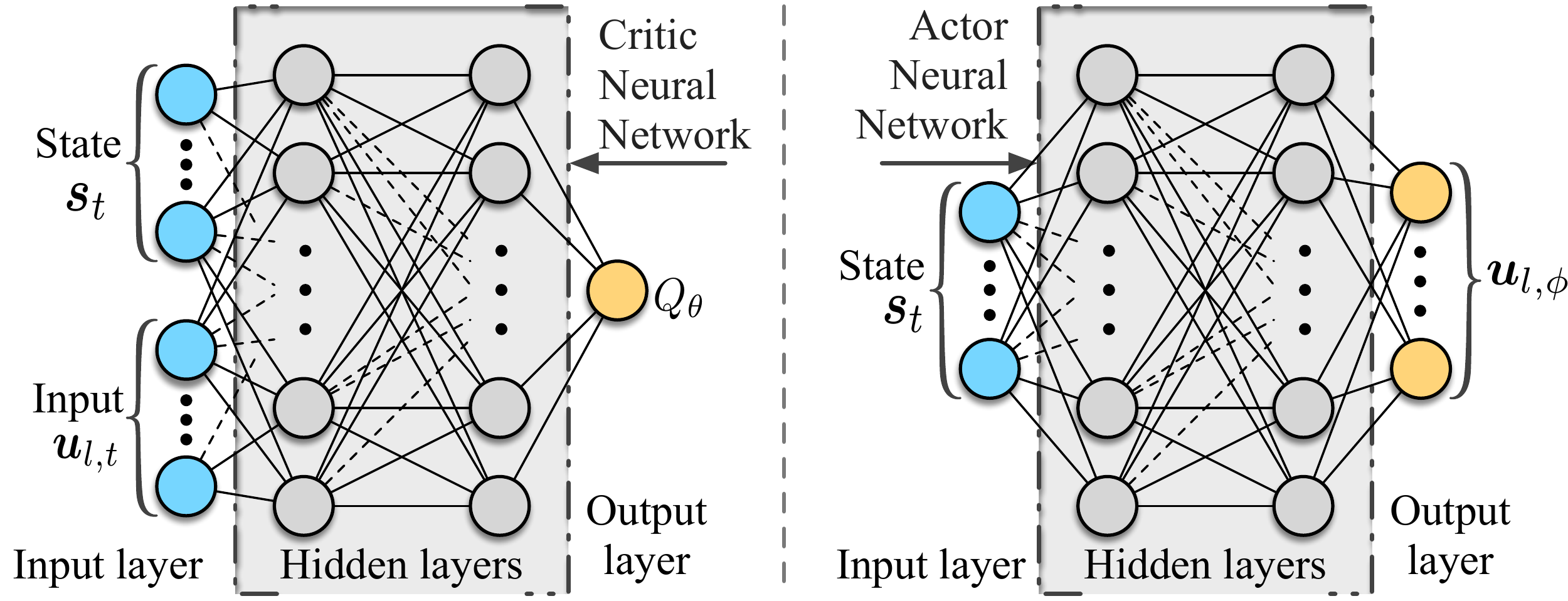}
    \caption{Approximation of $Q_{\theta}$ and $\boldsymbol{u}_{l,{\phi}}$ using MLP's}
    \label{fig:ACNN}
\end{figure}
In RL, the value function $Q_{\boldsymbol{\pi}}\left(\boldsymbol{s}_{t}, \boldsymbol{u}_{l, t}\right)$, which contains future information as shown in (\ref{eq: Action-Value Func}), is not known in advance. Similarly, the control policy $\boldsymbol{\pi}$ is unknown as well. Hence, a feasible solution is to approximate both the value function $Q_{\boldsymbol{\pi}}\left(\boldsymbol{s}_{t}, \boldsymbol{u}_{l, t}\right)$ and the control policy $\boldsymbol{\pi}$ using deep neural networks.  In this paper, the deep neural networks used to approximate both the value function $Q_{\boldsymbol{\pi}}\left(\boldsymbol{s}_{t}, \boldsymbol{u}_{l, t}\right)$ and the policy $\boldsymbol{\pi}\left(\boldsymbol{u}_{l,t}|\boldsymbol{s}_{t}\right)$ are chosen to be  fully connected multiple layer perceptrons (MLP) with rectified linear unit (ReLU) nonlinearities as the activation functions. The ReLU nonlinearities are defined by
\begin{equation*}
    \underline{relu}\left(z\right) = \max\left\{z\text{, }0\right\}
\end{equation*}
The ReLU activation function outperforms other activation functions like sigmoid functions \cite{Dahl2013ICASSP}. For a vector $z=[z_1\text{, }\ldots,\text{, }z_n]^T\in\mathbb{R}^{n}$, there exists $\underline{relu}\left(z\right)=[\underline{relu}\left(z_1\right)\text{, }\ldots\text{, }\underline{relu}\left(z_n\right)]^T$. As an example, a MLP with ``ReLU'' as the activation functions and two hidden layers is 
\begin{equation}
\underline{MLP}_{\mathrm{w}}^{2}\left(z\right) =\mathrm{w}_2\left[ \underline{relu}\left(\mathrm{w}_1 \left[\underline{relu}\left(\mathrm{w}_0\left[z^T\text{,}1\right]\right)^T\text{,}1\right]^T \right)^T\text{,} 1\right]^T\label{eq:MLP}
\end{equation}
where $\left[z^T\text{, }1\right]^T$ is a vector composed of $z$ and a bias $1$, the superscript ``$2$'' denotes the total number of hidden layers, the subscript ``$\mathrm{w}$'' denotes the parameter set to be trained in a MLP with $\mathrm{w}=\left\{\mathrm{w}_0\text{, }\mathrm{w}_1\text{, }\mathrm{w}_2\right\}$, and $\mathrm{w}_0$, $\mathrm{w}_1$,  and $\mathrm{w}_2$  are weight matrices with appropriate dimensions.

If there is a set of inputs $z=\left\{z_1\text{, } \ldots\text{, } z_L\right\}$ for the MLP in  (\ref{eq:MLP}) with $z_1$, $\ldots$, $z_L$ denoting vector signals, we have
\begin{equation}
\underline{MLP}_{\mathrm{w}}^{2}\left(z\right) =\underline{MLP}_{\mathrm{w}}^{2}\left(\left[z_1^T\text{, }  \ldots\text{, } z_L^T\right]^T\right) \label{eq:MLP_2}
\end{equation}
Besides, $\underline{MLP}_{\mathrm{w}}^{2}\left(z_1\text{, } z_2\right)=\underline{MLP}_{\mathrm{w}}^{2}\left(\left[z_1^T\text{, } z_2^T\right]^T\right)$ for two vector inputs $z_1$ and $z_2$. If $z_1=\left\{z_{11}\text{, } \ldots\text{, } z_{1L}\right\}$ is a set of vectors, $\underline{MLP}_{\mathrm{w}}^{2}\left(z_1\text{, } z_2\right)=\underline{MLP}_{\mathrm{w}}^{2}\left(\left[z_{11}^T\text{, }\ldots\text{, } z_{1L}^T, z_2^T\right]^T\right)$.

Let $Q_{\theta}\left(\boldsymbol{s}_{t}, \boldsymbol{u}_{l, t}\right)$ be the approximated Q-function using a MLP with a set of parameters denoted by $\theta$.  Following (\ref{eq:MLP}) and (\ref{eq:MLP_2}), the Q-function approximation $Q_{\theta}\left(\boldsymbol{s}_{t}, \boldsymbol{u}_{l, t}\right)$  is
\begin{equation}
Q_{\theta}\left(\boldsymbol{s}_{t}, \boldsymbol{u}_{l, t}\right) =\underline{MLP}_{\theta}^{{K}_{1}}\left(\boldsymbol{s}_{t}, \boldsymbol{u}_{l, t}\right) \label{eq:MLP_Q}
\end{equation}
where $\theta =\left\{\theta_0, \ldots, \theta_{{K}_{1}}\right\}$ with $\theta_i\in\mathbb{R}^{{\left(L\right)}\times {\left(L+1\right)}}$ for $0\leq i\leq K_1$ denoting the weight matrices with proper dimensions.  The deep neural network for $Q_{\theta}$ is illustrated in Figure \ref{fig:ACNN}. 

The control law $\boldsymbol{u}_{l}$ is also approximated using a MLP. The approximated control law of $\boldsymbol{u}_{l}$ with a parameter set $\phi$ is 
\begin{equation}
{\boldsymbol{u}}_{l,\phi}=\underline{MLP}_{\phi}^{{K}_{2}}\left(\boldsymbol{s}_{t}\right) \label{eq:MLP_u}
\end{equation}
The illustration of ${\boldsymbol{u}}_{l,\phi}$ is given in Figure \ref{fig:ACNN}. In SAC, there are two outputs for the MLP in (\ref{eq:MLP_u}). One is the control law $ {\boldsymbol{u}}_{l,\phi}$, the other one is ${\sigma}_{\phi}$ that is the standard deviation of the exploration noise \cite{Haarnoja2018SAC1}.  According to (\ref{eq:Policy}), the parameterized policy $\boldsymbol{\pi}_{{\phi}}$ in our model-reference RL is 
\begin{equation}
\boldsymbol{\pi}_{{\phi}} =  \mathcal{N}\left(\boldsymbol{u}_{l, \phi}\left(\boldsymbol{s}_t\right), \boldsymbol{\sigma}_{\phi}^2\right)    \label{eq:MLP_Pi}
\end{equation}
The deep neural network for $Q_{\theta}$ is called critic, while the one for $\boldsymbol{\pi}_{{\phi}}$ is called actor. 

 \subsection{Algorithm design and implementation}\label{subsec:Training}
\begin{figure}[tbp]
    \centering
    \includegraphics[width=0.42\textwidth]{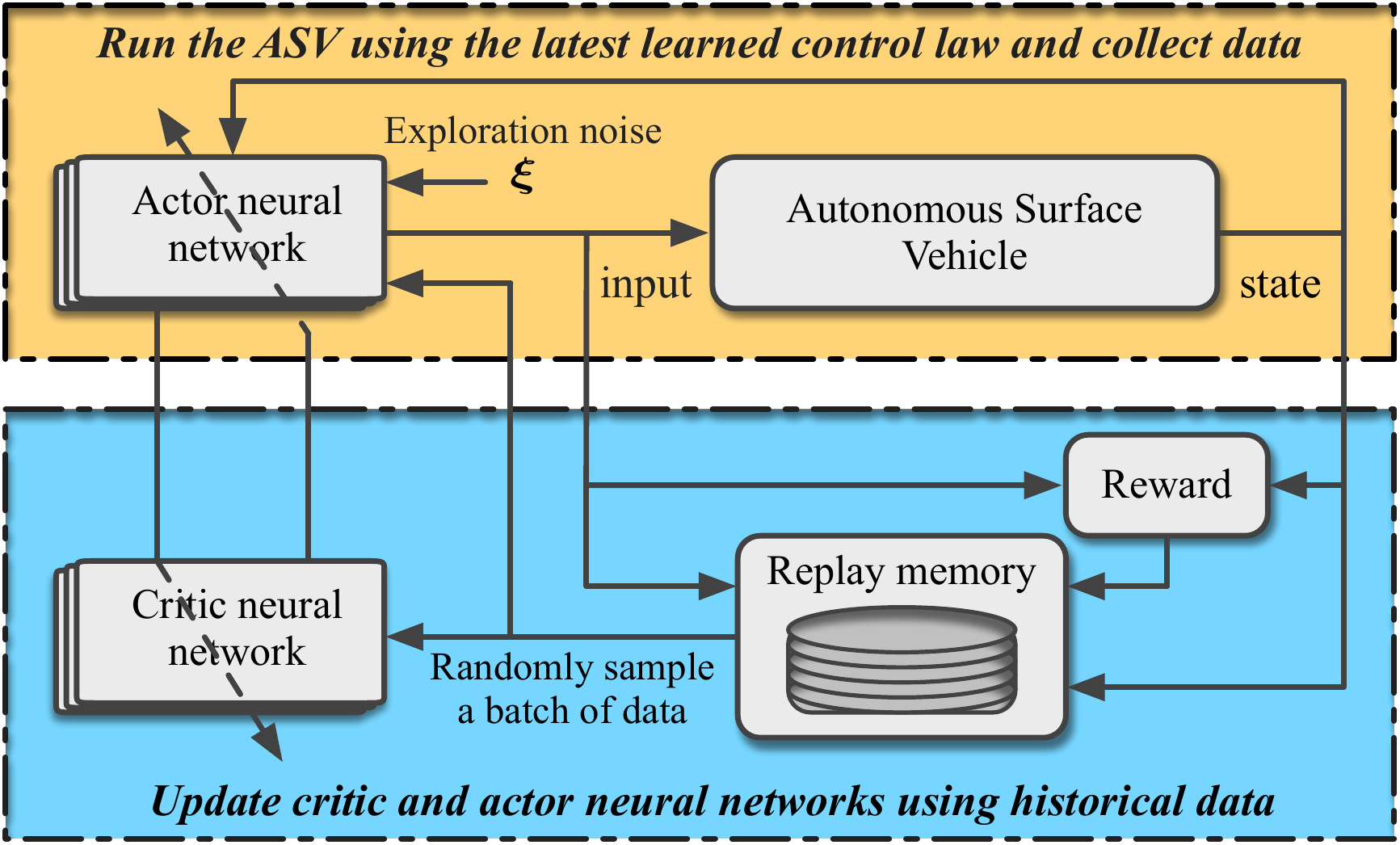}
    \caption{Offline training process of deep reinforcement learning}
    \label{fig:TrainRL}
\end{figure}

The algorithm training process is illustrated in Figure  \ref{fig:TrainRL}. The whole training process will be offline. We repeatedly run the system (\ref{eq:ASV_Dyn2}) under a trajectory tracking task. At each time step $t+1$, we collect data samples, such as an input from the last time step  $\boldsymbol{u}_{l, t}$, a state from the last time step $\boldsymbol{s}_{t}$, a reward $R_t$, and a current state $\boldsymbol{s}_{t+1}$. Those historical data will be stored as a tuple $\left(\boldsymbol{s}_{t}, \boldsymbol{u}_{l, t}, R_t, \boldsymbol{s}_{t+1}\right)$ at a replay memory $\mathcal{D}$ \cite{Mnih2015Nature}. At each policy evaluation or improvement step, we randomly sample a batch of historical data, $\mathcal{B}$, from the replay memory $\mathcal{D}$ for the training of the parameters $\theta$ and $\phi$. Starting the training, we apply the baseline control policy $\boldsymbol{u}_b$ to an ASV system to collect the initial data  $\mathcal{D}_0$ as shown in Algorithm \ref{alg:MRRL}. The initial data set $\mathcal{D}_0$ is used for the initial fitting of Q-value functions. When the initialization is over, we execute both $\boldsymbol{u}_b$ and the latest updated RL policy $\boldsymbol{\pi}_{{\phi}}\left(\boldsymbol{u}_{l, t}\vert \boldsymbol{s}_{t}\right)$ to run the ASV system.

At the policy evaluation step, the parameters $\theta$ are trained to minimize the following Bellman residual.
\begin{equation}
J_{Q}\left(\theta\right)= \mathbb{E}_{\left(\boldsymbol{s}_{t}, \boldsymbol{u}_{l, t}\right)\sim\mathcal{D}}\left[\frac{1}{2}\left(Q_{\theta}\left(\boldsymbol{s}_{t}, \boldsymbol{u}_{l, t}\right)-Y_{target}\right)^2\right] \label{eq:JQ_theta}
\end{equation}
where $\left(\boldsymbol{s}_{t}, \boldsymbol{u}_{l, t}\right)\sim\mathcal{D}$ implies that we randomly pick data samples $\left(\boldsymbol{s}_{t}, \boldsymbol{u}_{l, t}\right)$ from a replay memory $\mathcal{D}$, and 
\begin{equation*}
  Y_{target}=R_t+\gamma\mathbb{E}_{\boldsymbol{s}_{t+1}}\big[\mathbb{E}_{\boldsymbol{\pi}}\left[Q_{\bar{\theta}}\left(\boldsymbol{s}_{t+1}, \boldsymbol{u}_{l,t+1}\right)-\alpha\ln\left(\boldsymbol{\pi}_{\phi}\right)\right] \big]  
\end{equation*} 
where $\bar{\theta}$ is the target parameter which will be updated slowly. Applying a stochastic gradient descent technique (ADAM \cite{Kingma2014Adam} in this paper) to (\ref{eq:JQ_theta}) on a data batch $\mathcal{B}$ with a fixed size yields 
\begin{align*}
    \nabla_{\theta} J_{Q}\left(\theta\right) &= \sum  \frac{\nabla_{\theta}Q_{\theta}}{\vert\mathcal{B}\vert} \Big(Q_{\theta}\left(\boldsymbol{s}_{t}, \boldsymbol{u}_{l, t}\right)-Y_{target}\Big) 
\end{align*}
where $\vert\mathcal{B}\vert$ is the batch size. 
\begin{algorithm}[tbp]
   \caption{Model reference reinforcement learning control} \label{alg:MRRL}
\begin{algorithmic}[1]
   \STATE Initialize parameters $\theta_1$, $\theta_2$ for $Q_{{\theta}_{1}}$ and $Q_{{\theta}_{2}}$, respectively, and $\phi$ for the actor network  (\ref{eq:MLP_u}).
   \STATE Assign values to the target parameters $\bar{\theta}_1\leftarrow\theta_1$, $\bar{\theta}_2\leftarrow\theta_2$, $\mathcal{D}\leftarrow\emptyset$,  
   $\mathcal{D}_0\leftarrow\emptyset$,
  \STATE {Get data set $\mathcal{D}_0$ by running $\boldsymbol{u}_b$ on (\ref{eq:ASV_Dyn2}) with $\boldsymbol{u}_l=\boldsymbol{0}$}
  \STATE {Turn off the exploration and train initial critic parameters $\theta_1^0$, $\theta_2^0$ using $\mathcal{D}_0$ according to (\ref{eq:JQ_theta}). }
  \STATE {Initialize the replay memory $\mathcal{D}\leftarrow\mathcal{D}_0$}
  \STATE {Assign initial values to critic parameters $\theta_1\leftarrow\theta_1^0$, $\theta_2\leftarrow\theta_2^0$ and their targets $\bar{\theta}_1\leftarrow\theta_1^0$, $\bar{\theta}_2\leftarrow\theta_2^0$}
   \REPEAT
   \FOR{each data collection step}
    \STATE Choose an action $\boldsymbol{u}_{l, t}$ according to $ \boldsymbol{\pi}_{\phi}\left(\boldsymbol{u}_{l, t}\vert \boldsymbol{s}_{t}\right)$ 
    \STATE Collect $\boldsymbol{s}_{t+1}=\left\{\boldsymbol{x}_{t+1}, \boldsymbol{x}_{m, t+1}, \boldsymbol{u}_{b,t+1}\right\}$ based on the nominal system (\ref{eq:ASV_Dyn_nom}) and the full system (\ref{eq:ASV_Dyn2})
    \STATE $\mathcal{D}\leftarrow\mathcal{D}\bigcup \left\{\boldsymbol{s}_{t}, \boldsymbol{u}_{l, t}, R\left(\boldsymbol{s}_{t}, \boldsymbol{u}_{l, t}\right), \boldsymbol{s}_{t+1}\right\}$
    \ENDFOR
    \FOR{each gradient update step}
    \STATE Sample a batch of data $\mathcal{B}$ from $\mathcal{D}$
   \STATE $\theta_j\leftarrow\theta_j-\iota_Q \nabla_{\theta} J_{Q}\left(\theta_j\right) $,   and $j=1$, $2$
   \STATE $\phi\leftarrow\phi-\iota_\pi \nabla_{\phi}J_{\boldsymbol{\pi}}\left(\phi\right)$, 
   \STATE $\alpha\leftarrow \alpha - \iota_\alpha \nabla_{\alpha}J_{{\alpha}}\left(\alpha\right)$
   \STATE $\bar{\theta}_j\leftarrow\kappa\theta_j+\left(1-\kappa\right)\bar{\theta}_j$,  and $j=1$, $2$
   \ENDFOR
   \UNTIL{convergence (i.e. $J_{Q}\left(\theta\right)<$ a small threshold)}
   \STATE \textbf{Output} the optimal parameters $\phi^*$ and $\theta_j^*$,   and $j=1$, $2$
\end{algorithmic}
\end{algorithm}

At the policy improvement step, the objective function defined in (\ref{eq:PI_Q}) is represented  using data samples from the replay memory $\mathcal{D}$ as given in (\ref{eq:PI_Phi}).  
\begin{align}
    J_{{\pi}}\left(\phi\right)&=\mathbb{E}_{\left(\boldsymbol{s}_{t}, \boldsymbol{u}_{l, t}\right)\sim\mathcal{D}}\Big(\alpha\ln(\boldsymbol{\pi}_{{\phi}})  -Q_{{\theta}}\left(\boldsymbol{s}_{t}, \boldsymbol{u}_{l, t}\right)\Big) \label{eq:PI_Phi}
\end{align}
Parameter $\phi$ is trained to minimize (\ref{eq:PI_Phi})  using a stochastic gradient descent technique. Applying the policy gradient technique  to (\ref{eq:PI_Phi}), we can calculate the gradient of $J_{\boldsymbol{\pi}}\left(\phi\right)$ with respect to $\phi$ in terms of the stochastic gradient method as 
\begin{equation*}
    \nabla_{\phi}J_{{\pi}} = \sum \frac{\alpha\nabla_{\phi}\ln\boldsymbol{\pi}_{\phi}+\left(\alpha\nabla_{\boldsymbol{u}_{l}}\ln\boldsymbol{\pi}_{\phi}-\nabla_{\boldsymbol{u}_{l}}Q_{{\theta}}\right)\nabla_{\phi}\hat{\boldsymbol{u}}_{l,\phi}}{\vert\mathcal{B}\vert}\label{eq:StochG_Pi}
\end{equation*}
The temperature parameters $\alpha$ are updated by minimizing 
\begin{equation}
    J_{\alpha} =\mathbb{E}_{\boldsymbol{\pi}}\left[-\alpha\ln \boldsymbol{\pi}\left(\boldsymbol{u}_{l, t}\vert \boldsymbol{s}_{t}\right)-\alpha\bar{\mathcal{H}}\right]
\end{equation}
where $\bar{\mathcal{H}}$ is a target entropy.  In the final implementation, two critics are introduced to reduce the over-estimation issue in the training of critic neural networks \cite{Fujimoto2018TD3}. Under the two-critic mechanism, the target value $Y_{target}$ is modified to be
\begin{align}
  Y_{target} &=R_t+\gamma\min\Big\{Q_{\bar{\theta}_{1}}\left(\boldsymbol{s}_{t+1}, \boldsymbol{u}_{l,t+1}\right), \Big.\nonumber \\
  &\Big.Q_{\bar{\theta}_{2}}\left(\boldsymbol{s}_{t+1}, \boldsymbol{u}_{l,t+1}\right)\Big\}-\gamma\alpha\ln\left(\boldsymbol{\pi}_{\phi}\right)  \label{eq:TargetNew}
\end{align}
The entire process is summarized in Algorithm \ref{alg:MRRL}, in which $\iota_Q$, $\iota_\pi$, $\iota_\alpha>0$ are learning rates, and $\kappa>0$ is a constant scalar.

Once the training process is over, Algorithm \ref{alg:MRRL} will output the optimal parameters for the deep neural networks in (\ref{eq:MLP_Q}) and (\ref{eq:MLP_u}). Once the optimal parameters are obtained, the learned control law $\boldsymbol{u}_l$ is approximated by 
\begin{equation}
{\boldsymbol{u}}_{l}\simeq {\boldsymbol{u}}_{l,\phi^*}
\end{equation}
where $\phi^*$ is the optimal parameter set for the MLP in (\ref{eq:MLP_u}) and is obtained via Algorithm \ref{alg:MRRL}.

\section{Algorithm analysis} \label{sec:AlgAnalysis}
In this section, the performance of the proposed model-reference RL algorithm will be analyzed, including convergence and closed-loop stability for tracking control.
\subsection{Convergence analysis}
The general structure of a deep RL algorithm is summarized in Algorithm \ref{alg:PI_Tech}. The learning process will recursively execute the policy evaluation and policy improvement until convergence.  As we mentioned in Section \ref{subsec:MR_Form}, the baseline control $\boldsymbol{u}_b$ is assumed to stabilize the ASV without collision avoidance. Therefore, the following assumption is introduced for the convergence analysis.
\begin{assumption}\label{assump:BaseCntrl_Stab}
If no obstacles are considered, the trajectory tracking errors of the ASV are bounded using the baseline control $\boldsymbol{u}_b$.
\end{assumption}
\begin{algorithm}[tbp]
  \caption{Policy iteration technique} \label{alg:PI_Tech}
\begin{algorithmic}[1]
  \STATE Start from an initial control policy $\boldsymbol{u}_0$
  \REPEAT
  \FOR{Policy evaluation}
    \STATE Under a fixed policy $\boldsymbol{u}_l$, apply the Bellman backup operator $\mathcal{T}^{\pi}$ to the Q value function, $Q\left(\boldsymbol{s}_{t},\boldsymbol{u}_{l,t}\right)=\mathcal{T}^{\pi}Q\left(\boldsymbol{s}_{t},\boldsymbol{u}_{l,t}\right)$  given in (\ref{eq:BellmanOp})
    \ENDFOR
    \FOR{Policy improvement}
    \STATE Update policy $\boldsymbol{\pi}$ according to (\ref{eq:PI_Q})
  \ENDFOR
  \UNTIL{convergence}
\end{algorithmic}
\end{algorithm}

According to (\ref{eq:ASV_Reward_1}) and (\ref{eq:ASV_Reward_2}), both $R_{t,1}$ and $R_{t,2}$ are non-positive. With Assumption \ref{assump:BaseCntrl_Stab},  the reward function $R_{t,1}$ is ensured to be bounded. Additionally, the reward function $R_{t,2}$ is bounded by design as shown in Figure \ref{fig:Reward2} for a finite number of obstacles. Hence, the overall reward $R_{t}$ is bounded, namely
\begin{equation}
R_{t}\in\left[R_{min},\; 0\right] \label{eq:BoundedReward}
\end{equation} 
where  $R_{min}$ is the lowest bound for the reward function under the baseline control $\boldsymbol{u}_b$. 

In terms of (\ref{eq:BoundedReward}), we can present the following Lemma \ref{lem:Pi_Eval} and Lemma \ref{lem:Pi_Improve} for the convergence analysis of the entropy-regularized SAC algorithm \cite{Haarnoja2018SAC1,Haarnoja2018SAC2}.
\begin{lemma}[Policy evaluation] \label{lem:Pi_Eval}
Let $\mathcal{T}^{\pi}$ be the Bellman backup operator under a fixed policy $\boldsymbol{\pi}$ and $Q^{k+1}\left(\boldsymbol{s},\boldsymbol{u}_l\right)=\mathcal{T}^{\pi}Q^{k}\left(\boldsymbol{s},\boldsymbol{u}_l\right)$. The sequence $Q^{k+1}\left(\boldsymbol{s},\boldsymbol{u}_l\right)$ will converge to the soft Q-function $Q^{\boldsymbol{\pi}}$ of the policy $\boldsymbol{\pi}$ as $k\to\infty$.
\end{lemma}
\begin{proof}
Proof details are given in Appendix \ref{app:Lemma_Eval}.
\end{proof}

\begin{lemma}[Policy improvement] \label{lem:Pi_Improve}
Let $\boldsymbol{\pi}_{old}$ be an old policy and $\boldsymbol{\pi}_{new}$ be a new policy obtained according to (\ref{eq:KL_pi}). There exists $Q^{\boldsymbol{\pi}_{new}}\left(\boldsymbol{s},\boldsymbol{u}_l\right)\geq Q^{\boldsymbol{\pi}_{old}}\left(\boldsymbol{s},\boldsymbol{u}_l\right)$ $\forall \boldsymbol{s}\in\mathcal{S}$ and $\forall \boldsymbol{u}\in\mathcal{U}$.
\end{lemma}
\begin{proof}
Proof details are given in Appendix \ref{app:Lemma_Pi_Improve}.
\end{proof}

In terms of (\ref{lem:Pi_Eval}) and (\ref{lem:Pi_Improve}), we are ready to present Theorem \ref{thm:Converge} to show the convergence of the model-reference RL algorithm. In the sequel, the superscript $i$ denotes the $i$-th iteration of the policy iteration algorithm or the $i$-th policy improvement, where $i=0$, $1$, $\ldots$, $\infty$.
\begin{theorem}[\textbf{Convergence}] \label{thm:Converge}
Suppose $\boldsymbol{\pi}^{i}$ is the policy obtained at the $i$-th policy improvement with $\boldsymbol{\pi}^{0}$ denoting any initial policy in $\Pi$, and $i=0$, $1$, $\ldots$, $\infty$. If one repeatedly applies the policy evaluation and policy improvement steps as elaborated in Algorithm \ref{alg:PI_Tech}, there exists $\boldsymbol{\pi}^i\to\boldsymbol{\pi}^{*}$ as $i\to\infty$ such that  $Q^{\boldsymbol{\pi}^{*}}\left(\boldsymbol{s},\boldsymbol{u}_l\right) \geq Q^{\boldsymbol{\pi}^{i}}\left(\boldsymbol{s},\boldsymbol{u}_l\right)$ $\forall \boldsymbol{\pi}^i\in\Pi$,  $\forall \boldsymbol{s}\in\mathcal{S}$,  and $\forall \boldsymbol{u}_l\in\mathcal{U}$, where $\boldsymbol{\pi}^{*}\in\Pi$ denotes the optimal policy.
\end{theorem}
\begin{proof}
Proof details are given in Appendix \ref{app:Theorem_Convergence}.
\end{proof}

\subsection{Stability of the tracking control} \label{subsec:DeepRLControlStab}
In this subsection, we will show the closed-loop stability of the overall control law (baseline control $\boldsymbol{u}_b$ plus the learned control $\boldsymbol{u}_{l}$) for the tracking control without obstacles. The closed-loop stability  is analyzed under the general tracking performance without the consideration of collision avoidance, as the tracking control is the fundamental task. 
Before the closed-loop stability is analyzed, the  admissible control concept is introduced in  Definition \ref{def: FeasCntrl}, which is similar to the admissible control in adaptive dynamic programming \cite{Beard1997Auto,Al-Tamimi2008TSC,Jiang2015TAC}.
\begin{definition}\label{def: FeasCntrl}
A control law $\boldsymbol{u}_b$ is said admissible with respect to the system (\ref{eq:ASV_Dyn2}), if it can stabilize the system (\ref{eq:ASV_Dyn2}) and ensure that the state of (\ref{eq:ASV_Dyn2}) is uniformly ultimately bounded under system uncertainties.
\end{definition}
Note that the admissible control in \cite{Beard1997Auto,Al-Tamimi2008TSC,Jiang2015TAC} needs to provide the asymptotic stability for the system.  However, the admissible control in Definition \ref{def: FeasCntrl} doesn't necessarily ensure the system (\ref{eq:ASV_Dyn2}) to be asymptotically stable. Hence, the admissible control concept in this paper is less conservative than that in \cite{Beard1997Auto,Al-Tamimi2008TSC,Jiang2015TAC}.
 
Assume that the baseline control $\boldsymbol{u}_b$ developed using the nominal system (\ref{eq:ASV_Dyn_nom}) of the ASV (\ref{eq:ASV_Dyn2}) is an admissible control law for the uncertain system (\ref{eq:ASV_Dyn2}). Let $\Delta\left(t\right)$ be the overall uncertainties in (\ref{eq:ASV_Dyn2}). Without loss of generality, it is assumed that $\Delta\left(t\right)$ is bounded, namely $\Vert\Delta\left(t\right)\Vert_{\mathcal{L}_{\infty}}\leq \bar{\Delta}$ where $\Vert\cdot \Vert_{\mathcal{L}_{\infty}}$ denotes the $\mathcal{L}_\infty$ norm.  In this paper, the objective of the tracking control in obstacle-free environment is to ensure that the ASV system (\ref{eq:ASV_Dyn2}) can track its desired behaviour defined by its nominal system (\ref{eq:ASV_Dyn_nom}), namely $\Vert\boldsymbol{x}-\boldsymbol{x}_m\Vert_2\to{0}$ as $t\to\infty$.  Let $\boldsymbol{e}_t=\boldsymbol{x}_t-\boldsymbol{x}_{m,t}$ be  the tracking error of the ASV at the time instant $t$. The following assumption is made for a admissible baseline control $\boldsymbol{u}_b$ according to Definition \ref{def: FeasCntrl} and Theorem 4.18 in \cite{Khalil2002Book} (Chapter 4, Page 172).  
\begin{assumption}\label{assump:BaselineC}
The baseline control law $\boldsymbol{u}_b$ is feasible with respect to  the uncertain system (\ref{eq:ASV_Dyn2}), and there exists a continuously differentiable function $\mathbb{V}\left(\boldsymbol{s}_{t}\right)$ associate with $\boldsymbol{u}_b$ such that 
\begin{equation}
\begin{array}{c}
    \mu_1\left(\Vert\boldsymbol{e}_{t}\Vert_2\right) \leq \mathbb{V}\left(\boldsymbol{e}_{t}\right)\leq \mu_2\left(\Vert\boldsymbol{e}_{t}\Vert_2\right) \\
    \mathbb{V}\left(\boldsymbol{e}_{t+1}\right)-\mathbb{V}\left(\boldsymbol{e}_{t}\right)\leq - \mathbb{W}_1\left(\boldsymbol{e}_{t}\right)+\mu_3\left(\Vert\Delta\left(\boldsymbol{e}_t\right)\Vert_2\right) \\
    \mathbb{W}_1\left(\boldsymbol{e}_{t}\right) > \mu_3\left(\Vert\Delta\left(t\right)\Vert_2\right),\; \forall \Vert\boldsymbol{e}_t\Vert_2 > c_\Delta
\end{array}
\end{equation}
where $\mu_1\left(\cdot\right)$, $\mu_2\left(\cdot\right)$, and $\mu_3\left(\cdot\right)$ are class $\mathcal{K}$ functions, $\mathbb{W}\left(\boldsymbol{e}_{t}\right)$ is a continuous positive definite function, and $c_\Delta$ is a constant related to the upper bound of system uncertainty 
\end{assumption}

Assumption \ref{assump:BaselineC} is possible in real world. One can treat the nominal model (\ref{eq:ASV_Dyn_nom}) as a linearized model of the overall ASV system (\ref{eq:ASV_Dyn2}) around a certain equilibrium.  Assumption \ref{assump:BaselineC} presents the basic design requirements for the baseline control law.  With a baseline control law satisfying Assumption $1$, we  could obtain two advantages which makes the RL process more efficient. Firstly, it can ensure that the reward function $R_t$ is bounded, implying that both  $V_{\boldsymbol{\pi}}(\boldsymbol{s}_{t+1})$ and $Q\left(\boldsymbol{s}_{t}, \boldsymbol{u}_{l, t}\right)$ are bounded. Secondly, it could provide a ``warm'' start for the RL process.

With a feasible baseline control, the uncertain ASV system can be ensured to be stable during the entire learning process. In the stability analysis,  we ignore the entropy term $\mathcal{H}\left(\boldsymbol{\pi}\right)$, as it will converge to zero in the end and it is only introduced to regulate the exploration magnitude. Hence, exploration noises will be set to be zero. Now, we present Theorem \ref{thm:Stab} to demonstrate the closed-loop stability of the ASV system (\ref{eq:ASV_Dyn2}) under the composite control law (\ref{eq:entireCntrl}).
\begin{theorem}[\textbf{Stability of tracking control}] \label{thm:Stab}
Suppose Assumption \ref{assump:BaselineC} holds. The overall control law $\boldsymbol{u}^i = \boldsymbol{u}_b + \boldsymbol{u}_{l}^i$ can always stabilize the ASV system (\ref{eq:ASV_Dyn2}), where $\boldsymbol{u}_{l}^i$ represents the RL control law from $i$-th iteration, and $i=0$, $1$, $2$, ... $\infty$.
\end{theorem}
\begin{proof}
The details of proof can be found in Appendix \ref{app:Theorem_Stab_TrackCntrl}.
\end{proof}

\section{Simulation results} \label{sec:NumSim}
In this section, the proposed learning-based control algorithm is implemented to the trajectory tracking control of a supply ship model presented in \cite{Skjetnea2005Auto, Peng2013}. The ASV has two actuators that are a propeller and a rudder in the rear.  Hence, we mainly consider two control inputs in the design, which are $\tau_u$ for the surge speed control and $\tau_r$ for the heading control, respectively. By default, the sway speed is not controlled, which implies $\tau_v=0$ in the simulations. Model parameters are summarized in Table \ref{tab:HydroTab} in Appendix \ref{app:Sim_Config}.  The unmodeled dynamics in the simulations are given by $g_1 = 0.279uv^2 + 0.342 v^2r$, $g_2 = 0.912u^2v$, and $ g_3 = 0.156ur^2+0.278urv^3$, respectively. 
The based-line control law $\boldsymbol{u}_b$ is designed based on a nominal model with the following simplified linear dynamics in terms of the backstepping control method \cite{Khalil2002Book, Zhang2018TIE}.
\begin{equation}
    \boldsymbol{M}_m\dot{\boldsymbol{\nu}}_m = \boldsymbol{\tau}-\boldsymbol{D}_m{\boldsymbol{\nu}}_m \label{eq:LinearModel}
\end{equation}
where $\boldsymbol{M}_m=diag\left\{M_{11},\;M_{22},\;M_{33}\right\}$. $\boldsymbol{D}_m=diag\left\{-X_{v},-Y_{v},\;-N_{r}\right\}$.

In the simulation, a motion planner is employed to generate the reference trajectories. The motion planner is expressed as
\begin{equation}
       \dot{\boldsymbol{\eta}}_r = \boldsymbol{R}\left(\boldsymbol{\eta}_r\right)\boldsymbol{\nu}_r,\quad
       \dot{\boldsymbol{\nu}}_r= \boldsymbol{a}_r \label{eq:MotionPlanner}
\end{equation}
where ${\boldsymbol{\eta}}_r=\left[x_r,y_r,\psi_r\right]^T$ is the generalized reference position vector, ${\boldsymbol{\nu}}_r=\left[u_r,0,r_r\right]^T$ is the generalized reference velocity vector, and $\boldsymbol{a}_r=\left[\dot{u}_r,0,\dot{r}_r\right]^T$.  

Four simulations scenarios are performed in this section. In the first scenario, our algorithm is implemented to an obstacle-free environment to show the closed-loop stability of the tracking control. In the second scenario, some fixed obstacles are added to the environment to demonstrate both the closed-loop stability and the collision avoidance capability of our proposed algorithm.  In the third scenario,  the proposed algorithm is applied to the environment with both still and moving obstacles. The efficiency of our algorithm is demonstrated via the comparison with the RL without baseline control. In the last scenario, the simulation was conducted at different values of the parameters $c$ in the reward function $R_{2,t}$ to illustrate the impact of $c$ on the collision avoidance performance.

\subsection{Trajectory tracking control without obstacles} \label{subsec:Traj_NoObs}
\begin{figure}[bp]
    \centering
    \includegraphics[width=0.38\textwidth]{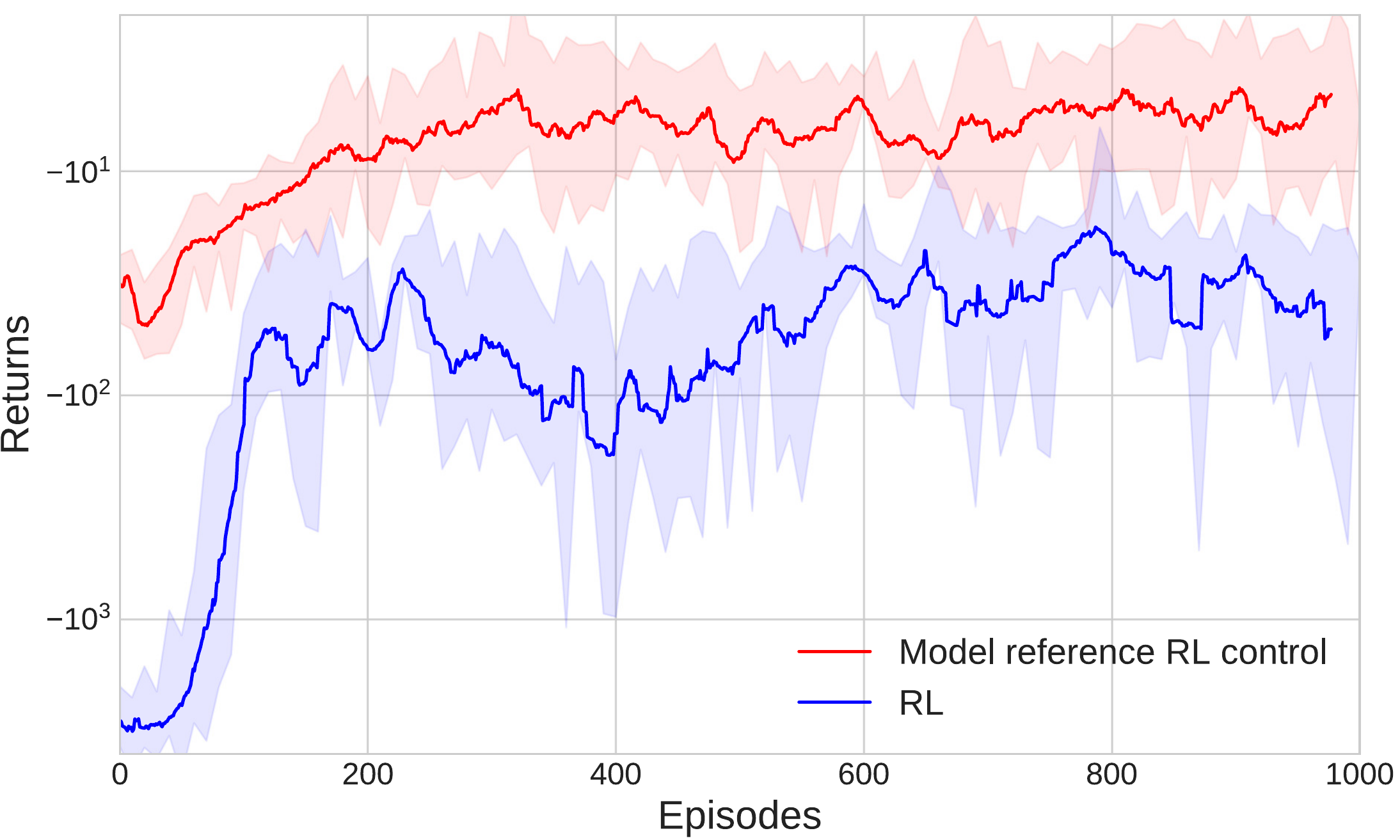}
    \caption{Learning curves of two RL algorithms at training (One episode is a training trial, and $1000$ time steps per episode) }
    \label{fig:learningCurve}
\end{figure}
\begin{figure*}[tbp]
 \centering
  \subfloat[Model reference RL control]{
	\begin{minipage}[c][1\width]{
	   0.32\textwidth}
	   \centering
	   \includegraphics[width=1\textwidth]{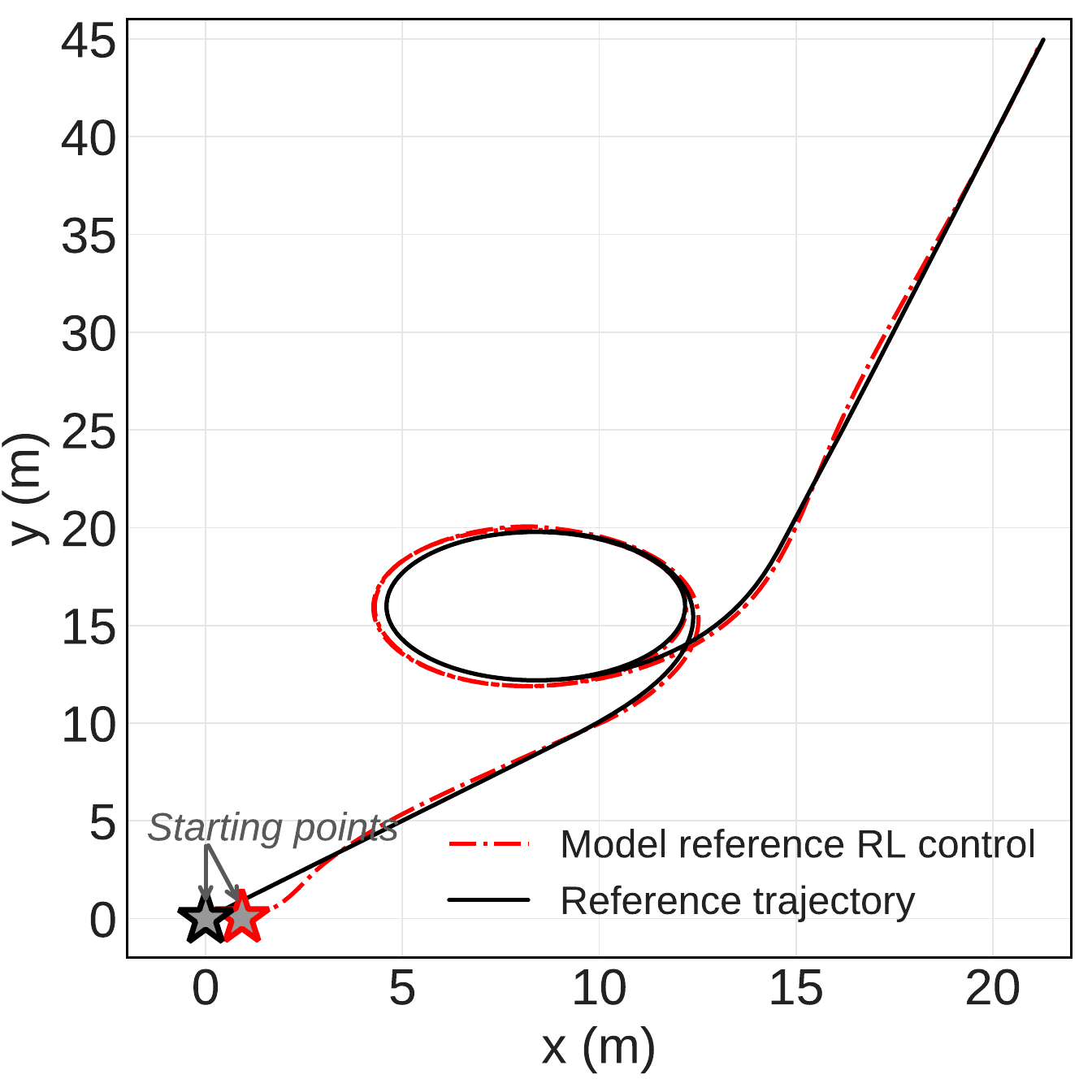}\vspace{-4mm}
	\end{minipage}}
 \hfill 	
  \subfloat[Only deep RL]{
	\begin{minipage}[c][1\width]{
	   0.32\textwidth}
	   \centering
	   \includegraphics[width=1\textwidth]{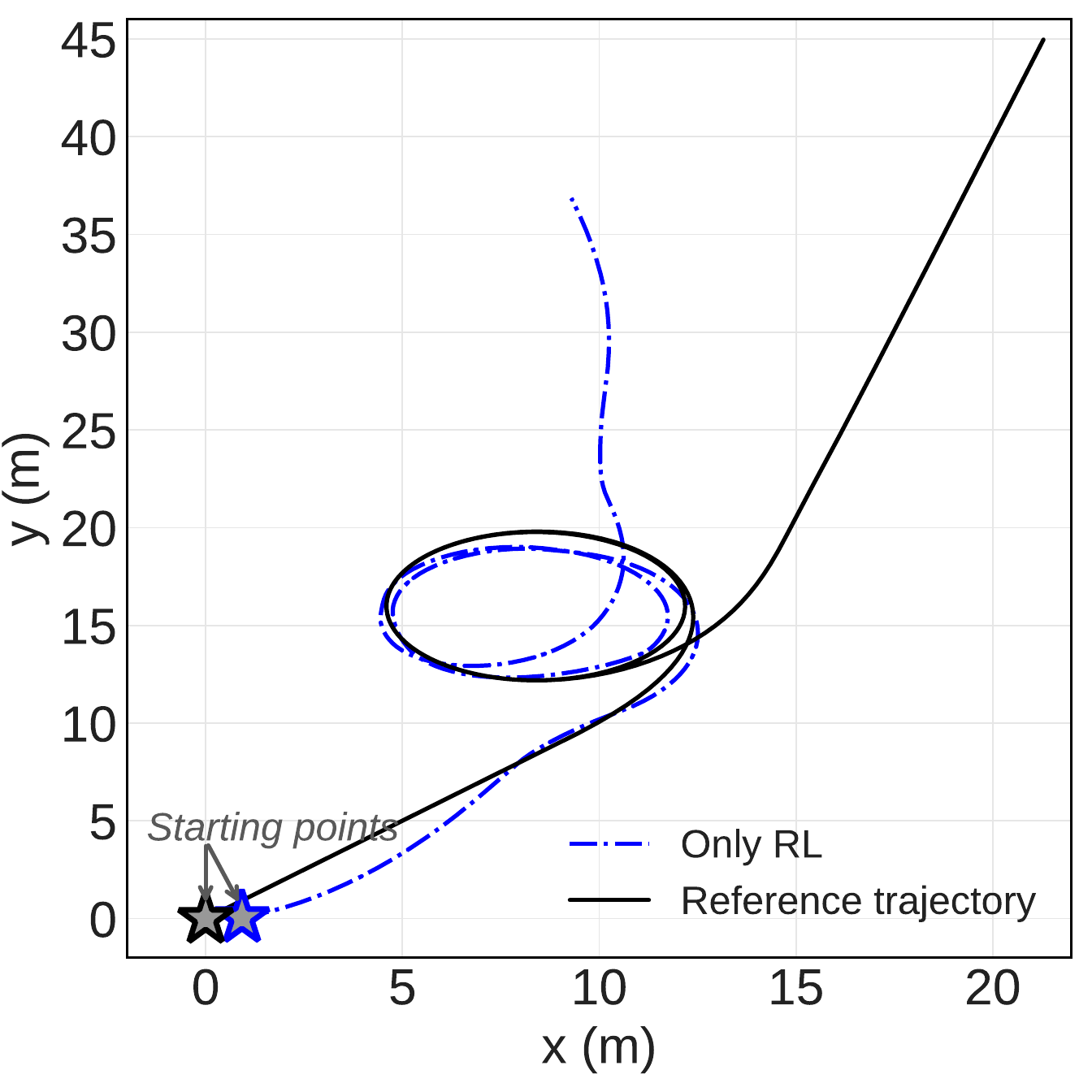}\vspace{-4mm}
	\end{minipage}}
 \hfill	
  \subfloat[Only baseline control]{
	\begin{minipage}[c][1\width]{
	   0.32\textwidth}
	   \centering
	   \includegraphics[width=1\textwidth]{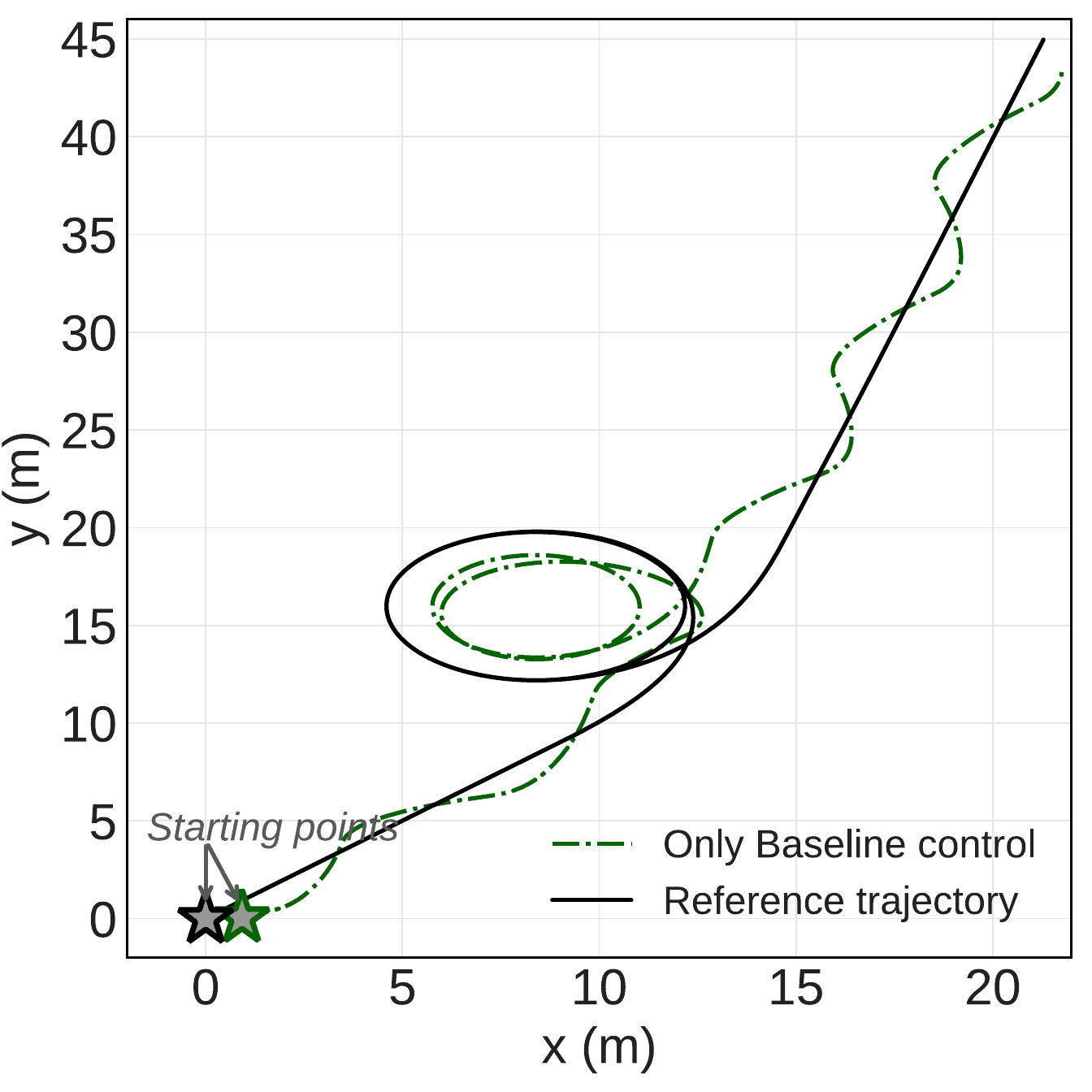}\vspace{-4mm}
	\end{minipage}}
\caption{Trajectory tracking results of the three algorithms}
\label{fig:TrajTracking_Case2}
\end{figure*}

In the first simulation, the initial position vector  ${\boldsymbol{\eta}}_r\left(0\right)$ is chosen to be ${\boldsymbol{\eta}}_r\left(0\right)=\left[0,0,\frac{\pi}{4}\right]^T$, and we set $u_r\left(0\right)=0.4$ $m/s$ and $r_r\left(0\right)=0$ $rad/s$. The reference acceleration $\dot{u}_r$ and angular rates are  chosen to be
\begin{eqnarray}
    \dot{u}_r&=&\left\{\begin{array}{cll}
    0.005& m/s^2 &\text{if }  t<20\; s \\
    0    & m/s^2 & \text{otherwise}
    \end{array}
    \right.  \label{eq:RefAcc_1}\\
    \dot{r}_r&=&\left\{\begin{array}{cll}
    \frac{\pi}{600}& rad/s^2 \; & \;  \text{if } 25\; s \leq t<50 \;s \\
    0    & rad/s^2  & \text{otherwise}
    \end{array} 
    \right. \label{eq:RefRate_1}
\end{eqnarray}
The reference signals ${\boldsymbol{\eta}}_r$ and ${\boldsymbol{\nu}}_r$ are calculated using the reference motion planner (\ref{eq:MotionPlanner}) based on the aforementioned initial conditions and the reference acceleration and angular rates given in (\ref{eq:RefAcc_1}) and (\ref{eq:RefRate_1}), respectively.

At the training stage, we uniformly randomly sample $x\left(0\right)$ and $y\left(0\right)$ from $\left(-1.5, 1.5\right)$,  $\psi\left(0\right)$ from $\left(0.1\pi, 0.4\pi\right)$ and $u\left(0\right)$ from $\left(0.2, 0.4\right)$, and we choose $v\left(0\right)=0$ and $r\left(0\right)= 0$. The proposed control algorithm is compared to two benchmark designs: the baseline control $\boldsymbol{u}_0$ and the RL control without $\boldsymbol{u}_0$. Configurations for the training and neural networks are found in Table \ref{tab:RLTab} in Appendix \ref{app:Sim_Config}. The matrices $\boldsymbol{H}_1$ and $\boldsymbol{H}_2$ are chosen to be $\boldsymbol{H}_1=diag\left\{0.025,0.025, 0.0016, 0.005,0.001, 0\right\}$ and $\boldsymbol{H}_2=diag\left\{1.25e^{-3},1.25e^{-3}\right\}$, respectively. During the training process, we repeat the training processes for $1000$ times (i.e., $1000$ episodes). For each episode, the ASV system is run for $100$ $s$. Figure \ref{fig:learningCurve} shows the learning curves of the proposed algorithm (red) and the RL algorithm without baseline control (blue). The learning curves demonstrate that both of the two algorithms will converge in terms of the long term returns. However, our proposed algorithm results in a larger return (red) in comparison to the RL without baseline control (blue). Hence, the introduction of the baseline control helps to increase the sample efficiency significantly, as the proposed algorithm (blue) converges faster to a higher return value. 
\begin{figure}
    \centering
    \includegraphics[width=0.45\textwidth]{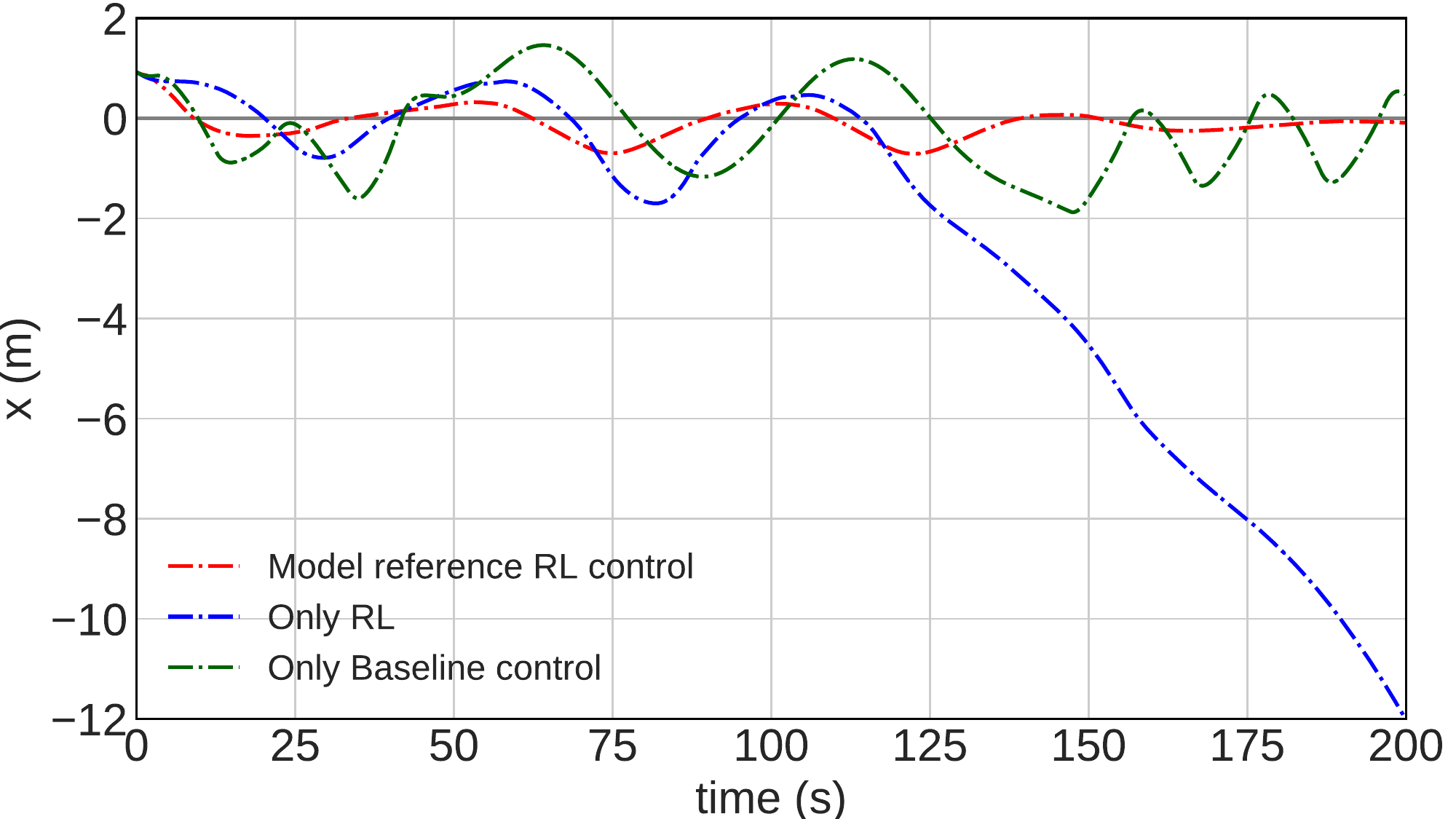}
    \caption{Position tracking errors, $e_x$}
    \label{fig:ErrorX_Case2}
\end{figure}
\begin{figure}
    \centering
    \includegraphics[width=0.45\textwidth]{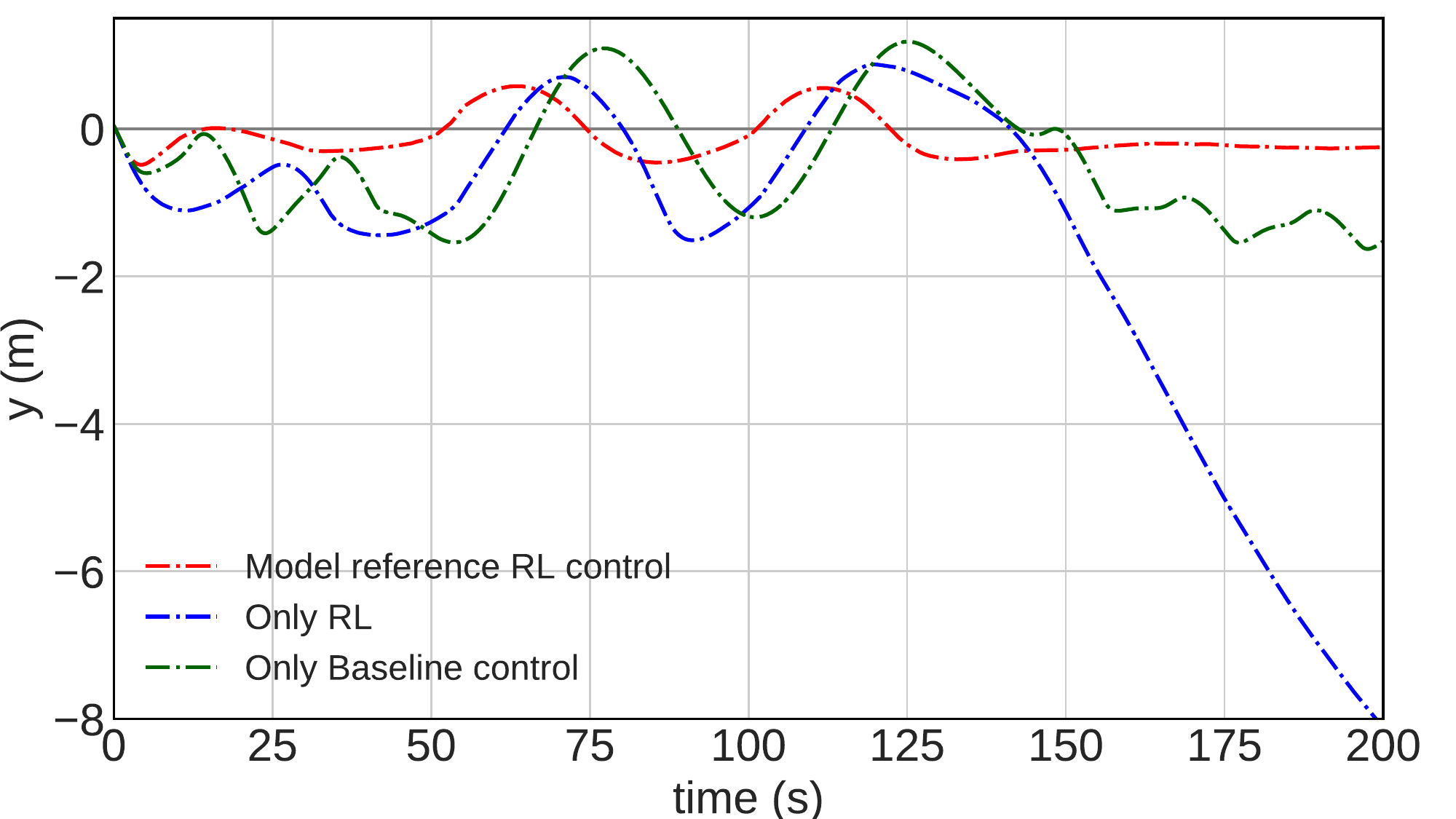}
    \caption{Position tracking errors, $e_y$}
    \label{fig:ErrorY_Case2}
\end{figure}
\begin{figure}
    \centering
    \includegraphics[width=0.45\textwidth]{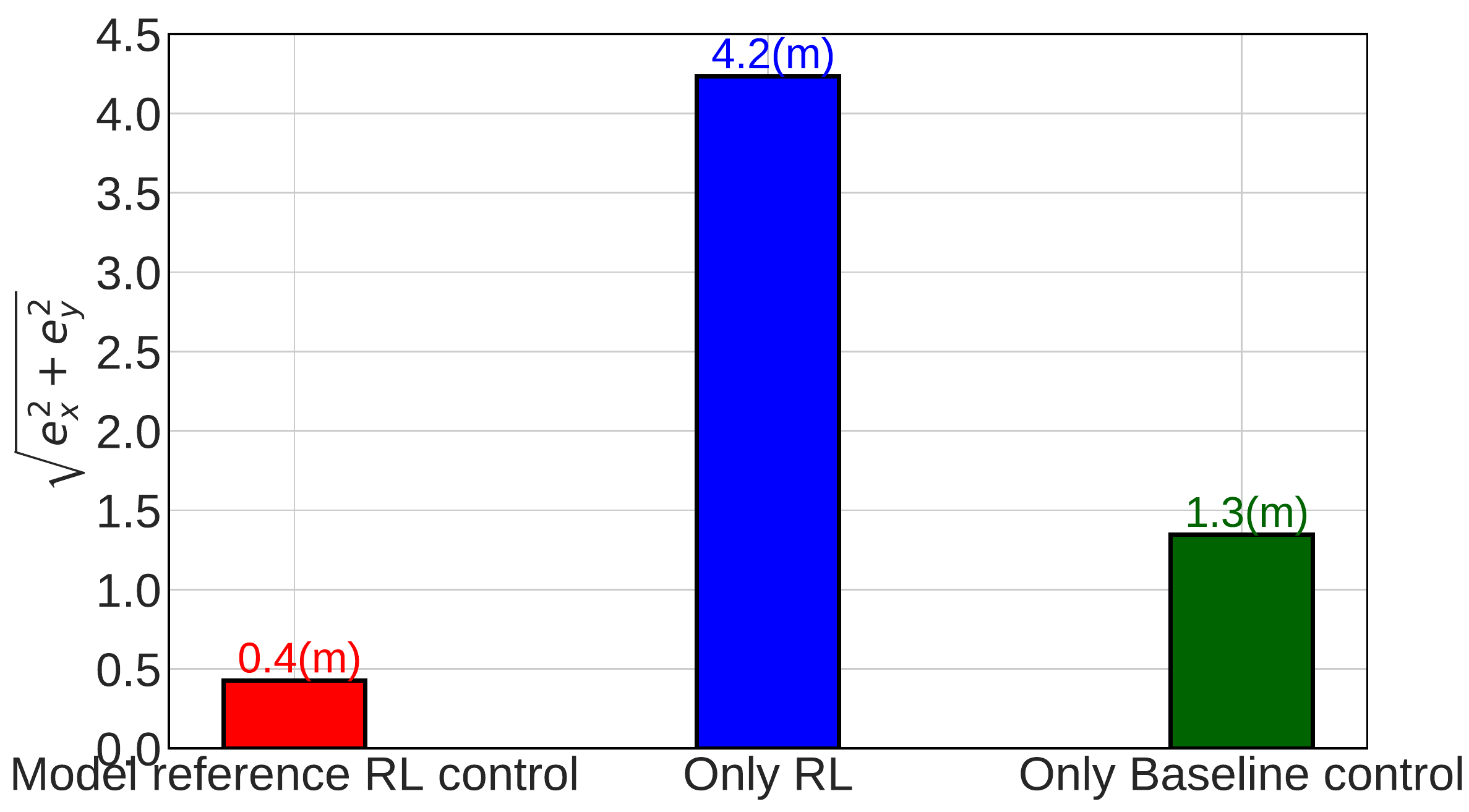}
    \caption{Mean absolute distance errors, $\sqrt{e_x^2+e_y^2}$}
    \label{fig:ErrorDist_Case2} 
\end{figure}
\begin{figure}
    \centering
    \includegraphics[width=0.45\textwidth]{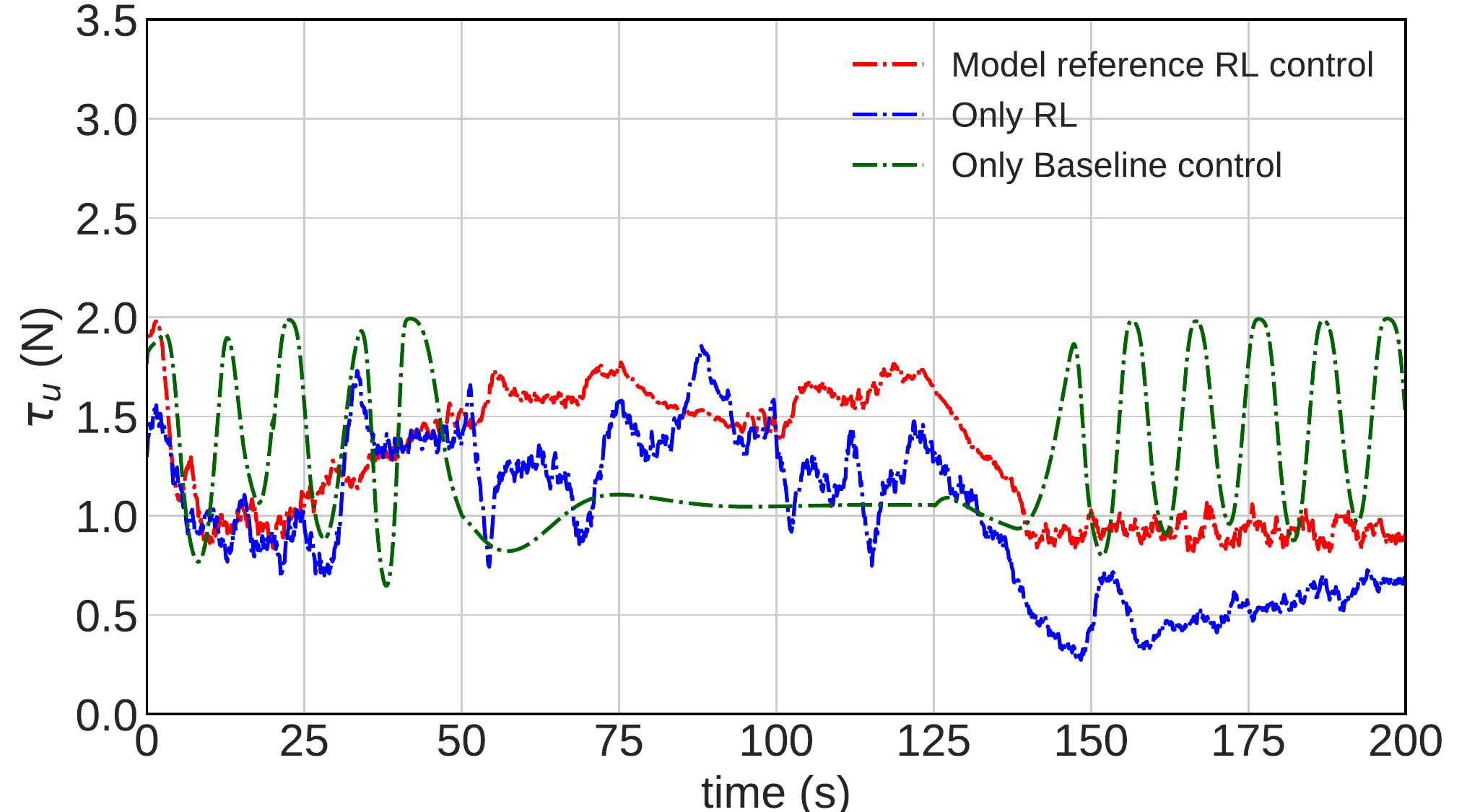}
    \caption{Control inputs, $\tau_u$}
    \label{fig:Tauu_Case2}
\end{figure}
\begin{figure}
    \centering
    \includegraphics[width=0.45\textwidth]{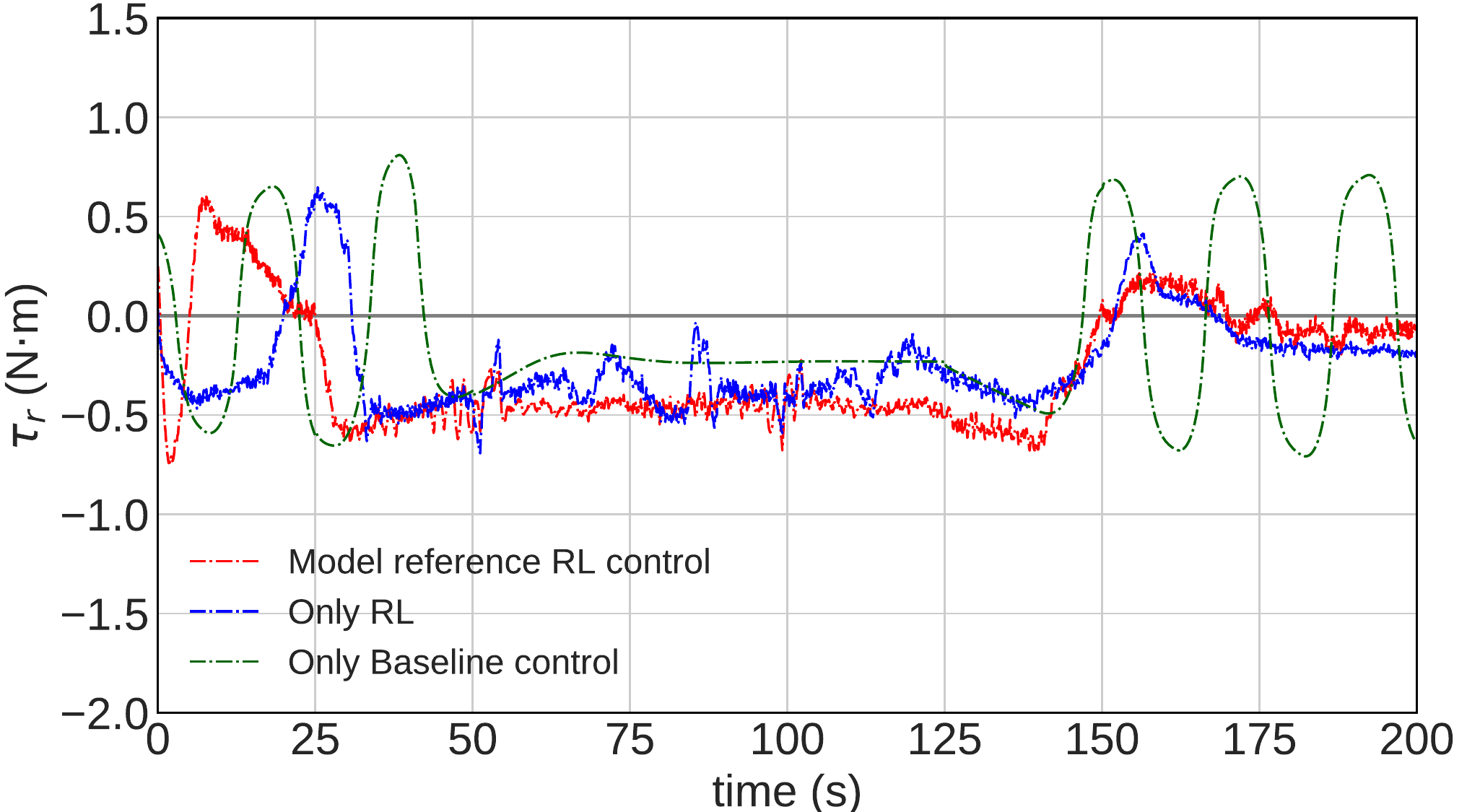}
    \caption{Control inputs, $\tau_r$}
    \label{fig:Taur_Case2}
\end{figure}

At the evaluation stage, we run the ASV system for $200$ $s$ and change the reference trajectory to demonstrate whether the control law can ensure stable trajectory tracking.  At the second evaluation, the reference angular acceleration is changed to
\begin{equation}
    \dot{r}_r=\left\{\begin{array}{cll}
    \frac{\pi}{600}& rad/s^2 &\text{if }\;\; 25\;s \leq  t<50\;s \\
    -\frac{\pi}{600}& rad/s^2 &\text{if } 125 \;s\leq  t<150\;s \\
    0    &rad/s^2  & \text{otherwise}
    \end{array}
    \right.
\end{equation}
The trajectory tracking performance of the three algorithms (our proposed algorithm, the baseline control $\boldsymbol{u}_0$, and only RL control) is shown in Figure \ref{fig:TrajTracking_Case2}. As observed in Figure \ref{fig:TrajTracking_Case2}.(b), the ASV trajectory by the control law learned merely using deep RL tends to drift away from the designed trajectory. It implies that only deep RL could not ensure the closed-loop stability. In addition, the baseline control itself fails to achieve acceptable tracking performance mainly due to the existence of system uncertainties. By combining the baseline control and deep RL,  the trajectory tracking performance is improved dramatically, and the closed-loop stability is guaranteed. The tracking errors in the $X$- and $Y$- coordinates of the inertial frame are summarized in Figure \ref{fig:ErrorX_Case2} and \ref{fig:ErrorY_Case2}, respectively. The ASV reaches its steady state after 80 $s$ as shown in Figures \ref{fig:ErrorX_Case2} and \ref{fig:ErrorY_Case2}.  Hence, we present the absolute average distance errors from 80 $s$ to 200 $s$ to compare the tracking accuracy of the three algorithms  in Figure \ref{fig:ErrorDist_Case2}. The introduction of the deep RL increases the tracking performance of the baseline control law substantially.  The control inputs are provided in Figures \ref{fig:Tauu_Case2} and \ref{fig:Taur_Case2}.

\subsection{Tracking control with fixed obstacles} \label{subsec:Sim_ColAvoid_StillObs}
\begin{figure}[tbp]
 \centering
  \subfloat[Model-reference RL control]{
	\begin{minipage}[c][1\width]{
	   0.245\textwidth}
	   \centering
	   \includegraphics[width=1\textwidth]{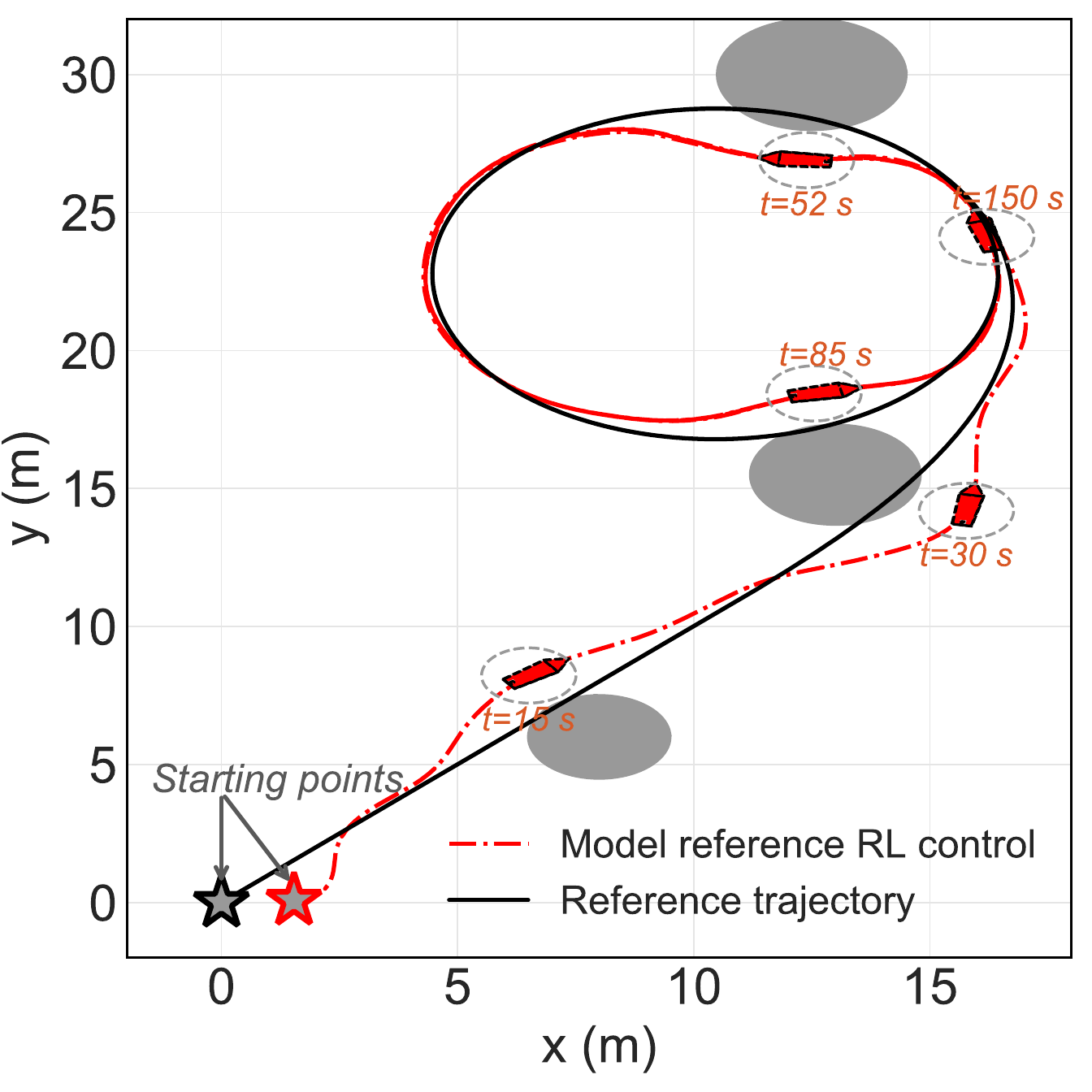}\vspace{-4mm}
	\end{minipage}}
  \subfloat[Only deep RL]{
	\begin{minipage}[c][1\width]{
	   0.245\textwidth}
	   \centering
	   \includegraphics[width=1\textwidth]{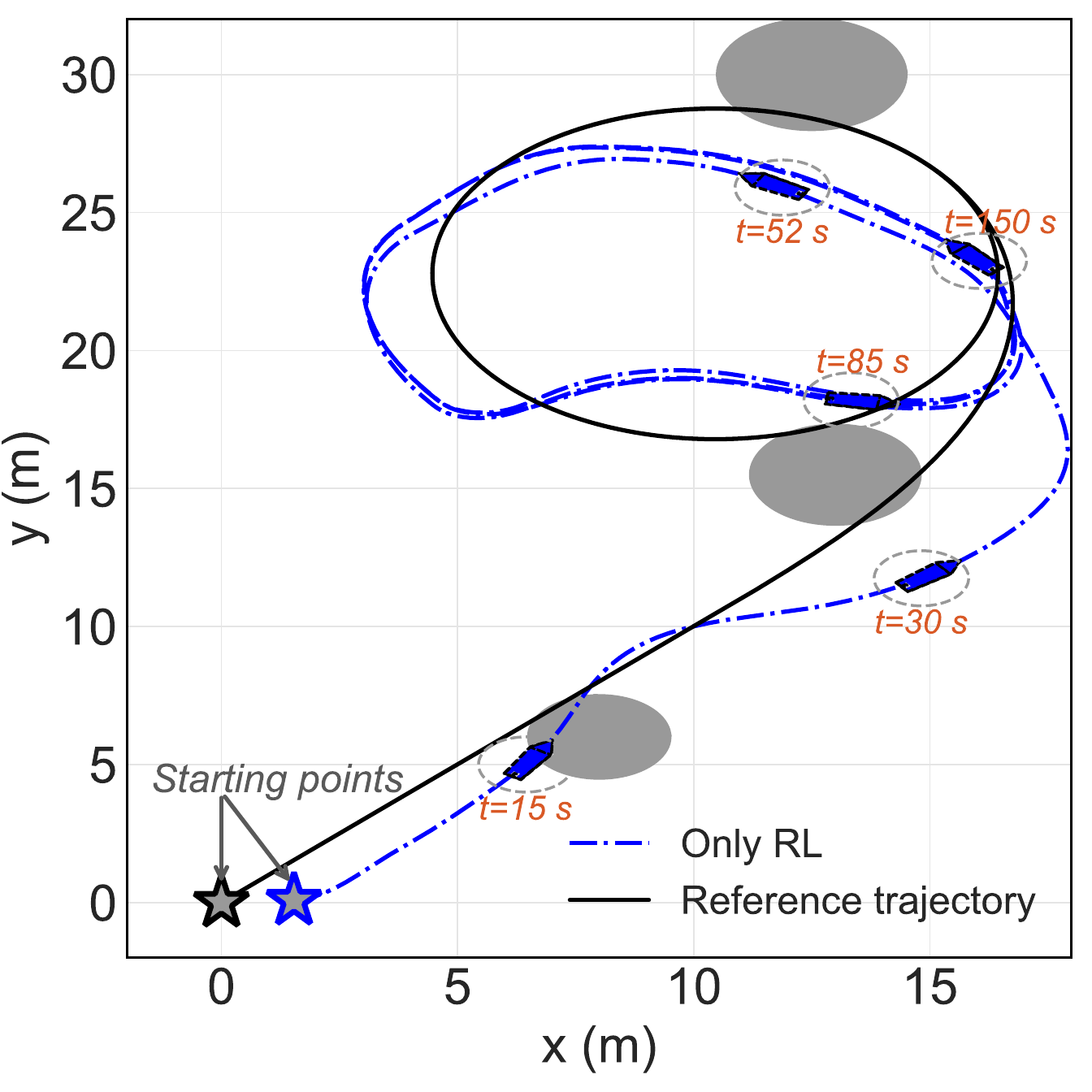}\vspace{-4mm}
	\end{minipage}}
\caption{Tracking control with fixed obstacles (Obstacle radii: $1.5$ $m$, $1.8$ $m$, and $2.0$ $m$ (\emph{from the lower to the upper}))}
\label{fig:TrajTracking_Obs}
\end{figure}

 In the second simulation, the initial position vector  ${\boldsymbol{\eta}}_r\left(0\right)$ is chosen to be  the same as the case in Section  \ref{subsec:Traj_NoObs}. We set $u_r\left(0\right)=0.7$ $m/s$ and $r_r\left(0\right)=0$ $rad/s$. The reference acceleration is set as $\dot{u}_r=0$ $m^2/s$. The angular rate is  
 \begin{equation}
    \dot{r}_r=\left\{\begin{array}{cll}
    \frac{\pi}{800}& rad/s^2 \; & \;  \text{if } 20\; s \leq t<50 \;s \\
    0    & rad/s^2  & \text{otherwise}
    \end{array}
    \right.
\end{equation}
Initial states of the ASV are randomly generated as summarized in Section \ref{subsec:Traj_NoObs}. Three fixed obstacles are added to the simulation environment as shown in Figure \ref{fig:TrajTracking_Obs}.a, which have a radius of $1.5$ $m$, $1.8$ $m$, and $2.0$ $m$, respectively. The detection radius for the ASV is $d_d=7.5$ $m$, and the radius of the ASV is $d_{a}=1$ $m$. The deep neural network configurations and training set-up for the collision avoidance scenario is the same as shown in Table \ref{tab:RLTab} in Appendix \ref{app:Sim_Config}.  We choose $q_{c,i}=1$ and $c_{i}=25$ for all obstacles in the simulation.
\begin{figure}[tbp]
    \centering
    \includegraphics[width=0.45\textwidth]{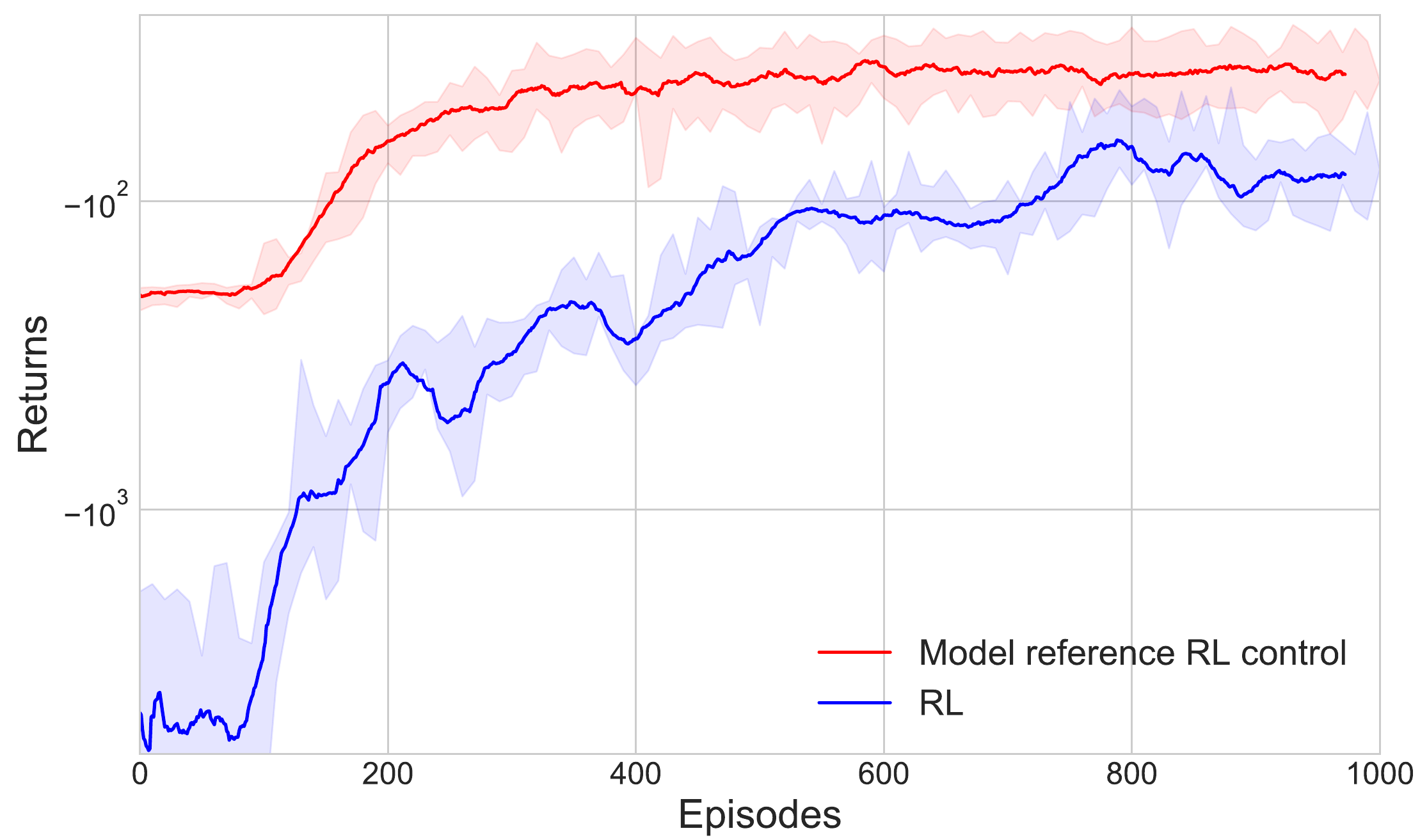}
    \caption{Learning curves of two RL algorithms at training (One episode is a training trial, and $1000$ time steps per episode) }
    \label{fig:learningCurve_Obs}
\end{figure}

At the training stage, $1000$ episodes of training are conducted. For  each episode, the ASV system is run for $100$ $s$. Figure \ref{fig:learningCurve_Obs} shows the learning curves of the proposed algorithm (red) and the RL algorithm without baseline control (blue).

\begin{figure}
    \centering
    \includegraphics[width=0.45\textwidth]{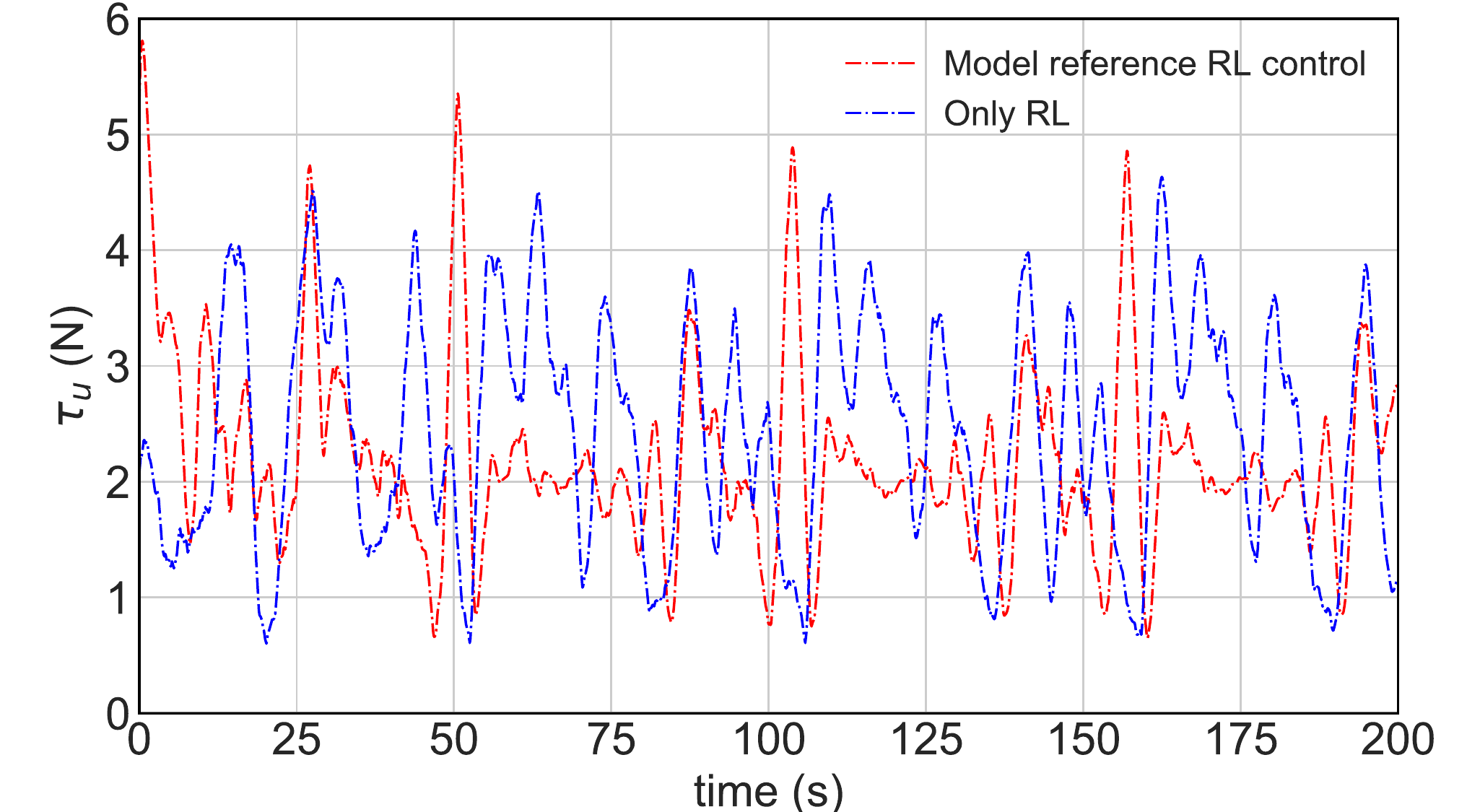}
    \caption{Control inputs, $\tau_u$}
    \label{fig:Tauu_Obs}
\end{figure}

\begin{figure}
    \centering
    \includegraphics[width=0.45\textwidth]{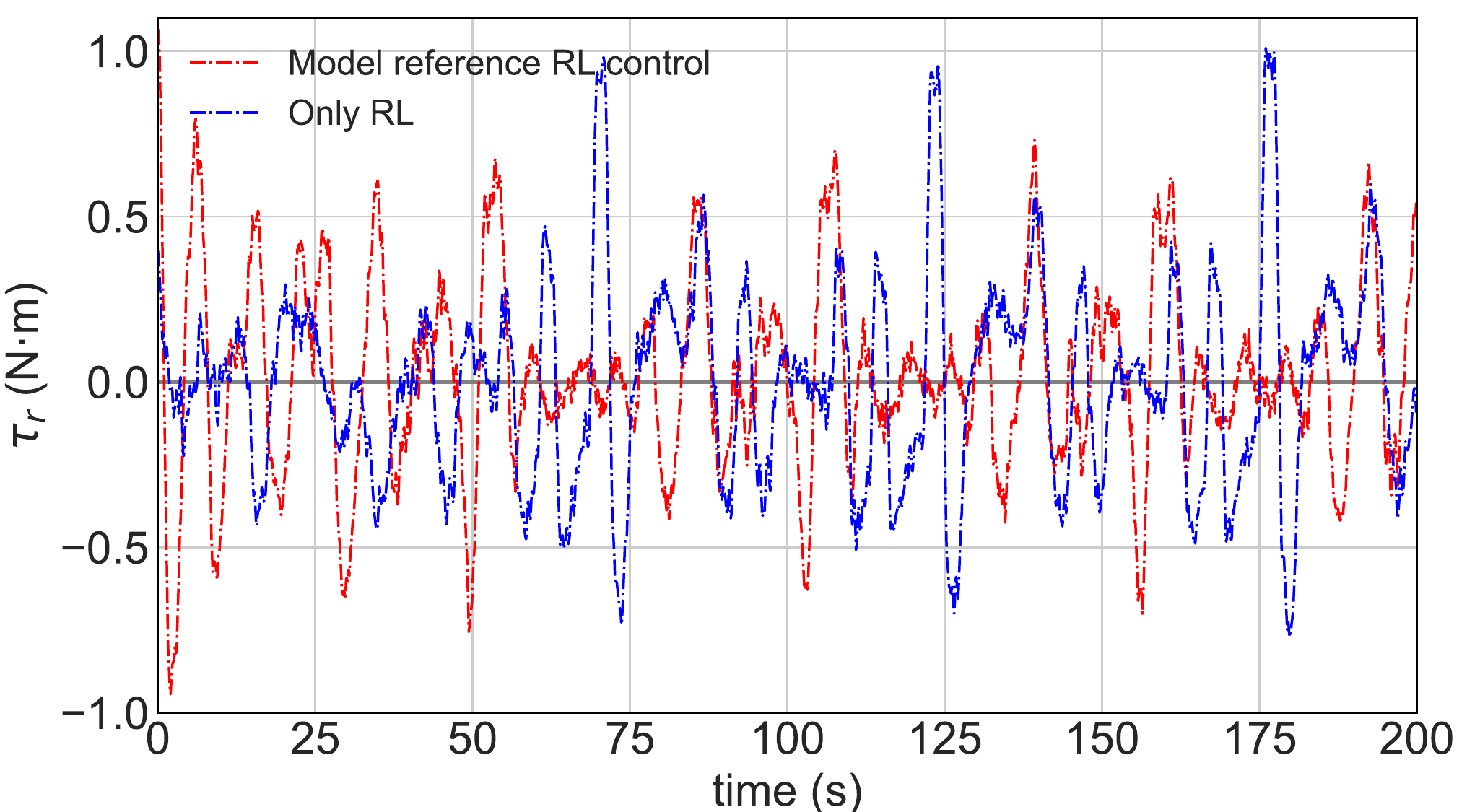}
    \caption{Control inputs, $\tau_r$}
    \label{fig:Taur_Obs}
\end{figure}

At the evaluation stage, we run the ASV system for $200$ $s$ to demonstrate whether the control law can ensure stable trajectory tracking and collision avoidance. The proposed algorithm is compared with the RL algorithm without baseline control. The simulation results of both our algorithm and the RL without baseline control are shown in Figure \ref{fig:TrajTracking_Obs}.  Although the RL without baseline control will converge in returns as shown  Figure \ref{fig:learningCurve_Obs}, the learned control law fails to avoid collision with some obstacle  as demonstrated in Figure \ref{fig:TrajTracking_Obs}.b. However, our algorithm can ensure both the trajectory tracking and the collision avoidance at the same time. The control inputs are shown in Figures \ref{fig:Tauu_Obs} and \ref{fig:Taur_Obs}.

\subsection{Tracking control with fixed obstacles and moving obstacles}
\begin{figure}[tbp]
    \centering
    \includegraphics[width=0.45\textwidth]{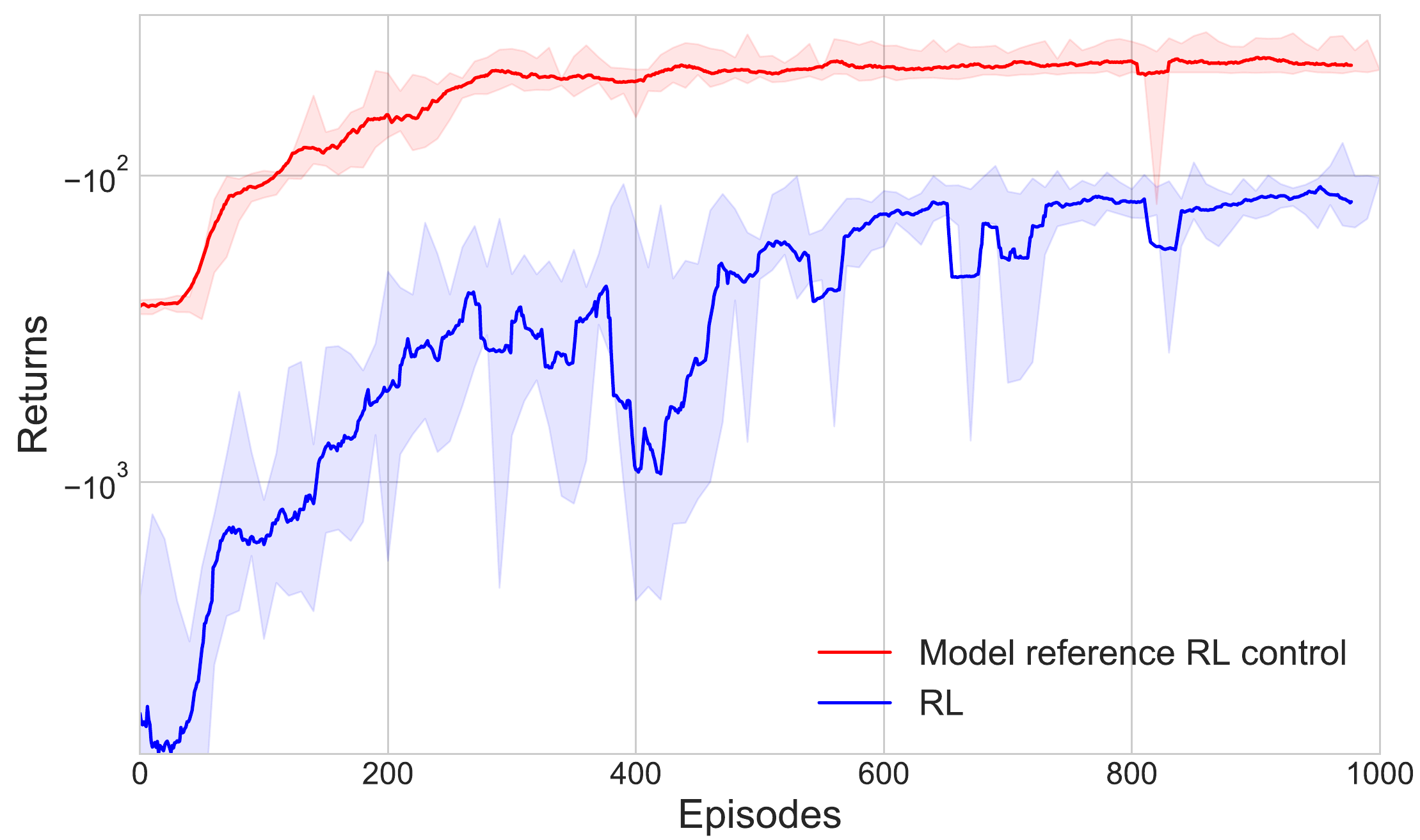}
    \caption{Learning curves of two RL algorithms at training (One episode is a training trial, and $1000$ time steps per episode) }
    \label{fig:learningCurve_MovingObs}
\end{figure}

\begin{figure}[bp]
 \centering
  \subfloat[Model-reference RL control]{
	\begin{minipage}[c][1\width]{
	   0.245\textwidth}
	   \centering
	   \includegraphics[width=1\textwidth]{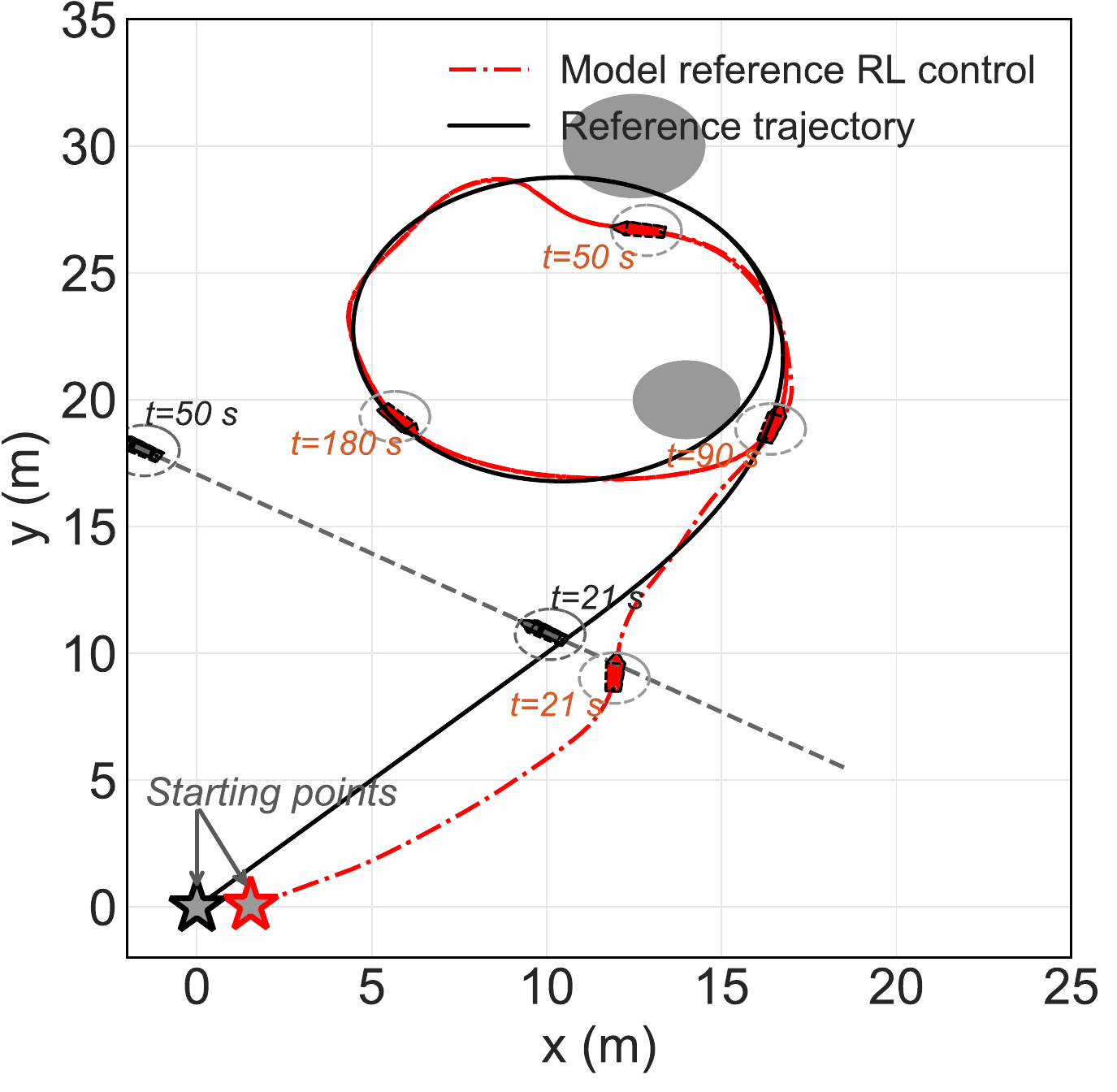}\vspace{-4mm}
	\end{minipage}}
  \subfloat[Only deep RL]{
	\begin{minipage}[c][1\width]{
	   0.245\textwidth}
	   \centering
	   \includegraphics[width=1\textwidth]{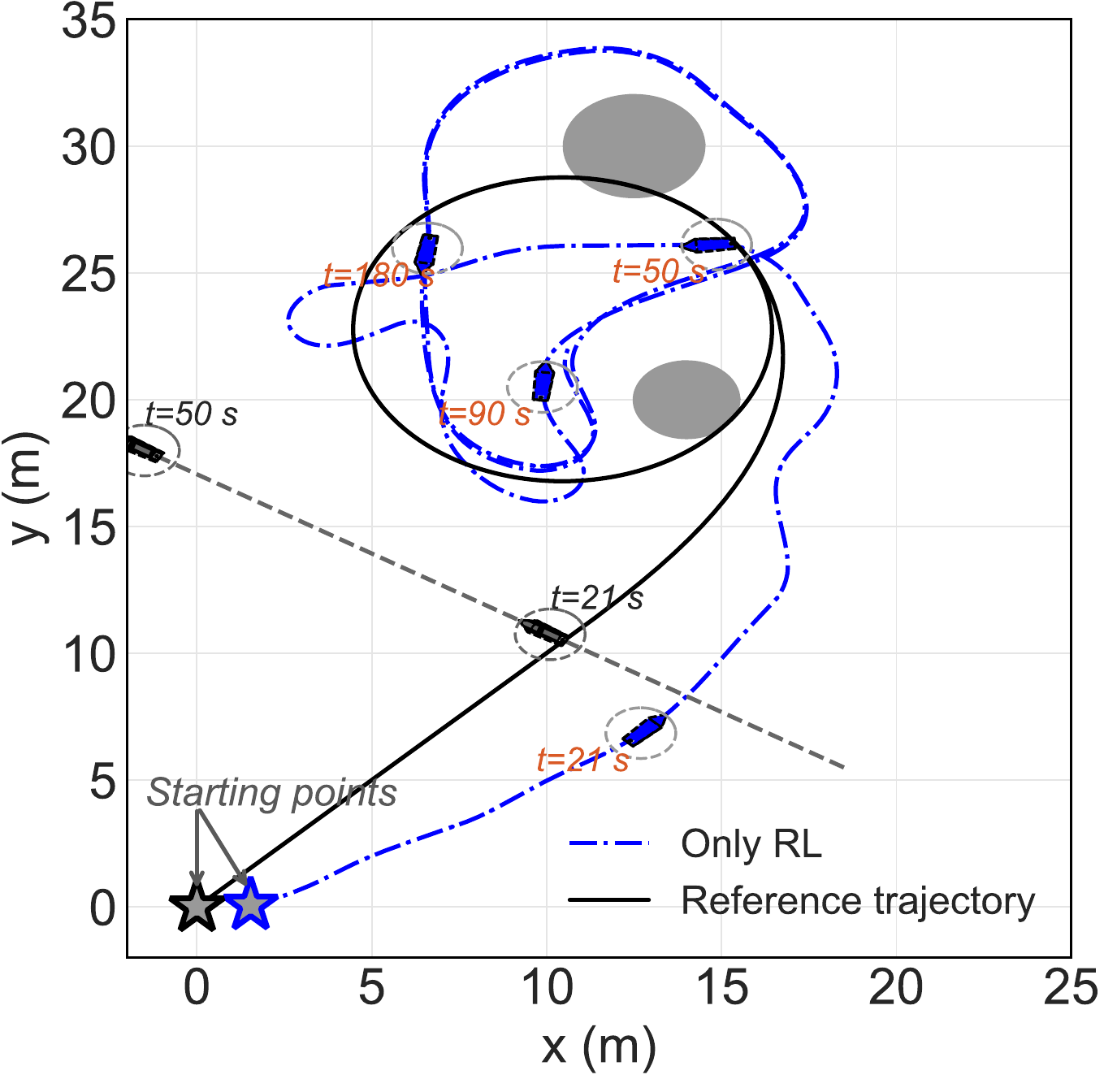}\vspace{-4mm}
	\end{minipage}}
\caption{Tracking control with both fixed and moving obstacles (Radii of fixed obstacles: $1.5$ $m$ and $2.0$ $m$ (\emph{from the lower to the upper}))}
\label{fig:TrajTracking_MovingObs}
\end{figure}

\begin{figure}
    \centering
    \includegraphics[width=0.45\textwidth]{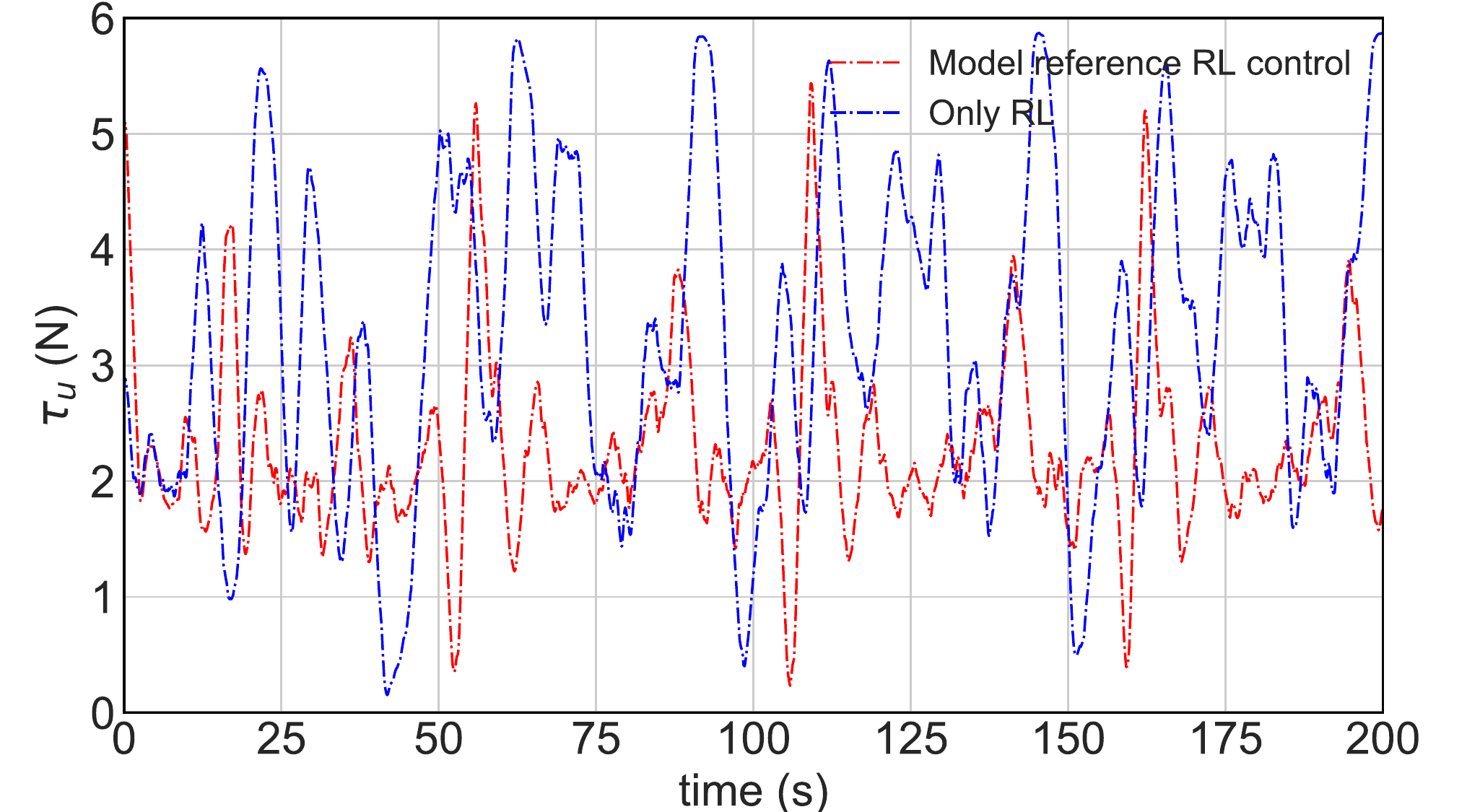}
    \caption{Control inputs, $\tau_u$}
    \label{fig:Tauu_MovingObs}
\end{figure}

\begin{figure}
    \centering
    \includegraphics[width=0.45\textwidth]{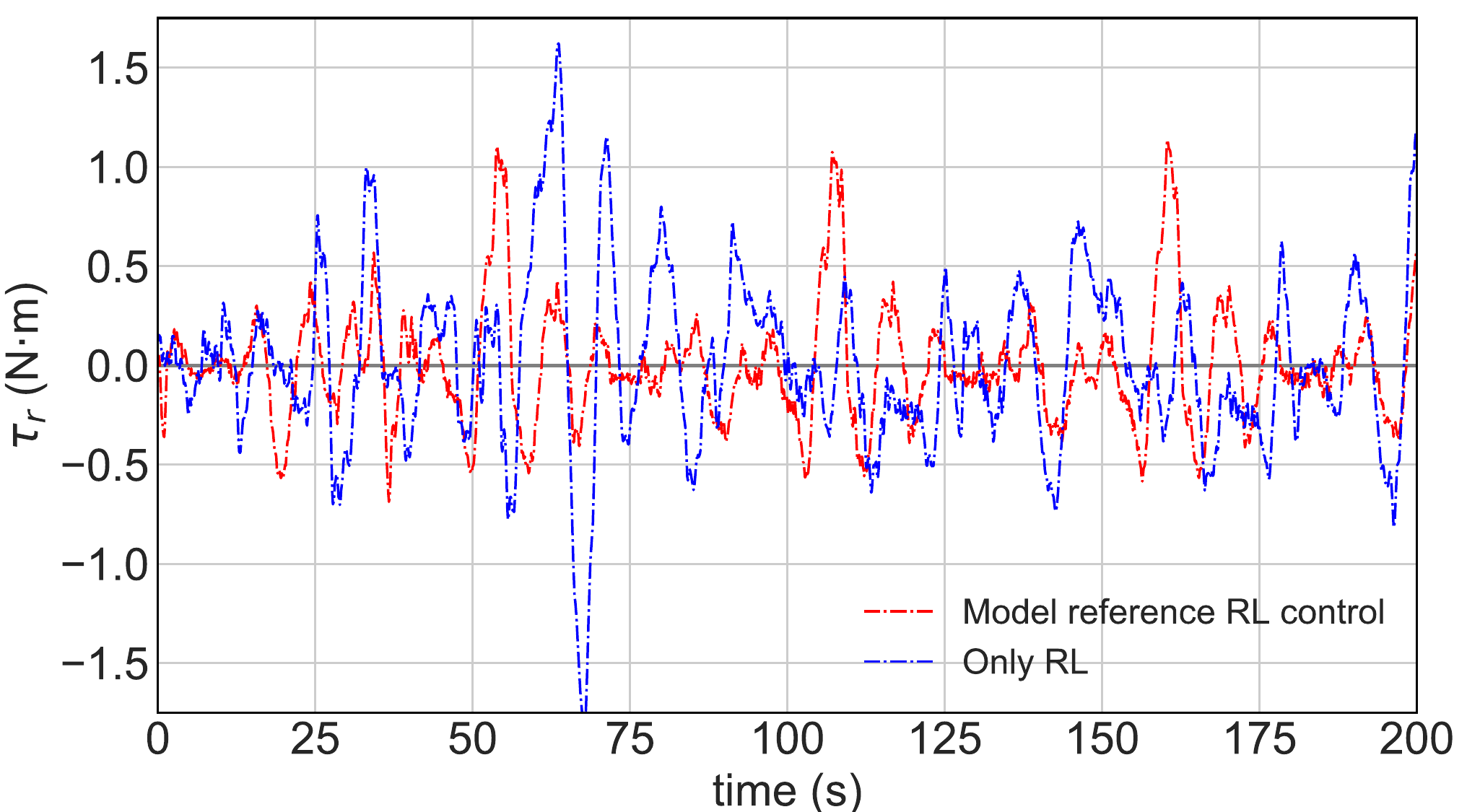}
    \caption{Control inputs, $\tau_r$ }
    \label{fig:Taur_MovingObs}
\end{figure}

In the third simulation, we show the collision avoidance with moving obstacles. The reference trajectory is the same as that in the second simulation in Section \ref{subsec:Sim_ColAvoid_StillObs}. In the simulation, there are two fixed obstacles and one moving obstacle (e.g., another ASV). The moving obstacle has a safe radius of $1$ $m$, and moving with a constant speed with $\boldsymbol{o}_{i,v}=\left[-0.4,\;0.25\right]$ $m/s$ in the simulation.  The training setup is the same as the case in Section \ref{subsec:Sim_ColAvoid_StillObs}. At the evaluation, the ASV system is run for 200 $s$. The learning curves are shown in Figure \ref{fig:learningCurve_MovingObs}.  The trajectory tracking performance of both our algorithm and the RL without baseline control is shown in Figure \ref{fig:TrajTracking_MovingObs}. Although both of the two algorithms can learn a control law with collision avoidance, our algorithm apparently has better tracking performance than the  RL without baseline control.  The control inputs are given in Figures \ref{fig:Tauu_MovingObs} and \ref{fig:Taur_MovingObs}.


\subsection{The impact of different choices of $c$}
\begin{figure*}[tbp]
 \centering
  \subfloat[$c=0.25$]{
	\begin{minipage}[c][1\width]{
	   0.32\textwidth}
	   \centering
	   \includegraphics[width=1\textwidth]{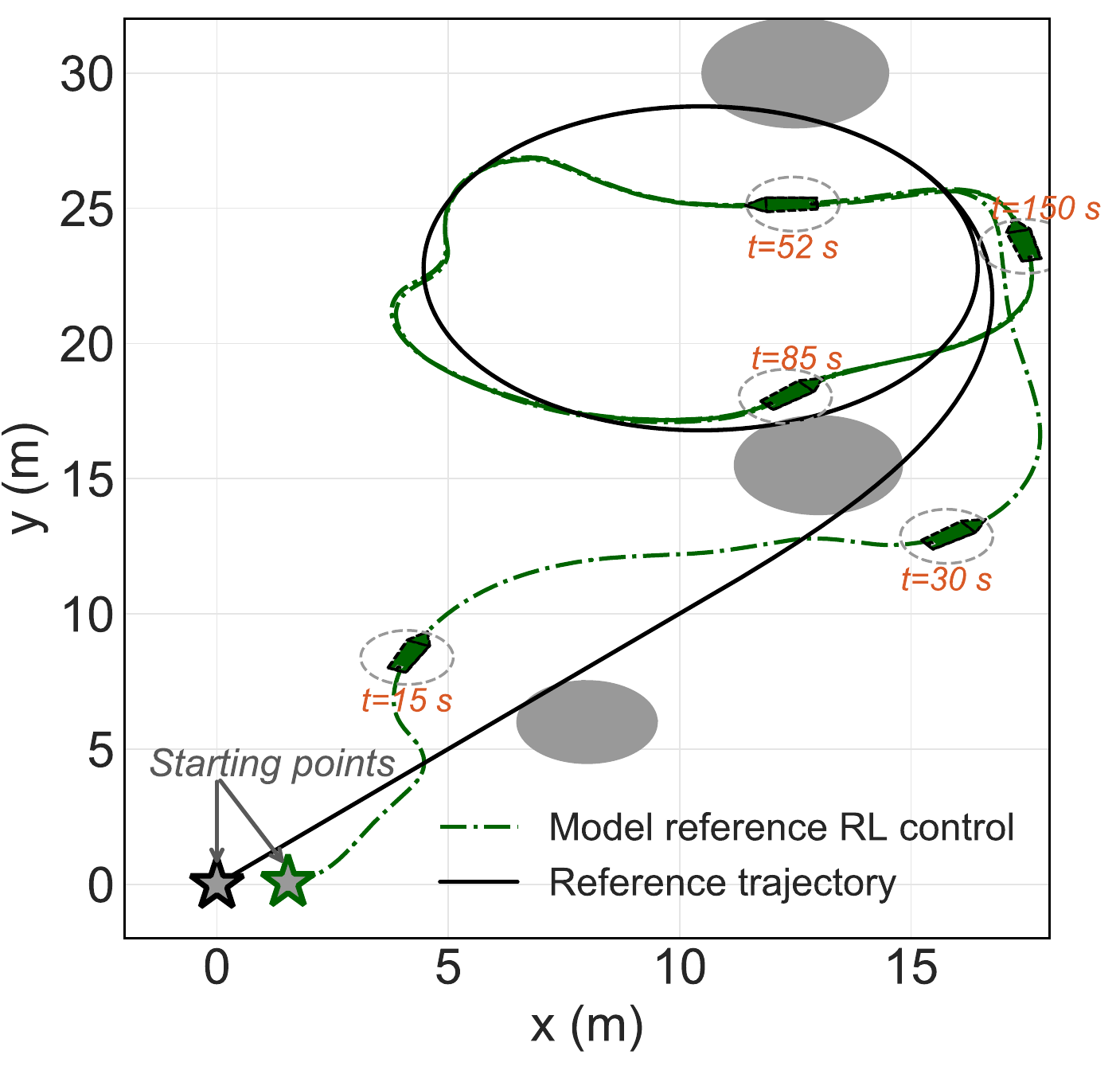}\vspace{-4mm}
	\end{minipage}}
 \hfill 	
  \subfloat[$c=2.5$]{
	\begin{minipage}[c][1\width]{
	   0.32\textwidth}
	   \centering
	   \includegraphics[width=1\textwidth]{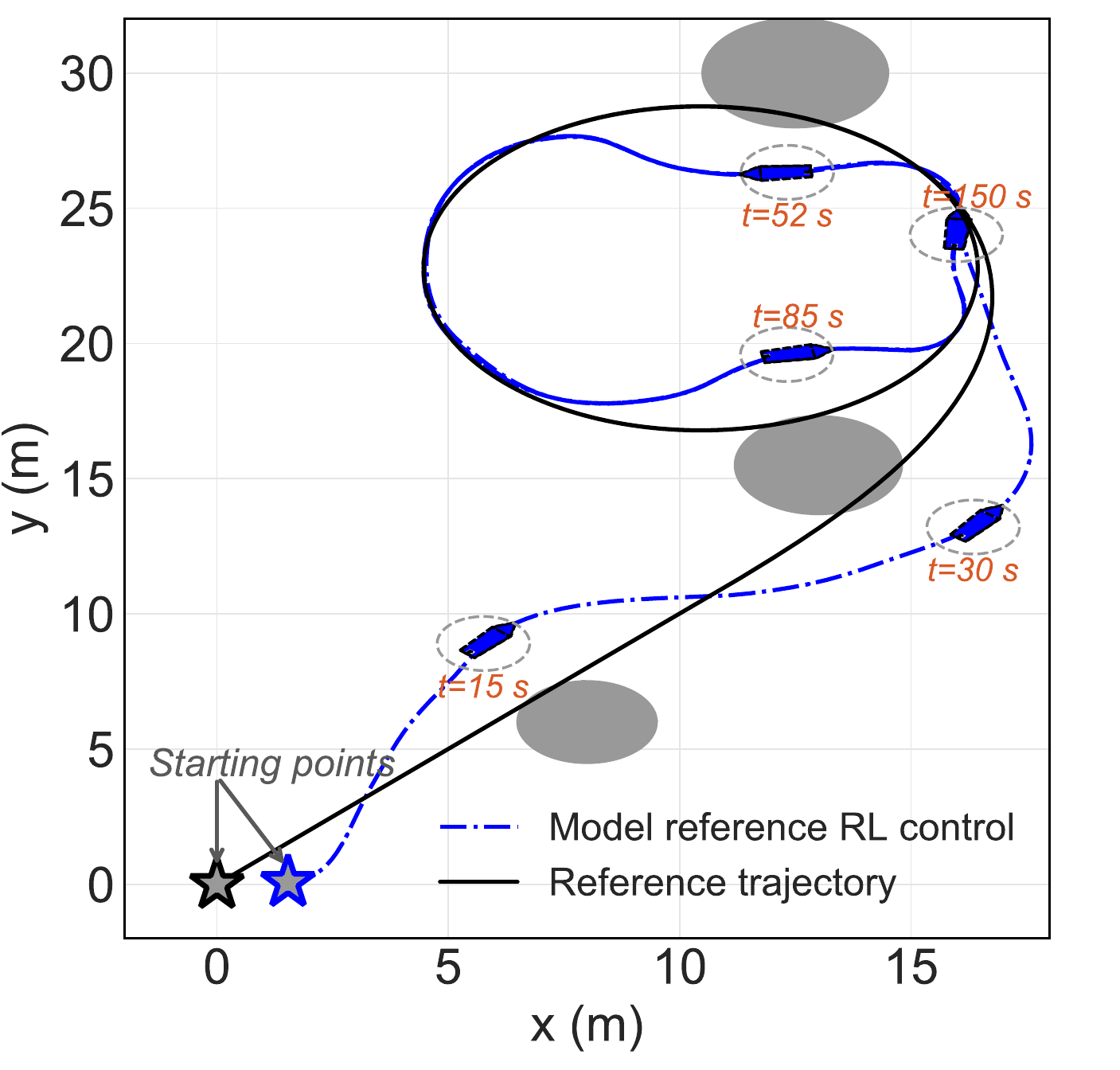}\vspace{-4mm}
	\end{minipage}}
 \hfill	
  \subfloat[$c=25$]{
	\begin{minipage}[c][1\width]{
	   0.32\textwidth}
	   \centering
	   \includegraphics[width=1\textwidth]{./Images/Trajectory_Control_RL_Obs3.pdf}\vspace{-4mm}
	\end{minipage}}
\caption{Collision avoidance performance at different values of $c$}
\label{fig:CollisionAvoidance_cs}
\end{figure*}

In this simulation, we train our model-reference reinforcement learning-based control at the different choices of $c$ for the collision avoidance reward $R_{2, t}$ in (\ref{eq:ASV_Reward_2}). Three choices are considered for $c$, which are $c=0.25$, $c=2.5$, and $c=25$, respectively.  Similar to the simulation environment in Section \ref{subsec:Sim_ColAvoid_StillObs}, three fixed obstacles are considered.  The trajectory tracking performance is summarized in Figure \ref{fig:CollisionAvoidance_cs}. When a smaller $c$ is chosen, the ASV will take more conservative actions to avoid collisions with obstacles as illustrated in Figure \ref{fig:CollisionAvoidance_cs}. This is because a small $c$ will make $R_{2, t}$ change slowly with respect to the distance between the ASV and an obstacle. The slow variation of  $R_{2, t}$ will make the ASV take more conservative actions to avoid collisions.
  
\section{Conclusions} \label{sec:Concl}
In this paper, we presented a novel learning-based control algorithm for ASV systems with collision avoidance. The proposed control algorithm combined a conventional control method with deep reinforcement learning to provide closed-loop stability guarantee, uncertainty compensation, and collision avoidance. Convergence of the learning algorithm was analyzed. We also presented the stability analysis of the tracking control. The proposed control algorithm shows much better performance in both tracking control and collision avoidance than the RL without baseline control. In the future works, we will further analyze the sample efficiency of the proposed algorithm, and extend the design to the scenario with extensive environment disturbances.

\appendix
\numberwithin{equation}{section}
\makeatletter
\renewcommand{\theequation}{A.\arabic{equation}} 
\renewcommand{\thesection}{A\arabic{section}}   
\renewcommand{\thetable}{A\arabic{table}}   
\renewcommand{\thefigure}{A\arabic{figure}}
\setcounter{equation}{0}
\setcounter{table}{0}

\subsection{Proof of Lemma \ref{lem:Pi_Eval}} \label{app:Lemma_Eval}
\begin{proof}
The following entropy-augmented reward function $\hat{R}_{t}$ is introduced.
\begin{equation}
\hat{R}_{t} = {R}_{t}-\gamma \mathbb{E}_{\boldsymbol{s}_{t+1}}\left\{\mathbb{E}_{\boldsymbol{\pi}}\left[\alpha\ln\left(\boldsymbol{\pi}\left(\boldsymbol{u}_{l,t+1}|\boldsymbol{s}_{t+1}\right)\right)\right]\right\}\label{eq:AugReward}
\end{equation}
Hence, the Bellman backup operation can be rewritten as
\begin{equation}
    \mathcal{T}^{\boldsymbol{\pi}}Q_{\boldsymbol{\pi}}\left(\boldsymbol{s}_t,\boldsymbol{u}_{l, t}\right)=\hat{R}_t+\gamma \mathbb{E}_{\boldsymbol{s}_{t+1},\boldsymbol{\pi}}\left[Q_{\boldsymbol{\pi}}\left(\boldsymbol{s}_{t+1},\boldsymbol{u}_{l,t+1}\right) \right] \label{eq:BellmanOpNew}
\end{equation}
With the choice of a finite coefficient $\alpha$,  the second term in (\ref{eq:AugReward}) is always bounded. According to (\ref{eq:BoundedReward}), there are two constants $\hat{R}_{min}$ and $\hat{R}_{max}$ such that $\hat{R}_{t}\in \left[\hat{R}_{min}, \hat{R}_{max}\right]$, and furthermore, $\vert \hat{R}_{t} \vert\leq \bar{R}$ with $\bar{R}=\max\left\{\vert\hat{R}_{min}\vert, \vert\hat{R}_{max}\vert\right\}$. In terms of (\ref{eq:V_Func}) and (\ref{eq: Action-Value Func}), one has $Q_{\boldsymbol{\pi}}\left(\boldsymbol{s}_t,\boldsymbol{u}_{l, t}\right)=\hat{R}_t+\gamma\sum_{t+1}^{\infty}\sum_{\boldsymbol{u}_{l,t+1}}\boldsymbol{\pi}\left(\boldsymbol{u}_{l,t+1}|\boldsymbol{s}_{t+1}\right)\sum_{\boldsymbol{s}_{t+1}}\mathcal{P}_{t+1|t}\hat{R}_{t+1} $, thus
\begin{equation}
\Vert Q_{\boldsymbol{\pi}}\left(\boldsymbol{s}_t,\boldsymbol{u}_{l, t}\right) \Vert_{\infty} \leq \frac{\bar{R}}{1-\gamma} 
\label{eq:MaxQinftyNorm}
\end{equation}
where  the  $\infty$-norm  in (\ref{eq:MaxQinftyNorm}) is defined to be $\Vert Q_{\boldsymbol{\pi}}\left(\boldsymbol{s},\boldsymbol{u}_{l}\right) \Vert_{\infty}=\max_{s, u_{l}} \vert Q_{\boldsymbol{\pi}}\left(\boldsymbol{s},\boldsymbol{u}_{l}\right) \vert$. Hence, the Q-value $Q_{\boldsymbol{\pi}}$ is bounded in $\infty$-norm based on the baseline control.  For two distinct Q values $Q_{\boldsymbol{\pi}}$ and $Q'_{\boldsymbol{\pi}}$, there exists
\begin{align}
&\Vert \mathcal{T}^{\boldsymbol{\pi}}Q_{\boldsymbol{\pi}}-\mathcal{T}^{\boldsymbol{\pi}}Q'_{\boldsymbol{\pi}} \Vert_\infty  =\Vert\hat{R}_t+\gamma \mathbb{E}_{\boldsymbol{s}_{t+1},\boldsymbol{\pi}}\left[Q_{\boldsymbol{\pi}}\left(\boldsymbol{s}_{t+1},\boldsymbol{u}_{l,t+1}\right) \right] \nonumber \\
&-\hat{R}_t-\gamma \mathbb{E}_{\boldsymbol{s}_{t+1},\boldsymbol{\pi}}\left[Q'_{\boldsymbol{\pi}}\left(\boldsymbol{s}_{t+1},\boldsymbol{u}_{l,t+1}\right)\right]   \Vert_\infty \nonumber \\
&\leq \gamma   \Vert Q_{\boldsymbol{\pi}}-Q'_{\boldsymbol{\pi}}\Vert_\infty
\end{align}
where $Q'_{\boldsymbol{\pi}}$ represents the Q-value function approximated at the last iteration, and $Q_{\boldsymbol{\pi}}$ is the Q-value function approximated at the current iteration. The Bellman backup operation (\ref{eq:BellmanOpNew}) is $\gamma$-contraction with $0\leq\gamma<1$. According to Banach's fixed-point theorem, $\mathcal{T}^{\boldsymbol{\pi}}$ possesses a unique fixed point. Hence, the sequence $Q^{k+1}\left(\boldsymbol{s},\boldsymbol{u}_l\right)$ will converge to the soft Q-function $Q^{\boldsymbol{\pi}}$ of the policy $\boldsymbol{\pi}$ as $k\to\infty$.
\end{proof}

\subsection{Proof of Lemma \ref{lem:Pi_Improve}} \label{app:Lemma_Pi_Improve}
\begin{proof}
Based on (\ref{eq:KL_pi}), we can  obtain 
\begin{align}
 &\mathbb{E}_{{\boldsymbol{\pi}}_{new}}\Big[\alpha\ln\left(\boldsymbol{\pi}_{new}\left(\boldsymbol{u}_{l,t}|\boldsymbol{s}_{t}\right)\right)-Q^{\boldsymbol{\pi}_{old}}\left(\boldsymbol{s}_{t}, \boldsymbol{u}_{l, t}\right)\Big]  \geq \nonumber \\
&\mathbb{E}_{{\boldsymbol{\pi}}_{old}}\Big[\alpha\ln\left(\boldsymbol{\pi}_{old}\left(\boldsymbol{u}_{l,t}|\boldsymbol{s}_{t}\right)\right)-Q^{\boldsymbol{\pi}_{old}}\left(\boldsymbol{s}_{t}, \boldsymbol{u}_{l, t}\right)\Big] \label{eq:NewPi_OldPi}
\end{align}
Let $V^{\pi}\left(\boldsymbol{s}_{t}\right)=\mathbb{E}_{\boldsymbol{\pi}}\left[Q_{\boldsymbol{\pi}}\left(\boldsymbol{s}_{t},\boldsymbol{u}_{l,t}\right) -\alpha\ln\left(\boldsymbol{\pi}\left(\boldsymbol{u}_{l,t}|\boldsymbol{s}_{t}\right)\right)\right]$. According to (\ref{eq:NewPi_OldPi}) and (\ref{eq:BellmanOp}), it yields
\begin{align}
 Q^{\boldsymbol{\pi}_{old}}\left(\boldsymbol{s}_{t}, \boldsymbol{u}_{l, t}\right)=& R_t+\gamma \mathbb{E}_{\boldsymbol{s}_{t+1}}\left[V^{{\pi}_{old}}\left(\boldsymbol{s}_{t}\right)\right] \nonumber\\
 \leq&  R_t+\gamma \mathbb{E}_{\boldsymbol{s}_{t+1}}\left[\mathbb{E}_{\boldsymbol{\pi}_{new}}\left[Q^{\boldsymbol{\pi}_{old}}\left(\boldsymbol{s}_{t+1}, \boldsymbol{u}_{l, t+1}\right)\right.\right.\nonumber\\
 &\left.\left.-\alpha\ln\left(\boldsymbol{\pi}_{new}\left(\boldsymbol{u}_{l,t+1}|\boldsymbol{s}_{t+1}\right)\right)\right]\right] \nonumber\\
\leq &R_t +\gamma \mathbb{E}_{\boldsymbol{s}_{t+1}}\Big[\mathbb{E}_{\boldsymbol{\pi}_{new}}\Big[R_{t+1}\Big.\Big.\nonumber\\
 &\Big.\Big.+\gamma \mathbb{E}_{\boldsymbol{s}_{t+2}}\left[\mathbb{E}_{\boldsymbol{\pi}_{new}}\left[Q^{\boldsymbol{\pi}_{old}}\left(\boldsymbol{s}_{t+2}, \boldsymbol{u}_{l, t+2}\right)\right.\right.\Big.\Big.\nonumber\\
 &\Big.\Big.\left.\left.-\alpha\ln\left(\boldsymbol{\pi}_{new}\left(\boldsymbol{u}_{l,t+2}|\boldsymbol{s}_{t+2}\right)\right)\right]\right]\Big.\Big.\nonumber\\
 &\Big.\Big.-\alpha\ln\left(\boldsymbol{\pi}_{new}\left(\boldsymbol{u}_{l,t+1}|\boldsymbol{s}_{t+1}\right)\right)\Big]\Big] \nonumber\\
 \vdots & \nonumber\\
 \leq &Q^{\boldsymbol{\pi}_{new}}\left(\boldsymbol{s}_{t}, \boldsymbol{u}_{l, t}\right)
 \end{align}
Hence, $Q^{\boldsymbol{\pi}_{new}}\left(\boldsymbol{s}_t,\boldsymbol{u}_{l,t}\right)\geq Q^{\boldsymbol{\pi}_{old}}\left(\boldsymbol{s}_t,\boldsymbol{u}_{l,t}\right)$ $\forall \boldsymbol{s}_t\in\mathcal{S}$ and $\forall \boldsymbol{u}_{l,t}\in\mathcal{U}$.
\end{proof}

\subsection{Proof of Theorem \ref{thm:Converge}} \label{app:Theorem_Convergence}

\begin{proof}
According to Lemma \ref{lem:Pi_Improve}, one has $Q^{\boldsymbol{\pi}^{i}}\left(\boldsymbol{s},\boldsymbol{u}_l\right)\geq Q^{\boldsymbol{\pi}^{i-1}}\left(\boldsymbol{s},\boldsymbol{u}_l\right)$, so $Q^{\boldsymbol{\pi}^{i}}\left(\boldsymbol{s},\boldsymbol{u}_l\right)$ is monotonically non-decreasing with respect to the policy iteration step $i$. In addition, $Q^{\boldsymbol{\pi}^{i}}\left(\boldsymbol{s},\boldsymbol{u}_l\right)$ is upper bounded according to the definition of the reward given in (\ref{eq:ASV_Reward}), so $Q^{\boldsymbol{\pi}^{i}}\left(\boldsymbol{s},\boldsymbol{u}_l\right)$ will converge to an upper limit $Q^{\boldsymbol{\pi}^{*}}\left(\boldsymbol{s},\boldsymbol{u}_l\right)$ with ${Q}^{\boldsymbol{\pi}^{*}}\left(\boldsymbol{s},\boldsymbol{u}_l\right) \geq Q^{\boldsymbol{\pi}^{i}}\left(\boldsymbol{s},\boldsymbol{u}_l\right)$ $\forall \boldsymbol{\pi}_i\in\Pi$,  $\forall \boldsymbol{s}\in\mathcal{S}$,  and $\forall \boldsymbol{u}_l\in\mathcal{U}$.
\end{proof}

\subsection{Proof of Theorem \ref{thm:Stab}} \label{app:Theorem_Stab_TrackCntrl}

\begin{proof}
In our proposed algorithm, we start the training/learning using the baseline control law $\boldsymbol{u}_b$. According to Lemma \ref{lem:Pi_Eval}, we are able to obtain the corresponding Q value function for the baseline control law $\boldsymbol{u}_b$. Let the Q value function be $ Q^{0}\left(\boldsymbol{s},\boldsymbol{u}_l^{0}\right)$ at the beginning of the iteration  where $\boldsymbol{u}_l^{0}$ is the initial RL-based control function. 
According to the definitions of the reward function in (\ref{eq:ASV_Reward}) and Q value function in (\ref{eq: Action-Value Func}), we can choose the Lyapunov function candidate as 
\begin{equation}
    \mathbb{V}^{0}\left(\boldsymbol{e}\right) = - Q^{0}\left(\boldsymbol{s},\boldsymbol{u}_l^{0}\right) \label{eq: Lyap}
\end{equation}
where $Q^{0}\left(\boldsymbol{s},\boldsymbol{u}_l^{0}\right)$ is the action value function of the initial control law $\boldsymbol{u}_l^{0}$. Note that the baseline control $\boldsymbol{u}_b$ is implicitly included in the state vector $\boldsymbol{s}$, as $\boldsymbol{s}$ consists of $\boldsymbol{x}$, $\boldsymbol{x}_m$, and $\boldsymbol{u}_b$ in this paper as discussed in Section \ref{sec:MR_DeepRL}. Hence, $\mathbb{V}\left(\boldsymbol{s}_{t}\right)$ in Assumption \ref{assump:BaselineC} is a  Lyapunov function for the closed-loop system of (\ref{eq:ASV_Dyn2}) with the baseline control $\boldsymbol{u}_b$.

Since ASVs have deterministic dynamics and exploration noises are not considered, we have $Q^0\left(\boldsymbol{s}_t,\boldsymbol{u}_{l,t}\right)=V^0\left(\boldsymbol{s}_t\right)$ and $Q^0\left(\boldsymbol{s}_t,\boldsymbol{u}_{l,t}\right)=R_t^0+\gamma Q^0\left(\boldsymbol{s}_{t+1},\boldsymbol{u}_{l,t+1}\right)$ where $R_t^0=R(\boldsymbol{s}_t,\boldsymbol{u}^{0}_{l,t})$. With the consideration of $\mathbb{V}^{0}\left(\boldsymbol{e}\right) = - Q^{0}\left(\boldsymbol{s},\boldsymbol{u}_l\right)$, there exists $\mathbb{V}^{0}\left(\boldsymbol{e}_t\right) = -R_t^0+\gamma \mathbb{V}^{0}\left(\boldsymbol{e}_{t+1}\right)$. 

If Assumption \ref{assump:BaselineC} holds, there exists $\mathbb{V}^{0}\left(\boldsymbol{e}_{t+1}\right)-\mathbb{V}^{0} \left(\boldsymbol{e}_{t}\right)\leq  - \mathbb{W}_1\left(\boldsymbol{e}_{t}\right)+\mu_3\left(\Vert\Delta\left(\boldsymbol{s}_t\right)\Vert_2\right)$  and $\mathbb{W}_1\left(\boldsymbol{e}_{t}\right) > \mu_3\left(\Vert\Delta\left(t\right)\Vert_2\right)$, $\forall \Vert\boldsymbol{e}_t\Vert_2 > c_\Delta$. Hence,
\begin{equation}
    \left(1-\gamma\right)\mathbb{V}^{0}\left(\boldsymbol{e}_{t+1}\right)+R_t^0\leq  - \mathbb{W}_1\left(\boldsymbol{e}_{t}\right)+\mu_3\left(\Vert\Delta\left(t\right)\Vert_2\right)\label{eq:Lyap_diff}
\end{equation}
In the policy improvement, the control law is updated by
\begin{equation}
   \boldsymbol{u}^{1}_l = \min_{\boldsymbol{\pi}} \left(-R_t^0+\gamma \mathbb{V}^{0}\left(\boldsymbol{e}_{t+1}\right)\right) \label{eq: iter1_min}
\end{equation}
Note that $\boldsymbol{u}_l$ is implicitly contained in both $R_t$ and  $\mathbb{V}^{0}\left(\boldsymbol{e}_{t+1}\right)$ according to (\ref{eq:ASV_Reward_1}) and (\ref{eq: Lyap}).  In the policy evaluation, the following update is conducted.
\begin{equation}
    \mathbb{V}^{1}\left(\boldsymbol{e}_{t}\right) = -R_t^1+\gamma \mathbb{V}^{0}\left(\boldsymbol{e}_{t+1}\right) \label{eq: iter1_V}
\end{equation}
where $R_t^1=R(\boldsymbol{s}_t,\boldsymbol{u}^{1}_{l,t})$. Hence, for $\boldsymbol{u}_l^{1}$, there exists
\begin{align*}
\mathbb{V}^{1}\left(\boldsymbol{e}_{t+1}\right)-\mathbb{V}^{1}\left(\boldsymbol{e}_{t}\right) =& \mathbb{V}^{1}\left(\boldsymbol{e}_{t+1}\right)+R_t^1-\gamma \mathbb{V}^{0}\left(\boldsymbol{e}_{t+1}\right) \nonumber \\
=& \mathbb{V}^{1}\left(\boldsymbol{e}_{t+1}\right)-\mathbb{V}^{0}\left(\boldsymbol{e}_{t+1}\right)+R_t^1 \nonumber\\
&-R_t^0+R_t^0+\left(1-\gamma\right)\mathbb{V}^{0}\left(\boldsymbol{e}_{t+1}\right) \nonumber \\
\leq&  - \mathbb{W}_1\left(\boldsymbol{e}_{t}\right)+\mu_3\left(\Vert\Delta\left(t\right)\Vert_2\right) +R_t^1\nonumber \\
& -R_t^0 + \mathbb{V}^{1}\left(\boldsymbol{e}_{t+1}\right)-\mathbb{V}^{0}\left(\boldsymbol{e}_{t+1}\right)
\end{align*}
According to (\ref{eq: iter1_min}) and (\ref{eq: iter1_V}), one has $\mathbb{V}^{1}\left(\boldsymbol{e}_{t+1}\right)\leq\mathbb{V}^{0}\left(\boldsymbol{e}_{t+1}\right)$ and $R_t^1\leq R_t^0$. Therefore, $\mathbb{V}^{0}\left(\boldsymbol{e}_{t+1}\right)-\mathbb{V}^{1}\left(\boldsymbol{e}_{t+1}\right)+R_t^0-R_t^1+\mathbb{W}_1\left(\boldsymbol{e}_{t}\right)\geq \mathbb{W}_1\left(\boldsymbol{e}_{t}\right)$.
As $\mathbb{W}_1\left(\boldsymbol{e}_{t}\right) > \mu_3\left(\Vert\Delta\left(t\right)\Vert_2\right)$, $\forall \Vert\boldsymbol{e}_t\Vert_2 > c_\Delta$, there must exist a new constant $c^1_\Delta\leq c_\Delta$ such that $\mathbb{V}^{0}\left(\boldsymbol{e}_{t+1}\right)-\mathbb{V}^{1}\left(\boldsymbol{e}_{t+1}\right)+R_t^0-R_t^1+\mathbb{W}_1\left(\boldsymbol{e}_{t}\right)>\mu_3\left(\Vert\Delta\left(t\right)\Vert_2\right)$, $\forall \Vert\boldsymbol{e}_t\Vert_2 > c^1_\Delta$. 

The new control law $\boldsymbol{u}^{1}_{l}$ can also ensure the closed-loop ASV system to be uniformally ultimately bounded. In the worst case, $c^1_\Delta=c_\Delta$, which implies that $\boldsymbol{u}^{1}_{l}$ will have the same control performance with $\boldsymbol{u}^{0}_{t}$, namely guaranteeing the same ultimate boundaries for the tracking errors. If there exists $c^1_\Delta>c_\Delta$, it implies that $\boldsymbol{u}^{1}_{l}$ will result in smaller tracking errors than $\boldsymbol{u}^{0}_{l}$. 

Following the same analysis,  we can show that $\boldsymbol{u}^{2}_l$ also stabilizes the ASV system (\ref{eq:ASV_Dyn2}) in terms of  $\mathbb{V}^{1}\left(\boldsymbol{s_{t}}\right)$ and replacing $\boldsymbol{u}^{0}_l$ in  (\ref{eq: iter1_min}) and (\ref{eq: iter1_V}) with $\boldsymbol{u}^{1}_l$. Repeating (\ref{eq: iter1_min}) and (\ref{eq: iter1_V}) for all $i=1$, $2$, $\ldots$, we can prove that all $\boldsymbol{u}^{i}_l$ can stabilize the ASV system (\ref{eq:ASV_Dyn2}), if Assumption \ref{assump:BaselineC} holds.  It implies that the  ASV system (\ref{eq:ASV_Dyn2}) will be stabilized by the overall control law $\boldsymbol{u}^i = \boldsymbol{u}_b + \boldsymbol{u}_{l}^i$. 
\end{proof}

\subsection{Simulation configurations}  \label{app:Sim_Config}

\begin{table}[h]
    \centering
    \caption{Model parameters} \vspace{-2mm} \label{tab:HydroTab}
    \begin{tabular}{cc|cc}
    \toprule
       Parameters  & Values & Parameters  & Values \\ \toprule
        $m$ & $23.8$ & $Y_{\dot{r}}$ & $-0.0$ \\
        $I_z$ & $1.76$ & $Y_{{r}}$ & $0.1079$ \\
        $x_g$ & $0.046$ & $Y_{\vert{v}\vert {r}}$ & $-0.845$ \\
        $X_{\dot{u}}$ & $-2.0$ & $Y_{\vert{r}\vert {r}}$ & $-3.45$ \\
        $X_{{u}}$ & $-0.7225$ & $N_{{v}}$ & $-0.1052$ \\
        $X_{\vert{u}\vert {u}}$ & $-1.3274$ & $N_{\vert{v}\vert {v}}$ & $5.0437$ \\
        $X_{{u}{u}{u}}$ & $-1.8664$  & $N_{\vert{r}\vert {v}}$ & $-0.13$ \\
        $Y_{\dot{v}}$ & $-10.0$ & $N_{\dot{r}}$ & $-1.0$ \\
        $Y_{{v}}$ & $-38.612$ & $N_{{r}}$ & $-1.9$ \\
        $Y_{\vert{v}\vert {v}}$ & $-36.2823$ & $N_{\vert{v}\vert {r}}$ & $0.08$ \\
        $Y_{\vert{r}\vert {v}}$ & $-0.805$ & $N_{\vert{r}\vert {r}}$ & $-0.75$  \\ \toprule
    \end{tabular}
\end{table}

\begin{table}[h]
    \centering
    \caption{Reinforcement learning configurations} \vspace{-2mm} \label{tab:RLTab}
    \begin{tabular}{cc}
    \toprule
      Parameters  & Values  \\ \toprule
        Learning rate $\iota_Q$ &  $0.001$ \\
        Learning rate   $\iota_\pi$ &  $0.0001$ \\
        Learning rate $\iota_\alpha$ &  $0.0001$ \\
        $\kappa$ & $0.01$ \\
        actor neural network & fully connected with two hidden layers \\
        &  (128 neurons per hidden layer) \\
        critic neural networks & fully connected with two hidden layers  \\
        &  (128 neurons per hidden layer) \\
        Replay memory capacity & $1\times 10^{6}$ \\
        Sample batch size &$128$ \\
        $\gamma$ & $0.998$ \\
        Training episodes & $1000$ \\
        Steps per episode & $1000$  \\
        time step size $\delta t$ & $0.1$  \\\toprule
    \end{tabular}
\end{table}

\bibliographystyle{IEEEtran}

\end{document}